\documentclass[11pt]{article}
\usepackage[]{graphicx}
\usepackage[]{epsfig}
\def\[[{\left[}
\def\]]{\right]}

\newcommand {\cto} {\right)}
\newcommand {\ato} {\left(}

\newcommand {\bi} {\bibitem}
\newcommand {\be} {\begin{equation}}
\newcommand {\bea} {\begin{eqnarray} \nonumber }
\newcommand {\ee} {\end{equation}}
\newcommand {\eea} {\end{eqnarray}}
 \newcommand {\eps} {\epsilon}
 \newcommand {\si} {\sigma}
\newcommand {\de} {\delta}

\newcommand {\ga} {\gamma}

 \newcommand {\al} {\alpha}
 
 \newcommand {\N} {{\cal N}}

\newcommand {\ba} {\overline}
\newcommand {\lan} {\langle}
\newcommand {\ran} {\rangle}

\newcommand {\cH}  {{\cal H}}

\newcommand {\cN}  {{\cal N}}

\newcommand {\Tr} {\mbox{Tr}}
\newcommand {\for} {\ \ \ \mbox{for}\ \ }
\newcommand {\sign} {\mbox{sign}}

\newcommand {\cp} {\right)}
\newcommand {\op} {\left(}
\def \(  {\left(}
\def \)  {\right(}

\def \form#1 {eq. (\ref{#1}) }
\def \parziale#1#2  {{\partial {#1} \over \partial {#2}}}

\def \bi{\bibitem}

 \def \(({\left(}
 \def \)){\right)}
\def \bi{\bibitem}

\def \b{\beta}

\def \z{\zeta}

\def \nn{\nonumber}
\def \beqna{\begin{eqnarray}}
\def \eeqna{\end{eqnarray}}
\def  \beq{\begin{equation}}
\def  \eeq{\end{equation}}
\def \x {{\bf  x}}
\def \y{{\bf  y}}
\def \z{{\bf  z}}
\def \ln{{\rm ln}}
\def \th{{\rm th}}
\def \ch{{\rm ch}}

\def \b{\beta}

\newcommand {\sig} {{\bf \sigma}}

\def \ln{{\rm ln}}
\def \ra{\rangle}
\def \ab2{\alpha\beta^2}
\def \la{\langle}

\begin{document}
\title{Glasses, replicas and all that\\
}
\author{Giorgio Parisi\\
Dipartimento di Fisica,  SMC and Udrm1 of INFM, INFN,\\
Universit\`a di Roma ``La Sapienza'',\\
Piazzale Aldo Moro 2,
I-00185 Rome (Italy)
}
\date{\empty}

\maketitle 
\begin{abstract} In these lectures I will review the approach to glasses based on the replica formalism.  
Many of the physical ideas are very similar to those of older approaches.  The replica approach has the advantage of 
describing in an unified setting both the behaviour near the dynamic transition (mode coupling transition) and near the 
equilibrium transition (Kauzman transition) that is present in fragile glasses.  The replica method may be used to solve 
simple mean field models, providing explicit examples of systems that may be studied analytically in great details and 
behave similarly to the experiments.  Finally, using the replica formalism, it is possible to do analytic explicit 
computations of the properties of realistic models of glasses and the results are in reasonable agreement with numerical 
simulations.
	\end{abstract}
	\newpage
	
	\tableofcontents 
\vfill
\section{Introduction}
\subsection{General considerations}
In these recent years many progresses in the study of glasses have been done using the replica formalism.  There are 
many indications that, if we would follow the evolution of a glass at a microscopical level, we would discover that at 
low temperatures the glass freezes in an equilibrium (or quasi equilibrium) configuration that is highly non-unique.  
This essential non-uniqueness of the ground state is present in many others systems where the energy landscape is highly 
corrugated: e.g. it is widely believed to be present in spin glasses, i.e. magnetic systems with a random mixture of 
 ferromagnetic and antiferromagnetic interactions \cite{EA,mpv,BOOK,CINQUE}.  This property is responsible of the peculiar 
behaviour of glassy systems and, at the same time, it is the most difficult to control theoretical.

The replica approach to spin glasses was developed in the seventies and it was shown to be the most effective and 
sophisticated tool to study the behavior of systems characterized by the presence of many equilibrium states.  The use 
of the same replica techniques to study not only spin glasses, but also glasses, was not immediate because there was a a 
strong phycological barrier to be crossed: disorder is present in the Hamiltonian for spin glasses, but it not present 
in the Hamiltonian for glasses and it was believed that the presence of disorder was one of the prerequisites for using 
the replica method.  However at a certain moment it became clear that disorder in the Hamiltonian was not necessary for 
using the replica techniques and the study of glasses using the replica method started to develop rather fast.

In the picture glasses may freeze in many microscopically different configurations when we decrease the temperature.  
This statement is common to many other approaches \cite{AdGibbs,St}, however the replica approach gives us a panoplia of 
sophisticated physical and mathematical tools that strongly increase our ability to describe, study and compute 
analytically the properties of glasses.

These tools have been used to compute analytically in a detailed way the properties of toy models for the glass 
transition.  Although theses models are very far from reality (the range of the forces is infinite) they display a very 
rich behaviour\cite{KTW}: for example there is a soluble mean field model without quenched disorder  where there is 
an equilibrium glass-liquid transition (Kauzman transition \cite{kauzman}), a dynamical transition\cite{CuKu,FM,pspin} (mode coupling 
transition(\cite{MCT}) and, at 
an higher temperature, a liquid-crystal transition that cannot be seen in the dynamic of the system (starting from the 
liquid phase) unless we cool the system extremely slowly\cite{MPR}.  The existence of these soluble models is very precious to us; 
they provide a testing ground of new physical ideas, concepts, approximation schemes that are eventually used in more 
realistic cases.

The aim of these lectures is to present an introduction to the replica tools, to describe some of the technical details 
and to stress the physical ideas.  The amount of work that has been done in the field is extremely large and here I have 
been forced to concentrate my attention only on a few points.

\subsection{Glassiness, metastability and hysteresis}
An essential feature of a glass system at the microscopic level is the existence of a corrugated free energy landscape.  
One may wonder which are the macroscopic counterparts of this property.  A very important consequence is the presence of 
metastability in an open region of parameter space, a new and unusual phenomenon that can be experimentally studied in a 
carefully way.
\begin{figure} 
\includegraphics[width=0.6\textwidth]{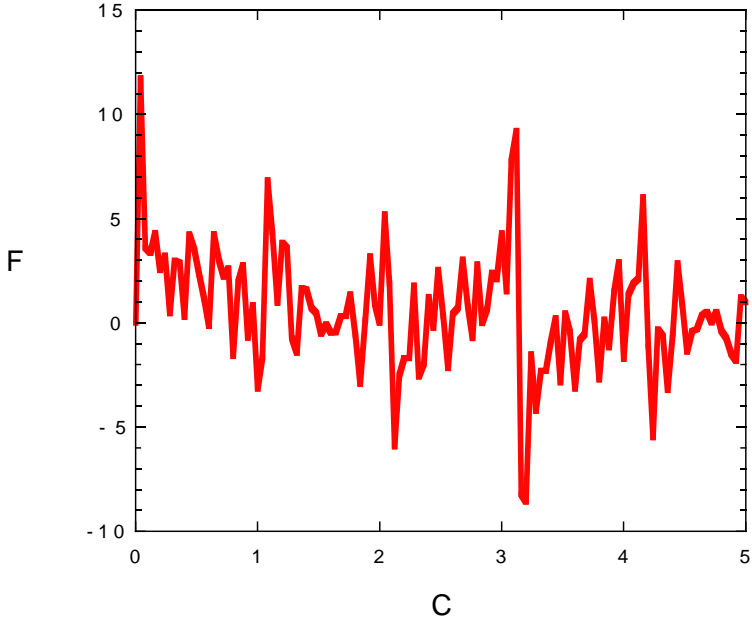}
\caption{An artistic view of the free energy of a system with corrugated free energy landscape as function of the 
configuration space.}
\label{CORRUGATED}
\end{figure}

Let us describe a non-glassy system where we have metastability.  The simplest case is  a system that 
undergoes a first order phase transition when we change a parameter. When the first order transition happens by
changing the temperature, if we cool the systems sufficiently slowly, the high temperature phase survives also below 
the 
critical temperature up to the spinodal temperature.

In order to present a familiar example I will consider a system where the control parameter is the magnetic field $h$: 
the simplest case is the ferromagnetic Ising model.  At low temperature the equilibrium magnetization $m(h)$ is given by 
$m(h)=m_{s}\ \sign(h)+O(h)$ for small $h$ ($m_{s}$ being the spontaneous magnetization): the magnetization changes 
discontinuously at $h=0$ in the low temperature phase where $m_{s} \ne 0$.

Let us consider a system that evolves with some kind of local dynamics.  If we slowly change the magnetic field from 
positive to negative $h$, we enter in a metastable region where the magnetization is positive, and the magnetic field is 
negative.  The system remains in this metastable state a quite large time, given by $\tau(h) \propto \exp( 
A/|h|^{\alpha})$, where $\alpha=d-1$ \cite{METAISI}.  When the observation time is of order of $\tau(h)$ the system 
suddenly jumps into the stable state.  This phenomenon is quite common: generally speaking we always enter into a 
metastable state when we cross a first order phase transition by changing some parameters.

If we start with the  state where $m>0$ at $h=0$ and we 
add a  {\em positive} magnetic field $h$ at time 0, the linear response susceptibility is  equal to
\begin{equation}
\chi_{LR}= {\lim_{t \to \infty}} {\partial\over\ \partial h} m(t,h),
\end{equation}
$ m(t,h)$ being the magnetization at time $t$. 
Using general arguments we can show that 
\begin{equation}
\beta^{-1} \chi_{LR}=\lim_{h\to 0^+}\sum_{i}\lan \si(i)\si(0)\ran^{c} \equiv 
\lim_{h\to 0^+}\sum_{i}(\lan \si(i)\si(0)\ran-\lan \si(i) \ran \lan \si(0)\ran).
\end{equation}

The linear response susceptibility is not equal to 
the equilibrium susceptibility that at $h$ exactly equal to zero  is infinite:
\begin{equation}
\chi_{eq}= {\partial\over \partial h} {\lim_{t \to \infty}}  m(t,h){\biggr |}_{h=0} m_{s}
= {\partial\over \partial h}\ \sign(h){\biggr |}_{h=0}=\infty.
\end{equation}
Indeed 
\be
\chi_{eq}(h)=\chi_{LR}(h)+ m_{s}\delta(h) \ ,
\ee
and the two susceptibilities $\chi_{eq}$ and $\chi_{LR}$ differs only at $h=0$.

This is the usual stuff that is described in books \cite{PARISISTAT}.  We claim that in glassy systems the situation is different.  For 
example in the case of glassy magnetic systems (e.g. spin glasses) there should be an open region in the space of 
parameters, 
where, if we change a parameter of the system (e.g. the magnetic field $h$) by an amount $\Delta h$, we have that 
$\chi_{LR}\ne \chi_{eq}$.  We expect that for $|h|<h_{c}(T)$ we stay in the glassy phase 
\footnote{The function $h_{c}(T)$ increases when we decrease the temperature; $h_{c}(T)$ vanishes at the critical point.}.

In this region
\bea
\Delta m(t) =\chi_{LR} \Delta h & \for &  1<<t<<\tau(\Delta h),\\
\Delta m(t) =\chi_{eq} \Delta h & \for &  \tau(\Delta h) << t,
\eea
where  $\tau(\Delta h)$ may have a power like behaviour
(e.g. $\tau(\Delta h) \propto |\Delta h|^{-4}$).

It is convenient to define the \emph{irreversible} susceptibility by 
\be
\chi_{eq}=\chi_{LR}+\chi_{irr} \ .
\ee
The glassy phase is thus characterized by a non-zero value of $\chi_{irr}$ \cite{mpv}.  If we observe the system for a time less 
that $\tau(\Delta h)$, the behaviour of the system at a given point of the parameter space depends on the previous story 
of the system and strong hysteresis effects are present.  I stress that, using  the previous definitions, hysteresis 
and history dependence do not necessary imply glassiness.  Hysteresis may be present if the time scale for approaching
equilibrium is very large (larger than the experimental time), but \emph{finite}.  Glassiness implies an equilibration 
time that is \textit{arbitrarily large}.  In other words hysteresis can be explained in terms of \emph{finite} free energy barriers 
that may involve only a few degrees of freedom; glassiness implies the existence of \emph{arbitrarily large} barriers 
that may arise only as a \emph{collective} effect of many degrees of freedom.

From my point of view the aim of the theoretical study of glasses  is to get a 
theoretical understanding of these effects  and to arrive to a qualitative and quantitative control
of these systems.

In these lecture I will address to various aspects of disorder systems using the replica approach.  The replica 
formalism is a language for studying different properties of disordered systems: as any other language has some 
advantages and some disadvantages.  The replica language has the advantage of being very compact and of putting the 
dirty under the carpet (thais may be also a disadvantage in some situations).  In other words you can do computations in 
a very simple and effective manner, but sometimes it hard to understand  the meaning of what 
are you doing.  The formalism of replicas and replica symmetry breaking seems to be the most adequate to discuss from a 
thermodynamic point of view the situation where $\chi_{irr}\ne 0 $.
   
The main physical problem I would like to understand is the characterization of the low temperature phase in structural 
glasses and in spin glasses.  As you will shall see in these lectures and in Cugliandolo's lectures, the main 
characteristic is the presence of aging for the response\cite{POLI,B,BCKM}.  Let us consider an aging experiment 
where the system is cooled at at time 0.  

The response function $R(t,t_{w})$ is the variation of the 
magnetization when we add an infinitesimal field at time $t_{w}$.  Aging implies that the function $R(t,t_{w})$ is not a 
constant in the region where $t$ and $t_{w}$ are \emph{both} large; indeed we have
\bea
R(t,t_{w})= R_{S} \for t<<t_{w} \\
R(t,t_{w})= R_{E} \for t>>t_{w}
\eea
By definition  $R_{S}=\chi_{LR}$ and the identification of $R_{E}$ with $\chi_{eq}$ follows from general 
arguments that will discussed in Cugliandolo's lectures.

As we shall see later, replica symmetry is broken as soon $R_{S}\ne R_{E}$; we will assume that this happens in many 
glassy systems.  This is what is experimentally seen in experiments done by humans using a value of $t_{w}$ that is much 
shorter of their life time.  These experiments are done for times that are much larger (15 or 20 order of magnitudo) 
that the microscopic time and many people do believe that aging in the response, i.e. $R_{S}\ne R_{E}$, survive in the 
limit where the waiting time goes to infinity.

One could also take the opposite point of view that aging is an artefact coming from doing the experiments at too short 
times and that aging would fades in the limits $t_{w}\to \infty$.  This would be a quite different interpretation of the 
experiments; in these lecture I will stick to the hypothesis that aging is present and that $R_{S}$ remains different 
from $R_{E}$ also in the infinite time limit.  Of course one could use experiments to decide which of the two hypothesis 
is correct, however this must be done by extrapolating the experimental data and the discussion would be too long to be 
presented here: moreover, we should remember that quite often, due to practical limitations, the experiments, like the 
God whose siege is at Delphi, neither say nor hide, they hint.

\subsection{The organization of these lectures}

These lectures are organized as follows. 

In section II I will present the simplest model of a glassy system: the random energy model (REM); I will describe the 
analytic solution of this model using both a direct approach and the replica formalism.

In the next section I will study models with correlated disorder (e.g. $p$-spin models); the analytic solution of these 
models is much more complicated than that of the REM and it can be done using the replica formalism; it this framework 
the meaning of spontaneously broken replica symmetry is elucidated.

In section IV I will introduce a key concept of this (and of others) approaches: complexity (aka configurational entropy).  A 
particular attention is given to the definition of this quantity: there is an intrinsic, albeit small, 
ambiguity in the value of the complexity that is present in short range model, but is absent in infinite range mean 
field models (if this ambiguity is neglected contradictory results may be obtained).  Various methods, both analytic and 
numeric, are introduced to compute this fundamental quantity.

In the next section I will present some general structural properties: equilibrium stochastic stability and its 
dynamic extension that can be used to prove  the generalized fluctuation dissipation relations that are described in 
more details in Cugliandolo's lectures.

In section VI I will present a very short introduction to the physics of structural glasses.

In the next section I will describe an analytic approach that can be used to do explicit first principles computations 
of the properties of glasses in the low energy phase.  Different approximation techniques are illustrated.

In section VIII I will present the results obtained using the techniques of the previous section for computing the 
properties of systems of interacting particles with a realistic potential. These analytic results  compare in a 
favorable way to 
the  numerical simulations.

Finally in the next section one finds a discussion  of the open problems.
Four appendices dedicated to technical problems close these lectures.

\section{The random energy model}

\subsection{The definition of the model}

The random energy model \cite{NICOLA,REM} is the simplest model for glassy systems.  It has various advantages: it is rather 
simple (its properties may be well understood with intuitive arguments, that may become fully rigorous) and it displays 
very interesting and new phenomena.

The random energy model is defined as following.  There are $N$ Ising spins ($\sigma_i$, 
$i=1,N$) that may take values $\pm1$; the total number of configurations is equal to 
$M\equiv 2^N$: each configuration  can be identified by a label $s$ in the interval $1-M$.

Generally speaking the Hamiltonian of the system is determined when we know all the values of the energies $E_s$, i.e. a 
value for each of the $M$ configurations of the system.  In many models there is an explicit expression for the 
energies 
as function of the configurations; on the contrary here the values of the $E_s$ are random, with a 
probability distribution $p(E)$ that is Gaussian:
\begin{equation}
p(E) \propto \exp \op-{E^2 \over 2 N} \cp \  .\label{PRO}
\end{equation}
The energies are random uncorrelated Gaussian variables with zero average and 
variance equal to $N$.

The partition function is simply given by
\begin{eqnarray}
Z_I=\sum_{s=1,M} \exp (-\beta E_s) = 2^{N}\int \rho_I(E) \exp (-\beta E),\\
\rho_I(E)\equiv 2^{-N}\sum_{s=1,M} \de(E-E_s). \nonumber
\end{eqnarray}
   
The values of the partition function and of the free energy density $(f_I=-\ln(Z_I)/(N\beta )$) depend on the instance 
$I$ of the system, i.e. by all the values of the energies $E_s$.  It can be proved that when $N 
\to \infty$ the dependance on $I$ of $f_I$ disappears with probability 1.  As usual, the most likely value of 
$f_I$ coincides with the average of $f_I$, where the average is done respect to all the possible instances of the system.  

In the following I will indicate with a bar the average over the instances of the system:
\begin{equation}
\ba{A_{I}} \equiv \int d\mu(I) A(I)\ ,
\end{equation}
where  $\mu(I)$ is the probability distribution of the instances of the system.
Using this notation we would like to compute 
\begin{equation}
f=\ba{f_{I}}
\end{equation}
and to prove that
\begin{equation}
\ba{(f_{I} -f)^{2}}\to_{N\to\infty} 0
\end{equation}

The model is enough simple to be studied in great details; exact expressions can be derived also for finite $N$.  In the 
next section we will give the main results without entering too much into the details of the computation.

\subsection{Equilibrium properties of the model}
The first quantity that we can compute is $\ba{Z_{I}}$. We have immediately
\begin{equation}
\ba{Z_{I}}= \sum_{s} \ba{\exp(-\beta E_{s})} = 2^{N} \exp(\frac{N}{2} \beta ^{2}) \ .  
\end{equation}
However the physically interesting quantity  is $\ba{\ln(Z_I)}$, that is related to the average of the free 
energy:
\begin{equation}
-N \beta f = \ba{\ln(Z_I)}
\end{equation}
If we make to bold assumption that
\begin{equation}
\ba{\ln(Z_I)}= \ln \op \ba{Z_I}\cp \label {BOLD}
\end{equation}
(this approximation is usually called the annealed approximation), we find that 
\begin{equation}
\ba{\ln(Z_I)}= N \((\ln(2)+\frac12 \beta^{2}\)) \ . \label{REPSYM}
\end{equation}
Generally speaking the relation eq.  (\ref{BOLD}) is not justified.  In this model it gives the correct results in the 
high temperature phase, but it fails in the low temperature phase.  For example, as we shall see later, at low 
temperature we must have that the quantity
\begin{equation}
G=-{\partial \over \partial \beta}\ln \op \ba{Z_I} \cp=
{\int d\mu(I) Z_{I} \lan E \ran_{I} \over \int d\mu(I) Z_{I}}
 \end{equation}
is different from the physically interesting quantity
\begin{equation}
E=-{\partial \over \partial \beta}\ba {\ln \op Z_I \cp} =\int d\mu(I)  \lan E \ran_{I}  \ ,
\end{equation}
where $\lan E \ran_{I}$ is the expectation of the energy in the instance $I$ of the 
problem:
\begin{equation}
\lan E \ran_{I}= {\sum_{s}\exp(-\beta E_{s}) E_{s}) \over \sum_{s}\exp(-\beta E_{s}) }
\end{equation}
In the computation of $G$ we give to the different systems a weight equal to the partition function $Z_{I}$ 
that is exponentially large (or small) and strongly fluctuates from system to system.  The correct results $E$ is 
obtained using the correct (flat) weight $d\mu(I)$.

One can argue (and it will be done later on in section \ref{LATERON}) that the previous derivation gives the correct 
form for the free energy in the low $\beta$ region.  However the previous results cannot be correct 
everywhere.  If we identify $G$ with $E$ and we compute the energy density  of the model  using (eq. \ref{REPSYM}), we find
\begin{equation}
e= \beta
\end{equation}
that superficially is not contradictory.  However, if we use this expression for the energy to compute the 
entropy density, we find
\begin{equation}
S(\beta)=-\beta^{2}{\partial f(\beta) \over  \partial \beta} =\ln(2) -\frac{1}{2\beta^{2}}
\end{equation}
This entropy density becomes negative for 
\begin{equation}
\beta>\beta_{c}=\sqrt{2 \ln(2)} \ .
\end{equation}
This is a major inconsistency in a model where the configuration space is discrete 
and the entropy must be a non-negative quantity.

The arguments we have used in deriving \form{REPSYM} are not solid.  We have computed the asymptotic 
behaviour of $Z_{n}$ for positive integer $n$ when the volume goes to infinity, but we cannot argue that the 
asymptotic behaviour is an analytic function of $n$.  A simple counterexample is  given by
\begin{equation}
\sinh(N n) \approx \exp(N |n|)
\end{equation}
The non-analyticity of the free energy density as functions of a parameter in the infinite volume limit is the essence 
of phase transitions.  It is quite possible that at low temperatures a phase transition in $n$ is present in the 
interval $0\le n \le 1$, so that the extrapolation of the free energy density from $n\ge 1$ to $n =0$ has nothing to do 
with the value of the free energy computed directly at $n=0$.

\begin{figure} 
    \includegraphics[width=.5\textwidth]{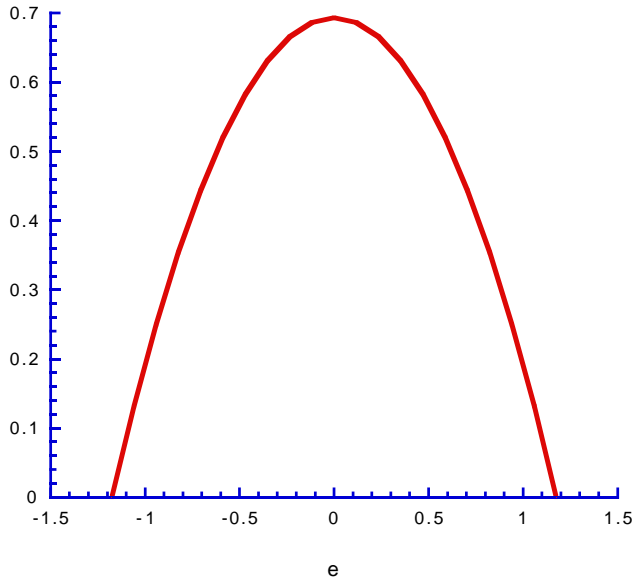}
    \caption{The quantity $\ln(N_0(e))/N$ as function of the energy density $e$ outside the 
    interval the function  $\N_0(e)$ is zero.
    }\label{N0}
    \end{figure}

There is a simple way to find the correct result \cite{NICOLA,REM}.   The probability of finding a configuration of energy in the interval 
$[E,E+dE]$ is given by
\begin{equation} 
\N_0(E) dE \ , 
\end{equation} 
where
\begin{equation}
\N_0(E)\equiv 2^N \exp \op -{E^2\over 2 N} \cp = \exp (N(\ln(2)- \frac12 e^2)),
\end{equation}
and $e\equiv E/N$ is the energy density.  In the case of a generic system (with probability 1 when $N \to \infty$) no 
configurations are present in the region where $\N_0(E)<<1$, i.e. for
\begin{equation} e^2<e_c^2\equiv 2 \ln(2).  \end{equation} 
Neglecting prefactors, the average total number of configurations of energy less that $e$ (when $e<0$) is given by

\begin{equation}
\int_{-\infty}^{Ne} dE\N_0(E) \approx \exp (N(\ln(2)- \frac12 e^2)) \ .
\end{equation}
This number becomes exponentially small as soon  $e<e_{c}$, implying that for a generic system there are no 
configurations for $e<e_{c}$.
We can thus write for a generic instance of the system the following relation: 
\begin{equation} \rho(E)\approx 
\N(E)\equiv \N_0(E)\theta(E_c^2-E^2).
\end{equation}
 The partition function can be written 
as 
\begin{equation} \int \N(E) \exp (-\beta E) .
\end{equation} 
Evaluating the integral with the saddle point method we find that in the high temperature region, i.e. for $ 
\beta<\beta_c\equiv e_c^{-1} $, the internal energy density is simple given by $-\beta$.  This behaviour must end 
somewhere because we know that the energy is bounded from below by $-e_{c}$.  In the low 
temperature region the integral is dominated by the boundary region $E\approx E_c$ and the energy 
density 
is exactly given by $-e_c$.
 \begin{figure}
     \includegraphics[width=.5\textwidth]{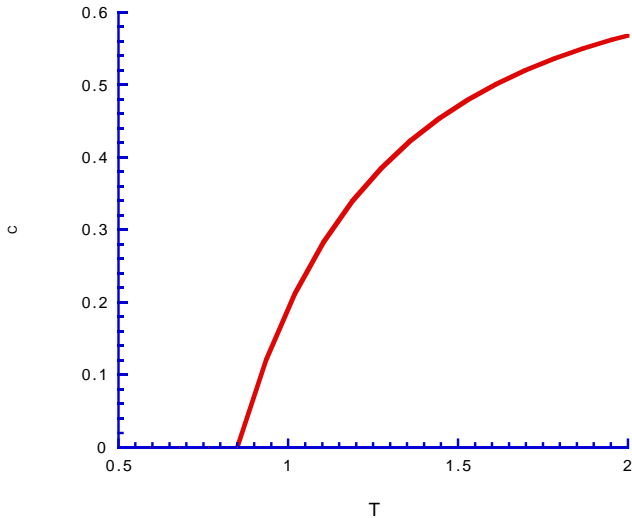}
   \caption{The entropy density as function of the temperature.
    }\label{S}
\end{figure}

The point where the high temperature behaviour breaks down is exactly the point where the entropy becomes negative.  
The 
entropy density is positive in the high temperature region, vanishes at $\beta_c$ and remains zero in the low 
temperature region.  In the high temperature region an exponentially large number of configurations 
contributes to the partition function, while in the low temperature region the probability is 
concentrated on a finite number of configurations \cite{REM}.

It is interesting to notice that in the low temperature phase the free energy is higher than the continuation from high 
temperature.  This is the opposite of the usual situation: in conventional mean field for non-disordered systems (e.g. 
ferromagnets) the low temperature free energy is lower than the continuation from high temperature \footnote{If two 
branches of the free energy are present usually the relevant one has the lower free energy.  We shall see later that 
this is not the case in glassy systems.}.  This correspond to a jump \emph{upward} of the specific heat when we decrease the 
temperature, while in the REM (and in real glasses) there is jump \emph{downward} of the specific heat.  The system has less 
configurations of what can be inferred from the behaviour at entropy at high temperature: some part of the predicted 
phase space is missing and consequently the free energy is higher than the analytic continuation of the high temperature 
results. This entropy crisis is the essence of the Kauzman transition\cite{kauzman}.

A more precise (and less hand-waving) computation can be done by using the representation
\begin{equation}
\ln(Z)=\int_{0}^{\infty}\frac{\exp(-t) -\exp(-tZ)}{t} 
\end{equation}
and writing
\begin{equation}
\ba{\exp(-tZ)}=\sum_{n=0,\infty}\frac{ (-tZ^{n})}{n!} \ .
\end{equation}
The quantities $\ba{Z^{n}}$ can be computed exactly and in this way we can also estimate the finite $N$ 
corrections to the asymptotic behaviour. The details of this computation can be found in  \cite{REM}.

\subsection{The properties of the low temperature phase}

It is worthwhile to study the structure of the configurations that mostly contribute to the partition 
function in the lower temperature phase.  The detailed computation of the finite $N$ corrections \cite{REM} shows that in 
the low temperature phase the average total entropy has a value that remains finite in the limit $N \to \infty$:
\begin{equation}
S_{T}=\Gamma'(1) -{\Gamma'(1-m)\over \Gamma(1-m)}
\end{equation}
where the quantity $m$ is given by
\begin{equation}
m=\beta_{c}/\beta=T/T_{c} \ .
\end{equation}
The parameter $m$ is always less the one in the whole temperature phase.

The finiteness of the total entropy implies that the probability distribution is concentrated 
on a finite number of configurations (see eq. \ref{FINITE}) also in the limit $N \to 
\infty$.  In order to study the contributions of the different configurations to the free energy, it is useful 
to sort the configurations with ascending energy.  We relabel the configurations and we 
introduce new labels such that $E_k<E_i$ for $k<i$.

  \begin{figure}
      \includegraphics[width=.6\textwidth]{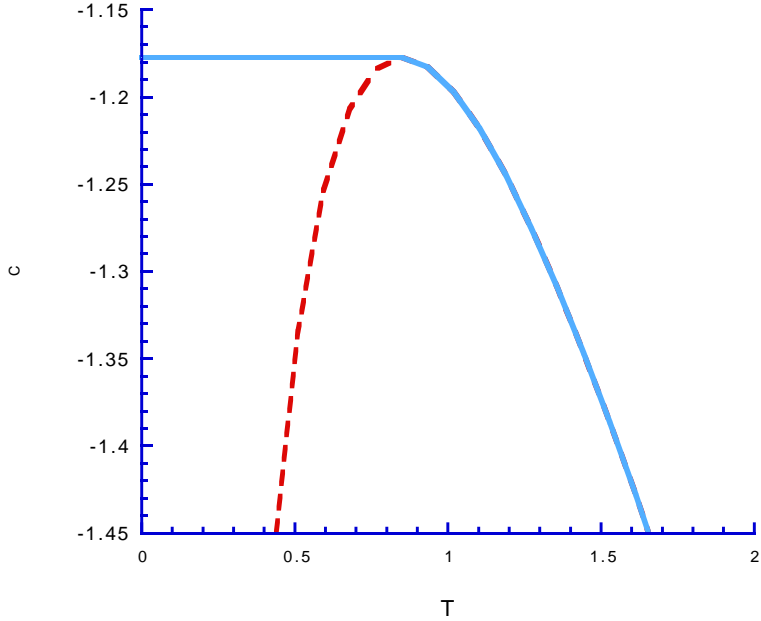}
   \caption{The free energy density as function of the temperature and the analytic 
    continuation of the high temperature result (dashed)
    }\label{F}
\end{figure} 

It is convenient to introduce the probability $w_{k}$ that the system is in the configuration $k$;
\begin{equation}
w_k\equiv {\exp(-\beta E_k) \over Z}\ .
\end{equation}
It is obvious that the $w_{k}$ form a decreasing sequence and that
\begin{equation}
\sum_{k=1,2^N}w_k =1 \ .
\end{equation}
Here the probability of finding an energy $E_{k}$ in the interval $[E,E+dE]$, not far from the ground state energy 
$E_{1}$ is given by
\begin{equation}
 \rho(E)\approx \exp (\beta _{c}(E-E^{*}))=\exp (\beta  m (E-E^{*})) . \label{WDIS}
\end{equation}
where $E^{*}$ is a reference energy not far from the ground state (the parameter $m$ controls the local exponential growth of 
the density of configurations).  It is not difficult to prove \cite{FPGS} that,  if 
we define
\begin{equation} 
\Delta= \beta_{c} (E_{1} -E^{*}) \ ,
\end{equation}
the probability distribution  of $\Delta$ is given by the Gumbel law:
\begin{equation}
P(\Delta)=\exp(\Delta -\exp(\Delta)).
\end{equation} 
It is somewhat more difficult \footnote{The proof sketched in one of the appendices.} to prove that probability of 
finding one of the variables $w$ in the interval $[w,w+dw]$ is given
\bea
\nu(w)  = C w^{-1-m}(1-w)^{-1+m} dw \ , \label{QUELLA}\\
C = {1 \over \Gamma(m) \Gamma(1-m)} \ .
\eea
where the proportionality factor can be easily found using the condition 
\begin{equation}
\int _{0}^{1} dw \nu(w) \  w =1 \ .
\end{equation}

Now it easy to compute
\begin{equation}
\int dw \nu(w) \ w^{k}= {\Gamma(k-m)\over \Gamma(1-m) \Gamma(k)} \ .
\end{equation}
As particular cases we get
\begin{equation}
\int dw \nu(w) w^{2} =1-m, \ \ \ \ \ \int\nu(w) w^{3}={(1-m)(2-m)\over 2} \ .
\end{equation}
We can check that the total entropy is given by
\begin{equation}
S_{T}=-\int dw \nu(w)  w \ln(w) \ , 
\end{equation} 
The integral
\begin{equation}
\int _{0}^{1} dw  \nu(w)
\end{equation}
is divergent, signaling that there are an infinite number of $w$'s.  

A detailed computation \cite{REM,mpv,parisibook2} shows that a finite number of terms dominates the sum over $k$.  
Indeed using the esplicite form of $\nu(w)$ near $w=0$ we get that
\begin{equation}
\sum_{k=1,L}  w_k =1-O(L^{-\lambda}), \ \ \ \lambda={1-m \over m}\ . \label{FINITE}
\end{equation}

In this model everything is clear: in the high temperature region the number of relevant configurations is infinite (as 
usual) and there is a transition to a low temperature region where only few configuration dominates.

The fact that the low temperature probability distribution is dominated by a few configurations can be seen also in the 
following way.  
We introduce a distance among two configurations $\al$ and $\ga$ as
\begin{equation}
d^2(\al,\ga) \equiv {\sum_{i=1,N} (\si^\al_i -\si^\ga_i)^2 \over 2 
n}.\label{DISTANZA}
\end{equation}
We also  introduce the overlap $q$ defined as
\begin{equation}
q(\al,\ga)\equiv {\sum_{i=1,N} \si^\al_i \si^\ga_i \over 2 N}=1-d^2(\al,\ga).
\end{equation}
The distance squared is normalized in such a way that it spans the interval 
$0-2$. It is equal to
\begin{itemize}
\item
 0, if the two configuration are equal ($q=1$).
\item
 1, if the configuration are orthogonal ($q=0$).
\item
 2, if $\si^\al_i=-\si^\ga_i$ ($q=-1$).
\end{itemize}
We now consider the function $Q(d)$ and $P(q)$, i.e. the probability that two equilibrium configurations are at distance 
$d$ or overlap $q$ respectively.  We find
\begin{itemize} 
\item For $T>T_c$ 
\begin{equation} 
Q(d)=\delta(d-1), \ \ \ P(q)=\de(q).  
\end{equation} 
\item For $T<T_c$ 
\begin{equation} Q(d)=(1-A) 
\delta(d-1)+A\delta(d), \ \ \ P(q)=(1-A) \delta(q)+A\delta(q-1).  
\end{equation}
where for each system ($I$) $A_{I}$ is equal to $\sum _{k=1,2^N}w_k^2$.  The average of $A_{I}$ over 
the different realizations of system is equal to $1-m$.
\end{itemize}

As soon as we enter in the low temperature region, the probability of finding two equal configurations is not zero.  
The 
transition is quite strange from the thermodynamic point of view.
\begin{itemize}
\item
It looks like a {\sl second} order transition because there is no latent heat.  It is characterized by a jump in the 
specific heat (that decreases going toward low temperatures).
\item
It looks like a {\sl first} order transition.  There are no divergent susceptibilities coming from above or from below 
(in short range models this result implies that there is no divergent correlation length).  Moreover the minimum value 
of $d$ among two equilibrium configurations jumps discontinuously (from 1 to 0).

\item If we consider a system composed by two replicas ($\si^1$ and $\si^2$)  
\cite{KPVI} and 
we write the Hamiltonian
\begin{equation}
H(\si^1,\si^2)=H(\si^1)+H(\si^2)+N \eps d^2(\si_1,\si^2)
\end{equation}
for $\eps=0$ the free energy is equal to that of the previous model (apart a factor 2) but we find a real first order 
thermodynamic transition, with a discontinuity in the internal energy, as soon as $\eps>0$.  The case $\eps=0$ is thus 
the limiting case of a \emph{bona fide} first order transitions.
\end{itemize}

These strange characteristics can be summarized by saying that the transition is of order one and a half, because it 
shares
some characteristics with both the first order and the second order transitions.
 
It impressive to note that the thermodynamic behaviour of real glasses near $T_c$ is very similar to the order one and a
half transition of REM. We will see later that this behaviour is also typical of the mean field approximation to glassy 
systems.
\subsection{A careful analysis of the high temperature phase} \label{LATERON}
Let us now try to justify more carefully the correctness of \form{BOLD} in the high temperature phase\cite{REM}.

We can firstly compute the quantity 
\begin{equation}
Z_{2} \equiv \ba{Z_{I}^{2}}=\ba{ \sum_{s,s'}\exp (-\beta E_{s}-\beta E_{s'})}
\end{equation}
If we consider separately the terms with $s=s'$ and those with $s\ne s'$, we easily get
\begin{equation}
Z_{2} =2^{N}(2^{N}-1) \exp( 2 A) +2^{N}\exp( 4 A) 
\approx 2^{2N} \exp( 2 A) +2^{N}\exp( 4 A)  \ ,
\end{equation}
where we have used the relation 
\begin{equation}
\int p(E) dE \exp ( m \beta E)= \exp(m^{2} A) \ ,
\end{equation}
with 
\begin{equation}
A=\frac{N}{2} \beta^{2}
\end{equation}
In a similar way we can compute $\ba{Z_{I}^{3}}$: we neglect terms that are proportional to $2^{-N}$ and we separate 
terms with all indices equal, terms with all the indices different and terms with two indices equal and one different; 
finally we find that
\begin{equation}
Z_{3} \equiv \ba{Z_{I}^{3}}
\approx 2^{3N} \exp( 3 A) +3 \ 2^{2N}\exp( (4+1) A)+ 2^{N}\exp(9 A) \ .
\end{equation}
Generally speaking, using the multifactorial formula and taking the leading contribution 
for each term
\begin{equation}
Z_{n}\equiv\ba{Z_{I}^{n}} = \sum_{l} \sum_{\{ m_i\}}\frac{n!}{ l! \prod_{i=1}^l m_i!} 
2^{l N} \exp( m_{i}^{2} A)\ , \label{MULTI}
\label{zk}
\end{equation}
where the constraint  $\sum_{i=1}^l m_i =n$ is satisfied. The previous formula can be 
derived by writing
\begin{equation}
Z_{I}^{n}=\sum_{s_{1}\ldots s_{n}} \exp(\beta \sum_{a=1,n} E_{s_{a}}) \ .
\end{equation}
We  now divide the $n$ indices in $l$ sets of size $m_{i}$ and estimate the 
contribution from each set, that is given by
\begin{equation}
\ba{Z_{I}^{m_{i}}}=2^{N}\exp(m_{i}^{2}A) \ .
\end{equation}
The final results \form{MULTI} is obtained by multiplying the contribution of each set and adding the a prefactor (i.e. 
$n!/ (l!  \prod_{i=1,l}m_i!)$) that has a combinatorial origine.  It is a simple exercise to check that the previous 
formula reduce to the one that we have written for $n=1,2,3$.

The previous formula can also written as
\begin{equation}
Z_{n}=
\sum_{l} \sum_{\{ m_i\}}\frac{n!}{ l! \prod_{i=1}^l m_i!} \ ,
\exp(N\sum_{i=1,l}G(m_{i}))
\end{equation}
where 
\begin{equation}
G(m_{i})= \ln(2)+A m_{i}^{2} \ .
\end{equation}

For small $\beta$ and for $n$ not too large the term $l=n$ and $m_{i}=1$  $\forall i$ is dominating: it 
gives that
\begin{equation}
Z_{n} \approx \exp (N n (\ln2+A)) \ .
\end{equation}
We can now use  the relation
\begin{equation}
\ba{\ln (Z) }= \lim_{n \to 0 } {\ln \left( \ba{Z^{n}}\right)\over n}
\end{equation} 
and  apply the previous formula also for non-integer $n$ up to $n=0$. In this way one finds 
that
\begin{equation}
{\ba{\ln (Z) }\over N}= \ln2+A=\ln2+\frac12 \beta^{2}\ ,
\end{equation}
and therefore 
\begin{equation}
f(\beta)=-{\ln(2)\over \beta}-\frac12 \beta
\end{equation}

The same results can be obtained {\sl manually} by writing
\begin{equation}
Z= 2^{N} -\beta \sum_{s}E_{s} +\frac{\beta^{2}}{2}\sum_{s}E_{s}^{2}+\ldots
\end{equation}
This gives 
\begin{equation}
\ln (Z) = N \ln2 -\beta {\sum_{s}E_{s}\over 2^{N}}-\frac12\beta^{2} \left( {\sum_{s} E_{s}\over 2^{N}} \right)^{2}+
\frac12\beta^{2} {\sum_{s}E_{s}^{2}\over 2^{N}} +\ldots
\end{equation}
Only the first and the last term contribute in the limit $N\to \infty$ (the energies have zero average and their number 
is $2^{N}$); a carefully evaluations shows that the remaining terms give a vanishing contribution.

\subsection{The replica method}
We may wonder if the previous results in the low energy phase may be obtained by starting from \form{zk} and by
partitioning the replicas using an appropriate choice of the variables $m_{i}$ that indicates the size of the different 
sets (or blocks).

The correct choice in the high temperature region was $m_{i}=1$.  An alternative choice (that is highly symmetric) is 
$m_{i}=m$.  The total number of partitions $l$ is given by $n/m$.  This saddle point  gives a contribution\cite{REM,mpv}
\begin{equation}
Z_{n}\approx \exp (Nn(\frac{\ln(2)}{m}+m A)) \ ,
\end{equation}
that correspond to 
\begin{equation}
-\beta \ba{f} =F(m)\equiv{G(m)\over m}= \frac{\ln(2)}{m}+\frac12\beta^{2}m  \ .
\end{equation}
The value of $m$ is arbitrary.  It is surprising that in the low temperature phase the correct result for the free 
energy is obtained by taking the maximum \footnote{If the correct results are given by the saddle point where the 
replicas are partitioned in equal size blocks (with $m \ne 1$) we will say that the replica broken at one step.} of 
$F(m)$ with respect to $m$, and the maximum is found in the interval $0-1$ (in the high temperature phase the correct 
result is given $F(1)$).

This prescription can looks rather strange, but we can  test some consequences of this approach for other 
quantities.  At this end we take two replicas of the same system and we compute the probability that these two 
configurations are the same using this approach and we compare with the known results.

Before going on we have to make a general remark.  The average over the samples of the statistical averages can be 
simply obtained by computing the average over an ensemble of $n$ replicas of the same system and taking the limit $n 
\to 
0 $ at the end.  The precise statement is the following:
\bea
\ba{\lan A(\sigma)\ran} \equiv \ba{\left( {\sum_{\sigma}A(\sigma) \exp (-\beta H(\sigma))\over
\sum_{\sigma} \exp (-\beta H(\sigma))} \right)}= \\
\lim_{n\to 0}
{\ba{\sum_{\sigma_{1}\ldots\sigma_{n}}A(\sigma_{1}) \exp (-\beta \sum_{a=1,n}H(\sigma_{a}))}\over \ba{
\sum_{\sigma_{1}\ldots\sigma_{n}}\exp (-\beta \sum_{a=1,n}H(\sigma_{a}))}} \ .
\eea
Indeed the last term is equal to 
\be
{ \ba{\sum_{\sigma}A(\sigma) \exp (-\beta H(\sigma))
\left(\sum_{\sigma} \exp (-\beta H(\sigma))\right)^{n-1}}
\over 
\ba{\left(\sum_{\sigma} \exp (-\beta H(\sigma))\right)^{n}}} \ .
\ee
In the limit $ n \to 0$ we recover the previous equation. A similar argument tells us that
\be
\ba{F}=-{\ba{\ln(Z)}\over \beta N} = \lim_{n \to 0}F^{(n)} \ ,
\ee
where
\be
F^{(n)}=-{\ln\left( \ba{Z^{n}}\right) \over \beta n N} \ .
\ee
Here to proof can be easily done by using $Z^{n}\approx 1+n \ln(Z)$ and the obvious relation $\ba{1}=1$.

We can follow the same strategy for computing quantities that depend on a pair of replicas. Let us define:
\begin{equation}\label{AVERA}
w^{(2)}(n)={\ba{\sum_{s_{1}\ldots s_{n}}\exp( -\beta\sum_{a=1,n}E(s_{a})) \delta_{s_{1},s{2}} } 
\over
\ba{\sum_{s_{1}\ldots s_{n}}\exp( -\beta\sum_{a=1,n}E(s_{a}))  }
} \ .
\end{equation}
Also in this case we find 
\be
\sum_{k=1,\infty}w^{(2)}_{k}\equiv w^{(2)}=\lim_{n \to 0}w^{(2)}(n) \ .
\ee
Indeed we can write that
\be
w^{(2)}(n)=
{\ba{\sum_{s}\exp( -\beta 2 E(s))  \left(\sum_{s}\exp( -\beta 2 E(s))\right )^{n-2}} 
\over
\left(\ba{\sum_{s}\exp( -\beta E(s))  }\right)^{n} \ .
}
\ee

Let us now compute $w_{2}(n)$ by considering the contribution of the saddle point where all the $m_{i}$ are equal to 
$m$\cite{mpv}.  In this case $w_{2}(n)$ is given by the probability of finding the replicas 1 and 2 inside the same block when we 
average over the ways in which one can divide the $n$ replicas into  $n/m$ blocks of size $m$.  The computation looks 
complex, but it greatly simplifies by noticing that this probability does not depends on the two replicas (1 and 2) 
that 
we are actually considering: it coincides with the probability that for a given partition, two random replicas stay 
in the same block.  A simple computation tell us that this probability is given by
\begin{equation}
 w_{2}(n)={n(m-1) \over n(n-1)}  \ .
\end{equation}
Indeed the first replica is arbitrary and the second replica is any of the remaining $m-1$ 
replicas of the block. The denominator is the total number of unordered pairs of replicas.
In the limit $n \to 0$ we recover the correct result:
\begin{equation}
w_{2} =1-m  \ .
\end{equation}
In the similar way we find
\begin{equation}
w_{3} =\lim_{n\to 0} {n(m-1)(m-2) \over n(n-1)(n-2)}={1-m)(2-m)\over 2} ={\Gamma(3-m)\over \Gamma(1-m)\Gamma(3)}
\end{equation}
The computation can be trivially done for any $k$ and from the moments we recover the 
distribution probability $\nu(w)$.

At this stage it is not clear why we have chosen this form of the $m_{i}$ and not a 
different one. However it is clear from the previous computations that the choice of the 
$m_{i}$ codes the probability distribution of the $w$ and the situation where the 
logarithms of the $w$'s are uncorrelated identically distributed variables with an 
exponential distribution correspond to $m_{i}=m$. Different situations may correspond 
to different forms of the $m_{i}$, but the one that is relevant here corresponds to 
constants $m_{i}$. For other purposes, i.e. for finding finite size corrections or 
exponentially vanishing probabilities, different choices of the $m_{i}$ are relevant \cite{CPV}. It 
may be interesting to remark that the one that we have taken here is in some sense the 
maximally symmetric one if we esclude the trivial choice $m_{i}=1$.

\subsection{Dynamical properties of the model}
The dynamical properties of the model can be easily investigated in a qualitative way.  Interesting behavior is present 
only in the region where the value of $N$ is large with respect to the time, but it will not be discussed here.

We consider here a single spin flip dynamics.  More precisely we assume that in a microscopic time scale scale the 
system explore all the configurations that differ from the original one by a single spin flip and it moves to one of 
them (or remain in the original one) with probability that is proportional to $\exp (-\beta H)$.  This behaviour is 
typical of many dynamical process, like Glauber dynamics, Monte Carlo, heath bath.

In this dynamical process each configuration $C$ has $N$ nearby configurations $C'$ to explore.  The energies of the 
configurations $C'$ are uncorrelated to the energy of $C$, so that they are of order $N^{1/2}$ in most of the case.  A 
simple computation gives that lowest energy of the configurations $C'$ would be of order $-(N\ln(N))^{1/2}$.  The 
important point is that the corresponding energy density ($-(\ln(N)/N)^{1/2}$) vanishes in the large $N$ limit.

If the configuration $C$ has an energy density $e$ less that zero, the time needed to do a transition to a nearby 
configuration will be, with probability one, exponentially large.  For finite times at large $N$ a configuration of 
energy $e<0$ is completely frozen.  Only at times larger that $\exp(\beta e N)$ it may jump to a typical configuration 
of energy zero.  At later times different behaviours are possible: the configuration comes back to the original 
configuration of of energy $e$ or, after some wandering in the region of configurations of energy density $\approx 0$, 
it fells in an other deep minimum of energy $e'$.  A computation of the probabilities for these different possibilities 
has not yet been done, although it should not too difficult.

The conclusions of this analysis are quite simple. 
\begin{itemize}
    \item
    Every configuration of energy $e<0$ is a deep local minimum of the Hamiltonian: if we flip one spin, the energy 
    increases of a large quantity ($\approx e N$).
\item
If we start from a random configuration, after a time that is finite when 
$N\to \infty$, the system goes to a configuration whose energy is of order 
$-\ln(N)^{1/2}$ and stops there.
\item If we start from a random configuration,  only at 
exponentially large times the system will reach an energy density that is different from zero.
\end{itemize}

\section{Models with partially correlated energy}
\subsection{The definition of the models}
The random energy model (REM) is rather unrealistic in that it predicts that the energy is completely upset by a single 
spin flip.  This feature can be eliminated by considering more refined models, e.g. the so called $p$-spins models 
\cite{GROMEZ,GARDNER}, where the energies of nearby configurations are also nearby.  We could say that energy density 
(as function of the configurations) is not a continuous function in the REM, while it is continuous in the $p$-spins 
models, in the topology induced by the distance, (\ref{DISTANZA}).  In this new case some of the essential properties 
of the REM are valid, but new features are present.

The Hamiltonian  of the $p$-spins models depends on some control variables $J$, that have a 
Gaussian distribution and play the same role of the random energies of the REM 
and by the spin variables $\si$. For $p=1,2,3$ the Hamiltonian is respectively given by
\begin{eqnarray}
H^1_J(\si)= \sum_{i=1,N} J_i \si_i \ , \\
H^2_J(\si)=  \sum_{i,k=1,N}' J_{i,k} \si_i \si_k \ , \\
H^3_J(\si)= \sqrt{p!/2} \sum_{i,k,l=1,N}' J_{i,k,l} \si_i \si_k \si_l \ , \nonumber
\end{eqnarray}
where the primed sum indicates that all the indices are different and ordered.  The variables $J$ must have a variance 
of $(N^{(1-p)/2})$ if the system has a non-trivial thermodynamical limit.  Here we will study the hard spin models 
\cite{KTW} where $\sigma_{i}=\pm 1$ \footnote{A different a well studied model is the spherical model 
\cite{pspin,CuKu,KPVI}, that has the same Hamiltonian: however the spins are real variables that satisfy the constraints 
$\sum_{i}\sigma_{i}^{2}=N$.  The spherical model has many features in common with the hard spin model and some 
computations are simpler.}.

It is possible to prove by an explicit computation that, if we send first $N \to \infty$ and later $p \to \infty$, one 
recover the REM \cite{GROMEZ}.  Indeed the energy is normalized in such a way that it remains finite when $p \to 
\infty$; however the differences in energy corresponding to one spin flip are of order $\sqrt{p}$ for large $p$ (they 
ar 
order $N$ in the REM), so that in the limit $p \to \infty$ the energies in nearby configurations become uncorrelated 
and 
the REM is recovered.

\subsection{The replica solution}

The main new property of the $p$ spin model is the presence of a correlation  among the energies of nearby 
configurations.  This fact implies that if $C$ is a typical equilibrium configuration, all the configurations that 
differ from it by a finite number of spin flips will differs energy by a bounded amount, also for very large $N$.  The 
equilibrium configurations are no more isolated (as in REM), but they belongs to valleys: the entropy restricted to a 
single valley is proportional to $N$ and it is an extensive quantity.

We can now proceed as before \footnote{In this case we must use some technical tools that are also used in the study of 
the infinite range ferromagnetic model, described in the appendix.}.  The computation of $Z_{n}=\ba{Z^{n}}$ can be done 
by introducing $n$ replicas \cite{mpv,GARDNER}.  After the Gaussian integration over the $J$ one gets
\begin{equation}
Z_{n}=\sum_{\Sigma}\exp \op \frac12  N \beta^{2} \sum_{a,b=1,n}Q_{a,b}(\Sigma)^{p}\cp \ ,
\end{equation}
where $\Sigma$ denotes the set of all $ nN$ $\sigma$ variables and
\begin{equation}
Q_{a,b}(\Sigma)=N^{-1}\sum_{i=1,N}\sigma_{a}(i)\sigma_{b}(i) \ .
\end{equation}

If we introduce the Lagrange multiplier $ \Lambda_{a,b}$ we find that the previous 
expression reduces to
\begin{equation}
Z_{n}=\int dq d\Lambda \sum_{\Sigma}\exp \op \frac12  N \beta^{2} \sum_{a,b=1,n}q_{a,b}^{p} 
+N \sum_{a,b=1,n}\Lambda_{a,b}(Q_{a,b}(\Sigma)- q_{a,b}) \cp \ ,
\end{equation}
where both $q$ and $\Lambda$ are $n \times n $ symmetric matrices and  the integrals $dq$ and  $d\Lambda$ runs over 
these matrices \footnote{The integrals over the variables $ \Lambda_{a,b}$ is along the imaginary axis and goes 
from $-i \infty $ to $+i \infty$.}
The sum over the spins can be done  in each point independently from the other point using the relation
\be
\exp (N \sum_{a,b=1,n}\Lambda_{a,b}(Q_{a,b}(\Sigma)) =\exp \op \sum_{i=1,N}\sum_{a,b=1,n}\Lambda_{a,b}
\sigma_{a}(i)\sigma_{b}(k)\cp \ .
\ee
Finally we get the result:
\begin{equation}
\int dq\exp \op N \op \frac12   \beta^{2} \sum_{a,b=1,n}q_{a,b}^{p}  +  G(\Lambda)
- q_{a,b}) \cp \cp
\end{equation}
where 
\begin{equation}
\exp (G(\Lambda))=\sum_{\sigma}\exp(\sum_{a,b=1,n}\Lambda_{a,b}\sigma_{a}\sigma_{b})
\label{LOCAL}
\end{equation}
In the limit $N \to \infty$ the previous integral is dominated by the saddle point.  We have to find out 
the solution of the equations
\bea
{p \beta^{2 }q_{a,b}^{p-1}\over 2}=\Lambda_{a,b}\ , \\
q_{a,b}=\lan \sigma_{a} \sigma_{b} \ran_{\Lambda} \ ,
\eea
where 
\begin{equation}
\lan \sigma_{c} \sigma_{d} \ran_{\Lambda}={\partial G(\Lambda) \over\partial  \Lambda_{c,d}}
=
{\sum_{\sigma}\exp(\sum_{a,b=1,n}\Lambda_{a,b}\sigma_{a}\sigma_{b})\sigma_{c}\sigma_{d}
\over
\sum_{\sigma}\exp(\sum_{a,b=1,n}\Lambda_{a,b}\sigma_{a}\sigma_{b})
} \label{eqQ} \ .
\end{equation}
The free energy density (per replica) is given by 
\be
f_{n}=\frac{-1}{-\beta n}\op \frac12   \beta^{2} \sum_{a,b=1,n}q_{a,b}^{p}  +  G(\Lambda)
- \sum_{a,b=1,n} \Lambda_{a,b} q_{a,b} \cp \ .
\ee
The internal energy density  is given by deriving  $\beta f_n(\beta)$ with respect to $\beta$. One finds:
\begin{equation}
E=\frac{1}{n}\beta \sum_{a,b=1,n}q_{a,b}^{p} 
\end{equation}

In the high temperature phase we have that the only solution to the saddle equations is
\bea
q_{a,b}=0 \for a\ne b \\
q_{a,b}=1 \for a= b \ .
\eea
The corresponding value of the internal energy is given 
by
\begin{equation}
E=- \beta \ .
\end{equation}
Also in this model the high temperature results cannot be true for all $\beta$ because it leads to 
a negative entropy at low temperatures.
\begin{equation}
\parziale{f}{m} =\parziale{f}{q} =0.
\end{equation}
  \begin{figure}
     \includegraphics[width=.6\textwidth]{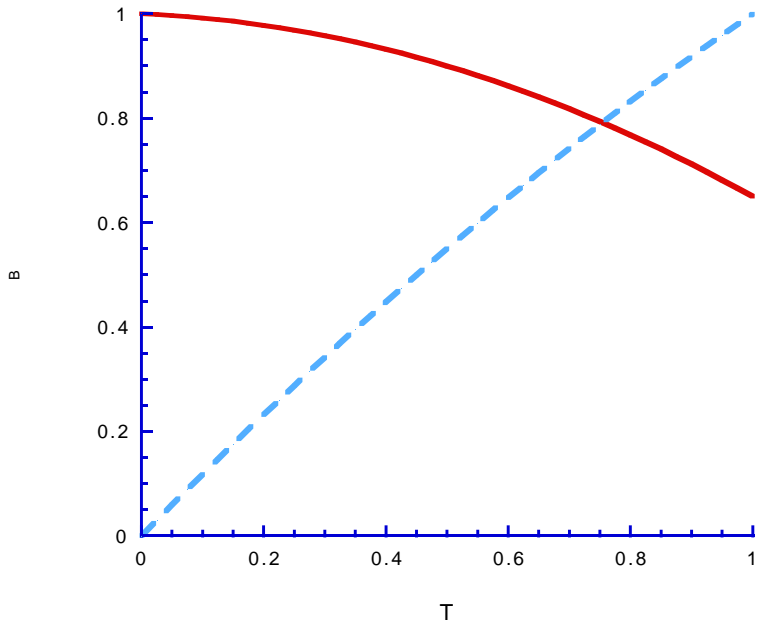}
       \caption{The qualitative behaviour quantity $q_{EA}$ and $m $ (dashed line) as functions of the temperature (in 
	 units of $T_{c}$) in the $p$ spin model for $p>2$.}
    \label{qp3}
\end{figure}

Following the approach of the previous section we divide the $n$ indices into sets of $m$ indices and we put $q_{a,b}=q$ 
if $a$ and $b$ belongs to the same set and $q_{a,b}=0$ if $a$ and $b$ do not belong to the same set.  (Obviously 
$q_{a,a}$ must be equal to $si^{2}=1$).  In the same way $ \Lambda_{a,b}=\Lambda,$ if $a$ and $b$ belongs to the same 
set, and $\Lambda_{a,b}=0$, if $a$ and $b$ do not belong to the same set.
Now the computation in \form{eqQ} can be done independently for each 
set of $m$ replicas.
The typical quantity we have to evaluate is
\bea
\sum_{\sigma}\exp \op \sum_{a,b=1,m} \frac12 \Lambda\sigma_{a}\sigma_{b}\cp=
\sum_{\sigma}\exp\op \frac12 \Lambda \op \sum_{a=1,m}\sigma_{a}\cp ^{2}\cp=
\\
=\int d\mu(h) \exp \op \sqrt{\Lambda }h(\sum_{a=1,m}\sigma_{a})\cp =
\int d\mu(h) \cosh(\sqrt{\Lambda } h)^{m}
\eea
where $h$ is a Gaussian variable with variance one:
\begin{equation}
d\mu(h)=
(2 \pi)^{-1/2} \exp(-h^{2}/2)
\end{equation}

We finally find the equations
\bea
{p \beta^{2 }q^{p-1}\over 2}=\Lambda \\
q={\int d\mu(h) \cosh(\sqrt{\Lambda } h)^{m}) \tanh^{2}(\sqrt{\Lambda } h)
\over
\int d\mu(h) \cosh(\sqrt{\Lambda } h)^{m}}
\eea

For each value of $m$ we can find a solutions (or more solutions) to the previous equations and we can compute the 
corresponding free energy.  If we use the same prescription than in the REM the correct solution is 
the maximum of the free energy as function of $m$ for $0\le m\le 1$. In other words we obtain a function $F(m)$ that we 
have to maximize.  It may looks strange to maximize the free energy, but this is what we need if the final free energy 
must be {\sl higher} of the one computed in the high temperature phase, i.e. $F(1)$.  Moreover there is a rigourous 
theorem, recently proved by Guerra \cite{GUERRANUOVO}, that the true free energy $f$ satisfies the relation
 \begin{equation}
 f\ge \max_{0\le m\le 1}F(m)
 \end{equation}
so \emph{maximization}, and not \emph{minimization}, of the free energy is the natural prescription in this situation 
\cite{mpv}.
 
In general, after having eliminated $\Lambda$ the free energy can be written as function of $m$ and $q$ ($f(q,m)$) and 
the value of these two parameters satisfy  the stationarity equations:

If we consider the case $p>2$, we finds that for $T<T_{c}$ there is a solution where the value of $q$ at the maximum of 
$f(q,m)$ (that we will call $q_{EA}$) is different from zero.  The value of $q_{EA}$ jumps at $T_{c}$ to a non-zero 
value.  For large values of $p$ the quantity $q_{EA}$ ($q_{EA}$ would be 1 in the REM) is of order $1-\exp(-A\beta p)$, 
while the parameter $m$ has the same dependence on the temperature as in the REM, i.e. it is equal to 1 at the critical 
temperature and has a linear behaviour al low temperature.  When $p $ is finite $m$ is no more strictly linear as 
function of the temperature.
The thermodynamical properties of the model are the same as is the REM: a 
discontinuity in the specific heat, with no divergent susceptibilities.

\begin{figure}
     \includegraphics[width=.6\textwidth]{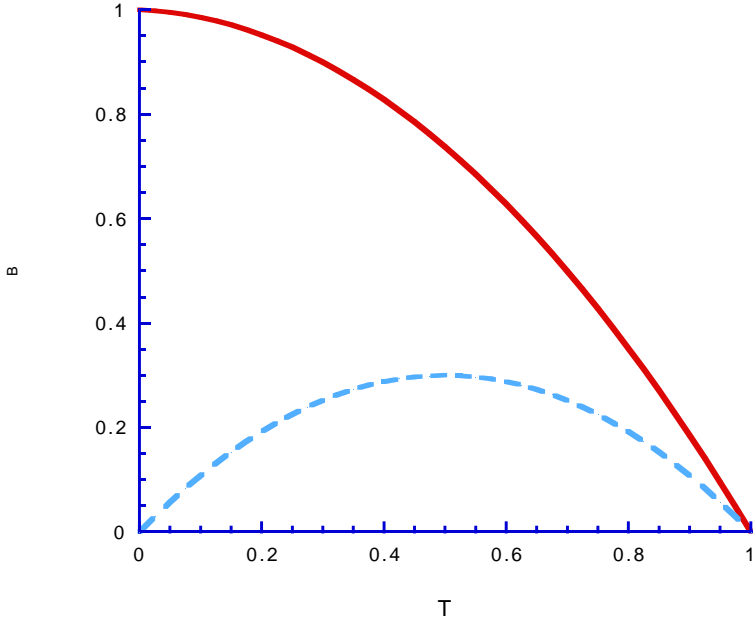}
     \caption{The quantity $q_{EA}$ and $m $ (dashed line) as functions of the temperature in the $p$ spin model (in 
		units of $T_{c}$) for $p=2$ (i.e. the SK model).}
    \label{qp2}
\end{figure}

The case $p=2$ (the Sherrington Kirkpatrick model \cite{SK} that is relevant for spin glasses) has a different behaviour especially 
near the critical temperature (see fig.  \ref{qp2}).

\subsection{The physical interpretation} \label{PHYINT}
Also in these models for each instance $I$  we can define the function $P_{I}(q)$ and 
compute its average over the different realization of the system.  The computation goes on exactly in the same way as 
in the REM. For example let us consider 
\begin{equation}
\ba{\int P_{I}(q)q}=\ba{\lan  {\sum_{i} \sigma_{i}^{a}\sigma_{i}^{b})\over N } \ran}
\end{equation}
where the values of the replica indices (i.e. $a$ and $b$) do not matter, as far as they 
are different.  In the same way as before we average over all the ways in 
which we can assign the two replicas to different blocks.  We  find 
\begin{equation}
\ba{\int P_{I}(q)q}= \ba {q_{a,b}} =(1-m) q_{EA} \ ,
\end{equation}
where $\ba {q_{a,b}}$ is the average of $q_{a,b}$ over the $n!$ permutations of 
the replicas.

In the same way
\begin{equation}
\ba{P_{I}(q)q^{s}}=\ba{\lan \op {\sum_{i} \sigma_{i}^{a}\sigma_{i}^{b})\over N} \cp ^{s}
\ran}= \ba {q_{a,b}^{s}} =(1-m) q_{EA}^{s} \ .
\end{equation}
Putting everything together one finds for the average over the different instances
\begin{equation}
\ba{P_{I}(q)}\equiv P(q)= m \delta (q)+(1-m)\delta(q-q_{EA}) \ .
\end{equation}
One can painfully reconstruct all the 
probability distributions of the $w's$by studying quantities like
\begin{equation}
\ba{\op \int P_{I}(q)q\cp^{2}}=q_{EA}^2 \ba{ \sum_{k}w_{k}^{2} }=\ba {q_{a,b}q_{c,d}}
\end{equation}
and doing the appropriate combinatorial estimates, at the ends one finds (exactly in the same way as in the REM) that 
they are proportional to $\exp(-\beta F_{k})$, where the $F_{k}$ are distributed according to the distribution 
\form{WDIS} .

In order to develop a formalism useful to discuss the physical meaning of these results it is convenient to introduce 
the concept of {\em pure states in a finite volume}\cite{mpv}.  This concept is crystal clear from a physical point of view, 
however it can be difficult to state it in a rigorous way (i.e to prove existence theorems).

We consider a system in a box of linear size $L$, containing a total of $N$ spins.  We partition the configuration space 
in regions, labeled by $\alpha$, and we define averages restricted to these regions \cite{PAR1,PAR2,CINQUE}: { these regions 
will correspond to our finite volume pure states or phases}.  It is clear that in order to produce something useful we 
have to impose sensible constraints on the form of these partitions.  We require that the restricted averages on these 
regions are such that connected correlation functions are small at large distance $x$ in a short range model (or when 
the points are different in an infinite range model).  This condition is equivalent to the statement that the 
fluctuation of intensive quantities \footnote{ Intensive quantities are defined in general as $ b =\frac{1}{N} 
\sum_{i=1}^N B_i  $ ,where the functions $B_i$ depend only on the value of $\si_i$ or from the value of the nearby 
spins.} vanishes in the infinite volume limit inside a given phase.

In a ferromagnet the two regions are defined by considering the sign of the total magnetization.  One region includes 
configurations with a positive total magnetization, the second selects negative total magnetization.  There are 
ambiguities for those configurations that have exactly zero total magnetization, but the probability that such a 
configuration can occur is exponentially small at low temperature.  

In order to present an interpretation of the results we assume that 
such decomposition exists also each instance of our problem.  
Therefore the {\em finite} volume Boltzmann-Gibbs measure can be 
decomposed in a sum of such finite volume pure states.  The states of 
the system are labeled by $\al$: we can write

\begin{equation}
  \lan\  \cdot\  \ran =\sum_{\alpha} w_{\al}\lan\  \cdot\  
  \ran_{\al} \ ,
  \protect\label{E-WSUM}
\end{equation}
with the normalization condition

\begin{equation}
  \sum_{\alpha}w_{\alpha}=1\ .
  \protect\label{E-WNOR}
\end{equation}
The function $P_{J}(q)$ for a particular sample is given by

\begin{equation}
  P_{J}(q)=\sum_{\al,\beta} w_\al w_\beta \delta(q_{\al,\beta}-q)\ ,
\end{equation}
where $q_{\al,\beta}$ is the overlap among two generic configurations 
in the states $\alpha$ and $\beta$.

Given two spin configurations ($\si$ and $\tau$)  we can introduce a natural concept of 
distance by

\begin{equation}
  d^{2}(\si,\tau)\equiv \frac{1}{N} \sum_{i=1}^N(\si_i-\tau_i)^2 \ ,
\end{equation}
that belongs to the interval [0 - 1], and is zero only if the two configurations are equal.  In the thermodynamical 
limit, i.e. for $N\to\infty$, the distance of two configurations is zero if the number of different spins remains 
finite.  The percentage of different $\si$'s, not the absolute number, is relevant in this definition of the distance.  
It is also important to notice that at a given temperature $\beta^{-1}$, when $N$ goes to infinity the number of 
 configurations inside a state is extremely large: it is proportional to $\exp (N { S}(\beta))$, where ${ S}(\beta)$ is 
the entropy density of the system).

We expect that finite volume pure states will enjoy the following properties that likely characterizes the finite volume pure states:

\begin{itemize}

	\item When $N$ is large each state includes an exponentially large number of configurations\footnote{We warn 
	the reader that in the case of a glassy system it is not possible to consider $N\to\infty$ limit of a given 
	finite volume pure state: there could be no one to one correspondence of states at $N$ and those at $2 N$..  }
        
        \item The distance of two different generic configurations
        $C_{\alpha}$ and $C_{\gamma}$ (the  first belonging  to state $\alpha$
        and the second to state $\gamma$) does not depend on the
        $C_{\alpha}$ and $C_{\gamma}$, but only on $\alpha$ and
        $\beta$. The distance
        $d_{\alpha,\beta}$ among the states $\alpha$ and $\beta$, is
        the distance among two generic configurations in these two
        states. The reader should notice that with this definition the
        distance of a state with itself is not zero. If we want we can define an alternative
        distance:

        \begin{equation}
          D_{\alpha,\beta} \equiv
          d_{\alpha,\beta} -
          \frac12\left(d_{\alpha,\alpha}+d_{\beta,\beta}\right)\ ,
        \end{equation}
        in such a way that the distance of a state with itself is zero
        ($D_{\alpha,\alpha}=0$). 
        \item The distance between two configurations belonging to the
        same state $\alpha$ is strictly smaller than the distance
        between one configuration belonging to state $\alpha$ and a
        second configuration belonging to a different state
        $\beta$. This last property can be written as

        \begin{equation} 
          d_{\alpha,\alpha} < d_{\alpha,\beta}\ .
        \end{equation} 
 This property forbids to have different states such that $D_{\alpha,\beta}=0$, and it is crucial in avoiding the 
 possibility of doing a too fine classification  \footnote{For example if in a ferromagnet at high 
 temperature we would classify the configurations into two states that we denote by $e$ and $o$, depending on if the 
 total number of positive spins is even or odd, we would have that $d_{e,e}=d_{e,o}=d_{o,o}$.}.
 
	\item The classification into states is the finest one that satisfies the three former properties.

\end{itemize}

The first three conditions forbid a too fine classification, while the
last condition forbids a too coarse classification.

For a given class of systems the classification into states depends on the temperature of the system.  In some case it 
can be rigorously proven that the classification into states is possible and unique \cite{KASTLE,KASROB,RUELLE} (in 
these cases all the procedures we will discuss lead to the same result).  
In usual situations in Statistical Mechanics the classification in phases is not very 
rich.  For usual materials, in the generic case, there is only one phase.
In slightly more interesting cases there may be two states.  For example, if we consider the 
configurations of a large number of water molecules at zero degrees, we can classify them as water or ice: here there 
are two states.  In slightly more complex cases, if we tune carefully a few external parameters like the pressure or 
the 
magnetic field, we may have coexistence of three or more phases (a tricritical or multicritical point).

In all these cases the classification is simple and the number of states is small.  On the contrary in the mean field 
approach to glassy systems the number of states is very large (it goes to infinity with $N$), and a very interesting 
nested classification of states is possible.  We note ``en passant'' that this behavior implies that the Gibbs rule 
\footnote{The Gibbs rule states that in order to have coexistence of $n$ phases ($n$-critical point), we must tune $n$ 
parameters.  Here no parameters are tuned and the number of coexisting phases is infinite! } is not valid for spin 
glasses.

The results we have obtained on the probability distribution of the overlap 
can be interpreted by saying that the system has many equilibrium states with weight that 
have the same probability distribution as a REM with the appropriate value of $m$ such that the 
average value of the local magnetization squared is
inside each state is just $q_{EA}$:
\begin{equation}
{\sum_{i=1,N}(m_{\alpha}(i))^{2}\over N}=q_{EA}
\end{equation}
Different states have magnetization that points toward orthogonal directions:
\begin{equation}
{\sum_{i=1,N}m_{\alpha}(i)m_{\gamma}(i)\over N}=0
\end{equation}
if $\alpha\ne \gamma$.

The only difference with the REM is that $q_{EA}$ is less than 1.  The 
states in the REM are composed by a single configuration, in the $p$ spin models local fluctuation are allowed and the 
states contain an 
exponentially number of configurations (local fluctuation are allowed) and have a non-zero 
entropy.

\subsection{The two susceptibilities}
Let us consider a model where the Hamiltonian is of the form
\be
H_{0}(\sigma) - h \sum_{i} \si_{i}
\ee
If the variables are interpreted as spins,  $h$ is the magnetic field.
A crucial question is what happens when we change the magnetic field. 

In order to simplify the analysis let us suppose that we stay in a system where, after we average on the disorder,
for $i\ne k$:
\be
\ba{\lan \si_{i} \si_{k} \ran }=\ba{\lan \si_{i}\ran \lan \si_{k} \ran } =0 \ . \label{GAUGE}
\ee
This property is valid in the $p$-spin models and also  in simple  models of spin glasses: e. g. in the Edwards Anderson 
model \cite{EA} where
\be
H_{0}(\sigma)=\sum_{i,k}J_{i,k}\si_{i}\si_{k} \ ,
\ee
where the sum over $i$ and $k$ runs over the nearest neighbours and  the $J$ are random variables with even probability 
distribution. 

A simple computation \cite{mpv} shows in general that the average equilibrium magnetic susceptibility is just given by
\begin{equation}
\chi_{eq}=\beta \int dq\ P(q) (1-q) \label{Chi} \ .
\end{equation}
Indeed 
\begin{equation}
N\chi_{eq}=
\beta  
 \sum_{i,k=1,N}\ba{ \lan \sigma_{i} \sigma_{k} \ran -\lan \sigma_{i}\ran \lan \sigma_{k} \ran } \ .
 \end{equation}
The terms with $i\ne k$ do not contribute after the average over the systems (as consequence of eq. \ref{GAUGE}): the 
only contribution comes from the terms where $i=k$.  We finally obtain
\begin{equation}
 N\chi_{eq}=  
 \sum_{i=1,N}\ba{( 1 -(\lan \sigma_{i}\ran \ )^{2}}=N \beta \op \ba{ 1-\sum_{\alpha,\gamma}
 w_{\alpha} w_{\gamma} q_{\alpha,\gamma}} \cp \ .
\end{equation}
In $p$ 
spin models using the form of the function $P(q)$ or using directly the properties  of the $w$'s and of the $q$'s we find 
 \be
 \chi_{eq}=\beta(1-(1-m)q_{EA}) = \beta(1-q_{EA}+m q_{EA})
 \ee
 
 It is interesting to note that  we can also write
 \begin{equation}
\chi_{eq}= \beta(1-q_{EA})+N^{-1}\beta \sum_{\alpha,\gamma}
 w_{\alpha} w_{\gamma} (M_{\alpha}-M_{\gamma})^{2}
 \end{equation}
 where $M_{\alpha}$ is the total magnetization is the state $\alpha$
 \begin{equation}
 M_{\alpha}=\sum_{i}\lan \sigma(i) \ran_{\alpha}
 \end{equation}

\begin{figure}
    \centering
  \includegraphics[width=.6\textwidth]{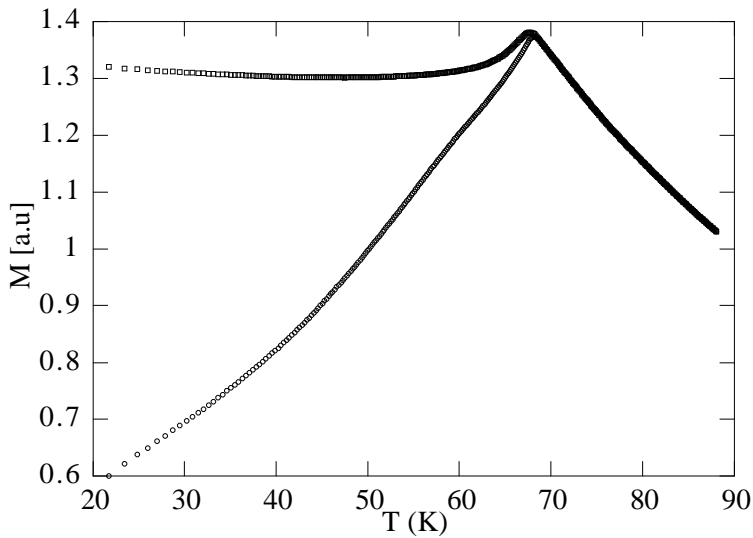} 
\caption{The experimental results for the FC (field cooled) and the ZFC (zero field cooled) magnetisation (higher and 
lower curve respectively) vs.  temperature in a spin glass sample ($Cu_{87}Mn_{13.5}$) for a very small value of the 
magnetic field
$H$ =1 Oe (taken from \cite{EXP}).  For a such a 
 low field non-linear effects can  be neglected  and the magnetization is proportional to the susceptibility.}
\label{2CHI}
\end{figure}

 The first term (i.e. $\beta(1-q_{EA})$) has a very simple interpretation: it is the 
 susceptibility if we restrict the average inside one state and it can be this identified
 with $\chi_{LR}$. The send term gives $\chi_{irr}$:
 \begin{equation}
 \chi_{irr}=\beta\sum_{\alpha,\gamma}
 w_{\alpha} w_{\gamma}(q_{EA}-q_{\alpha,\gamma}) =\beta m q_{EA} \ ,
 \end{equation}
 where the last equality is valid in  the $p$-spin models. In these models (for $p>2$) $\chi_{irr}$ jumps at the critical temperature
 ($\chi$ remains continuos): in spin glasses ($p=2$) $q_{EA}$ vanishes a the critical temperature and both $\chi$ and 
$ \chi_{irr}$ are continuos in agreement with the experimental results (fig. \ref{2CHI}).
 
The physical origine of $\chi_{irr}$ is clear. When we increase the magnetic field, the states with higher magnetization 
become more likely than the states with lower magnetization: this effect contributes to the increase in the 
magnetization.  However the time to jump to a state to an other state is very high (it is strictly infinite in the 
infinite volume limit and for infinitesimal magnetic fields where we can neglect non-linear effects are neglected): 
consequently  the time scales 
relevant for $\chi_{LR}$ and $\chi_{eq}$ are widely separated.

If we look to real systems (e.g. spin glasses) both susceptibilities are experimentally observable. 
\begin{itemize}
    \item The first susceptibly ($\chi_{LR}$) is what  we measure 
if we add an very small magnetic field at low temperatures.  The field should be small enough in 
order to neglect non-linear effects.  In this situation, when we change the magnetic field, the 
system remains inside a given state and it is not forced to jump from a state to an other state and 
we measure the ZFC (zero field cooled) susceptibility, that corresponds to $\chi_{LR}$.  
 \item
The second susceptibility ($\chi_{eq}$) can be approximately measured by doing cooling the system in presence of a small 
field: in this case the system has the ability to chose the state that is most appropriate in presence of the applied 
field.  This susceptibility, the so called FC (field cooled) susceptibility is nearly independent from the temperature 
(and on the cooling rate \footnote{The nearly independence of the field cooled magnetization on the cooling rate can be 
used to argue that field cooled magnetization is near to the equilibrium one; on the contrary, if the field cooled magnetization 
would have been strongly dependent on the cooling rate, the statement that it correspond to the equilibrium 
magnetization would be quite doubtful})
and corresponds to $\chi_{eq}$.
\end{itemize}

Therefore one can identify $\chi_{LR}$ and $\chi_{eq}$ with the ZFC susceptibility and with the FC susceptibility 
respectively.  The experimental plot of the two susceptibilities is shown in fig.  (\ref{2CHI}).  They are clearly 
equal 
in the high temperature phase while they differ in the low temperature phase.

The difference among the two susceptibilities is a crucial signature of replica symmetry breaking and, as far as I 
known, can explained only in this framework.  A small change in the magnetic field pushes the system in a slightly 
metastable state, that may decay only with a very long time scale.  This may happens only if there are many states that 
differs one from the other by a very small amount in free energy.

\subsection{The cavity method}
The cavity method is a direct approach that in principle can be used to derive in an 
esplicite way all the results that have been obtained with the replica method\footnote{ 
Not all the results have been actually derived.}.
The basic idea is simple: we consider a system with $N$ spins ($i=1,N$) and we construct a new system with 
$N+1$ spins by adding an extra spin (at $i=0$). We impose the consistency condition that the average
properties of the new spin are the same of that of the old ones \cite{mpv,MP12}.

In order to lighten the notation I will write down the formulae only for $p=2$, but it  easy
any to extend the computation to  other models.
The Hamiltonian of the new spin is;
\begin{equation}
\sigma_{0}\sum_{i=1,N}J_{0,i}\sigma_{i} \ .
\end{equation}
If we suppose that the spins $\sigma_{i}$ have vanishing correlations and we take care
that each individual term is small, we find that
\bea
m'_{0}\equiv \lan \sigma_{0}\ran =\tanh (\beta h)\ . \\
h=\sum_{i=1,N}J_{0,i}m_{i} \ ,
\eea
where $m_{i}$ denotes the magnetization of the spin $\sigma_{i}$ {\sl before} we add the 
spin 0. 
When the variables $J$ are random (or the variables $m_{i}$ are random), the central 
limit theorem implies that $h$ is a Gaussian 
random variable with variance
\begin{equation}
\ba{h^{2}}= q_{EA}\equiv {\sum_{i=1,N}m_{i}^{2}\over N} \ .
\end{equation}
If we impose the condition that the average magnetization squared of the new point is
equal to that of the old points, we arrive to the consistency equation:
\begin{equation}
q_{EA}=\ba {m_{0}^{2}}= \int d\mu_{q_{EA}}(h)\tanh^{2} (\beta h)
\end{equation}
where $d\mu_{q_{EA}}(h)$ denotes a normalized Gaussian distribution with variance
$q_{EA}$.
It is easy to check that the increase in the total free energy of the system is 
\begin{equation}
\Delta F(h) \equiv {-\ln (\cosh(\beta h)) \over \beta} \ .
\end{equation}
In this way we have derived the replica symmetric solution that corresponds to $m=0$.

If replica symmetry is broken the spins are uncorrelated within one states, but if we do 
not separate the states, we find strong correlations. 

The correct computation  goes as follows. We suppose that in the system with $N$ spins
we have a population of states
whose total free energies $F_{\al}$ are distributed (when $F_{\al}$ is not far from a 
given reference value $F^{*}$ that for lighten the notation we take equal to zero) as 
\begin{equation}
\cN_{N}(F_{N}) \propto \exp (\beta m F_{N})
\end{equation}
When we add the new spin, we will find a value of the field $h$ that depends on the 
state $\alpha$. We can now consider the conjoint probability distribution of the new free
energy and of the magnetic field.
We obtain
\begin{equation}
\cN_{N+1}(F,h)=\int dh P_{s}(h) \int dF_{N} \cN_{N}(F_{N}) \delta(F-F_{N}-\Delta F(h))
\end{equation}
where $P_{s}(h)$ is the probability distribution of the effective magnetic field produced on the spin at 0, for a generic state 
and it is still given by $d\mu_{q_{EA}}(h)$.  It is crucial to take into account that the new free energy will differs from 
the old free energy by an energy shift that is $h$ dependent.  If we integrate over $F_{N}$ and we use the esplicite 
exponential form for $\cN_{N}(F_{N})$ we find that
\begin{equation}
\cN_{N+1}(F,h)\propto \exp (\beta m F) \int dh P_{s}(h) \exp (\beta m \Delta F(h)))\propto
\exp (\beta m F) P_{N+1}(h) 
\end{equation}
The probability distribution of the field at \emph{fixed} value of the free energy is given by
\begin{equation}
P_{N+1}(h) \propto P_{s}(h) \exp (\beta m \Delta F(h)))=
d\mu_{q_{EA}}(h) \cosh(\beta h)^{m} \ .
\end{equation}
and it obviously different from $P_{s}(h)$ as soon as $m\ne 0$. In this way 
we find the consistency equation of the replica approach for $q_{EA}$.

A few comments are in order:
\begin{itemize}
    \item The probability distribution of $h$ at fixed value of the free energy of the $N$ 
    spins system ($P_{s}(h)$) is {\sl not} the probability distribution of $h$ at fixed value of the 
    free energy of the $N+1$ spins system $P_{N+1}(h)$: the two free energies differs by an $h$ 
    dependent addictive factor and they do not have a flat distribution (as soon as $m\ne 
    0$). The probability distribution of $h$ at a fixed value of the free energy of the $N$ 
    spins system is Gaussian, but the probability distribution of $h$ at fixed value of the 
    free energy of the $N+1$ spins system is not a Gaussian.
   
    \item Only in the case were $\cN_{N}(F_{N}) $ is an exponential distribution $\cN_{N+1}(F,h)$ factorizes into the 
    product of an $F$ and an $h$ dependent factor and the $\cN_{N+1}(F)$ has the same form of $\cN_{N}(F)$.  
    Self-consistency can be reached only in the case of an exponential distribution for $\cN_{N}(F_{N})$.
   
    \item The equations do not fix the value of $m$.  This in natural because (as we shall see later) we can write them 
    also in the case where the free energies densities we consider are different from that of the ground state ($F^{*}$ 
    is not near to the ground state).  In this case they do correspond to the distribution of the free energies inside 
    metastable states that are characterized by a different value of $m$ than the ground state. These equation will be 
    useful for the computation of the complexity.
 \end{itemize}
 
It is appropriate to add a last comment.  The computation we have presented relates the magnetization a spin of the 
systems with $N+1$ spins to the magnetizations of the system with $N$ spins: they are not a closed set of equations for 
a given system.  However we can also write the expression of magnetization at zero as function of the magnetizations of 
the system with $N+1$ spins, by computing the variations in the magnetization in a perturbative way.  Let us consider 
the case $p=2$ and let us denote by $m$ the magnetization of the old system ($N$ spins) and by $m'$ the magnetization of 
the new system ($N+1$ spins).
Perturbation theory tell us that
\be
m'_{i}\approx m_{i}+J_{0,i}m'_{0}{\partial m_{i}\over h_{i}}=
m_{i}+J_{0,i}m'_{0}\beta(1-(m'_{i})^{2} \ .
\ee
Using the previous formula we get the TAP equations\cite{TAP,mpv}:
\bea
m'_{0}=\tanh (\beta h)\\
h=\sum_{i=1,N}J_{0,i}m_{i} \approx \sum_{i=1,N}J_{0,i}m'_{i} -m'_{0} \sum_{i}J_{0,i}^{2}\beta (1-m'_{i})\\ 
\approx
\\sum_{i=1,N}J_{0,i}m'_{i} -m'_{0} \beta (1-q_{EA}) \nonumber
\eea
where $(N+1)q_{EA})=\sum_{i=0,N}m'_{i}$ and we have used the fact that $\ba{J_{0,i}^{2}}=N^{-1}$.
A detailed computation  show that the free energy corresponding to a solution of the TAP equations is given by the 
TAP free energy.
\begin{equation}
F[m]=\sum_{i<k}J_{i,k}m_{i}m_{k} -N \beta (1-q)^{2}- T \sum_{i}S(m_{i}) \ ,
\end{equation}
where $S(m)$ is the usual single spin entropy:
\begin{equation}
-S(m)={1+m \over 2} \ln({1+m \over 2}) +{1-m \over 2} \ln({1-m \over 2})  \  .
\end{equation}
It is important to note that the solutions of the TAP equations are also stationary points of the TAP free energy: 
using the relation 
\be
{\partial q_{EA}\over \partial m_{i}}= {m_{i}\over N}
\ee
the TAP equations can be written as 
\be
{\partial F[m] \over \partial m_{i}}=0 \ .
\ee
 \section{Complexity} \label{strategy}
It would be interesting to characterize better the free energy landscape of the models described in the previous 
section, especially in order to understand the dynamics.  Indeed we have already seen that in the REM the system could 
be trapped in metastable configurations.  In models where the energies are correlated the situation is more 
complicated; moreover in realistic finite dimensional models there are still further subtleties.

Although the word {\sl metastable configuration} has a strong intuitive appeal, we must  define what a metastable 
configuration is in a more precise way.  
There are two different (hopefully equivalent) definitions of a metastable state or valley:
\begin{itemize}
\item
From an equilibrium point of view a valley is a region of configuration space separated by the rest of the 
configuration space by free energy barriers that diverge when $N\to\infty$.  
More precisely the system, in order to go outside a valley by moving one spin (or one particle) at 
once, must cross a region where the free energy is higher than that of the valley by 
a factor that goes to infinity with $N$.
\item
From the dynamic point of view a valley is a region of configuration space where  the system remains for a time 
that goes to infinity with $N$.
\end{itemize}
The rationale for assuming that the two definitions are equivalent is the 
following.  We expect that for any reasonable dynamics where the system
evolves in a continuous way (i.e.  one spin flip at time), the system must cross a configuration of higher free energy 
when it goes from a 
valley to an other valley. The time for escaping from a valley is given by
\begin{equation}
\tau \simeq \tau_0 \exp (\beta \Delta F)
\end{equation}
where $\Delta F$ is the free energy barrier \footnote{In kinetically constrained models, where some local movements are 
forbidden, we can have dynamical valleys that do not correspond to valley from the equilibrium point of view.}.

It is crucial to realize that in infinite range models  valleys may have a free energy density higher that 
that of equilibrium states.  This phenomenon is definitely not present in short range models.  Two equilibrium states 
with infinite mean life must have the same free energy.  

The proof of this statement is simple.  Indeed let us suppose that the system may stay in two phases (or valleys) that 
we 
denote as $A$ and $B$.  If the free energy density of $B$ is higher than that of $A$, the system can go from $B$ to $A$ 
in a continuos way, by forming a bubble of radius $R$ of phase $A$ inside phase $B$ 
\footnote{If the surface tension among 
phase $A$ and $B$ is finite, has happens in any short range model, for large $R$ the volume term will dominate the free 
energy difference among the pure phase $B$ and phase $B$ with a bubble of $A$ of radius $R$.  This difference is thus 
negative at large $R$, it maximum will thus be finite.  A finite amount of free energy in needed in 
order to form a seed of phase $A$ where the spontaneous formation of phase $A$ will start. }.
For example, if we take a 
mixture of $H_2$ and $O_2$ at room temperature, the probability of a spontaneous temperature fluctuation in a small 
region, that leads to later ignition and eventually to the explosion of the whole sample, is greater than 
zero (albeit quite a small number), and obviously it does not go to zero when the volume goes to infinity. This 
argument does not work in mean field models where in some sense surface effects are as important as volume effects 
(when $D$ is large $R^{D}$, the volume, is not so different from $R^{D-1}$, the surface).

We have two possibilities open in positioning the predictions of mean field theory 
concerning the existence of real metastable states:
\begin{itemize}
	\item  We consider the presence of these metastable state with {\sl infinite} 
	mean life an artefact of the mean field approximation and we do not pay any
	attention to them.

\item We notice that in the real systems there are metastable states with very 
large (e.g.  much greater than one year) mean life.  We consider the {\sl 
infinite} time metastable states of the mean field approximation as precursors 
of these {\sl finite} mean life states.  We hope (with reasons) that the corrections to 
the mean field approximation will give a finite (but large) mean life to 
these states (how this can happen will be discussed in the next section).
\end{itemize}
In these notes I will explore the second possibility, that seems to be much more fruitful than the first one.

\subsection{The basic definitions}

Before discussing the difficulties related to the definition of the complexity in short 
range models, we must see the main definitions that are correct in the mean field approach.

The basic ideas are quite simple \cite{pspin,CuKu,FM,KPVI,KLEIN,FP,FRAPA,MONA}. In 
principle we  proceed in a way similar to the construction of equilibrium states described in section \ref{PHYINT}: we
partition the whole configuration  space  into valleys. If we call $Z_{\alpha}$ the contribution of each 
valley to the partition function,  the corresponding free energy is given by
\begin{equation}
Z_{\alpha}=\exp(-\beta F_\al)
\end{equation}

This definition does not give us a practical way to find the valleys.  An alternative approach, that should be 
hopefully 
equivalent, is the following.  In many case one can prove that the magnetization in a given valley should satisfy some 
equations, e.g. the TAP equations of the previous sections: valleys may be identified with solutions of the TAP 
equations and their free energy is given by the TAP free energy\footnote{At zero temperature one could try to identify 
valleys with the minima of the Hamiltonian, that are called inherent 
structures in the glass community \cite{St}.}..  Generally speaking in a system of $N$ spins we can 
introduce a free energy functional $F[m]$ that depends on the local magnetizations $m(i)$ and on the temperature.  
Only in the mean field case $F[m]$ is given by the TAP free 
energy.

We suppose that at sufficiently low temperature the functional $F[m]$ has many local minima (i.e. the number of 
minima goes to infinity with the number ($N$) of spins).  Exactly at zero temperature these local minima coincide with 
the local mimima of the potential energy as function of the coordinates of the particles.  Let us label then by an 
index 
$\alpha$.  To each of them we can associate a free energy $F_\al$ and a free energy density $f_\al= F_\al/N$.  In this 
way the valleys are associated to local minima of the free energy functional.

In this low temperature region we suppose that the total free energy of the system can be well 
approximated by the sum of the contributions to the free energy of each particular local minimum. We thus find:
\begin{equation}
Z\equiv \exp(-\beta N f_{S}) =\sum_\al \exp(-\beta N f_\al)\ .
\end{equation}

When the number of minima is very high, it is convenient to introduce the function $\cN(f,T,N)$, i.e. the density of 
minima whose free energy is near to $f$.  With this notation we can write the previous formula as
\begin{equation}
Z= \int df \exp (-\beta N f) \cN(f,T,N).
\end{equation}
In the region where $\cN$ is exponentially large we can write
\begin{equation}
\cN(f,T,N) \approx \exp(N\Sigma(f,T)),\label{CON}
\end{equation}
where the function $\Sigma$ is called the complexity or the configurational entropy (it is the 
contribution to the entropy coming from the existence of an exponentially large number of locally 
stable configurations).

The minimum (maximum) possible value of the free energy is given by $f_m(T)$ ($f_M(T)$).The relation (\ref{CON}) is 
valid in the region $f_m(T)<f<f_M(T)$.   Outside 
this region we have that $\cN(f,T)=0$.  It all cases known $\Sigma(f_m(T),T)=0$, and the function $\Sigma$ is 
continuous 
at $f_m$. On the contrary in mean field models it happens frequently that the function $\Sigma$ is discontinuous 
at $f_m$

For large values of $N$ we can write
\begin{equation}
\exp(-N \beta f_{S}) \approx \int_{f_m}^{f_M} df \exp (-N(\beta f- \Sigma(f,T)).\label{SUM}
\end{equation}
We can thus use the saddle point method and  approximate the 
integral  with the integrand evaluated at its maximum.
We find that
\begin{equation}
\beta f_{S}=\min_f\Phi(f) \equiv \beta f^* - \Sigma(f^*,T),
\end{equation}
where the potential $\Phi(f)$ (that will play a crucial role in this approach) is given by
\begin{equation}
\Phi(f)\equiv\beta f - \Sigma(f,T).
\end{equation}
(This formula is quite similar to the well known homologous formula for the free energy, i.e.  
$\beta 
f=\min_{E} (\beta E -S(E))$, where $S(E)$ is the entropy density as function of the energy density.)

If we call $f^*$ the value of $f$ that minimize $\Phi(f)$.  we have two possibilities:
\begin{itemize}
\item
The minimum $f^*$ is inside the interval and it can be found as solution 
of the equation $\beta=\partial \Sigma/\partial f$.  In this case we have
\begin{equation}
\beta \Phi=\beta f^* - \Sigma^*, \ \ \ \Sigma^*=\Sigma(f^*,T).
\end{equation}
The system may stay in one of the exponentially large number of possible minima.  The 
number of minima where is convenient for the system to stay is $\exp(N \Sigma ^*)$ .  
The entropy of the system is thus the sum of the entropy of a typical minimum and of 
$\Sigma^*$, i.e. the contribution to the entropy coming from the exponential large 
number of microscopical configurations.

\item
The minimum is at the extreme value of the range of variability of $f$.  We have that $f^*=f_m$ and $\Phi=f_m$.  In 
this 
case the contribution of the complexity to the free energy is zero.  The different states that contribute the free 
energy have a difference in free energy density that is of order $N^{-1}$ (a difference in total free energy of order 
1).  Sometimes we indicate the fact that the free energy is dominated by a few different minima by say the replica 
symmetry is spontaneously broken \cite{mpv,parisibook2}.
\end{itemize}

From this point of view the behaviour of the system will crucially depend on the free 
energy landscape \cite{ACP}, i.d. the function $\Sigma(f,T)$, the distance among the 
minima, the height of the barriers among them...

\subsection{Computing the complexity}\label{CC}

We have seen that complexity counts the number of metastable states.  It would be interesting to compute the complexity 
in a direct way without having to count all the metastable states (an impossible task for large $N$) An interesting 
route to the evaluation of complexity consists in introducing new artificial couplings and to consider the behaviour of 
the systems in these conditions.  This new approach works also in cases where the free energy functional is not known 
in 
an exact way, so that its minima cannot be computed.

The basic idea  is to start from an equilibrium configuration and to explore the configuration 
space phase around it \cite{FP,FRAPA}. If we can define in some way the entropy ($S_{v}$) of the valley around a 
given equilibrium configuration, we have  that 
\begin{equation}
S=\Sigma^{*}+S_{v} \label {SV} \ ,
\end{equation}
where $S$ is the total entropy of the system and $\Sigma^{*}$ is the equilibrium complexity.

More precisely, we study a system of $N$ interacting variables $\si_i$, $i=1,...,N$, with Hamiltonian $H(\sigma)$, 
$q(\sigma,\tau)$ is an overlap function, i.e. an intensive measure of similarity among the configurations $\sigma$ and 
$\tau$.  We can consider a reference equilibrium configuration $\sig$, that produce a fixed  
external potential on a replica $\tau$. The partition function of the 
second system is:
\begin{equation} 
Z(\sig,\eps)=
\sum_{\tau} \exp\left[-\b H(\tau)
+\b\eps N q(\tau,\sig)\right],
\label{zq}
\end{equation}
and 
\begin{equation}
\Gamma(\eps)=-(N\beta)^{-1} \lan \log Z(\sig,\eps)\ran_{\sig}
\end{equation}
is the $\epsilon$-dependent free energy ($\lan \cdot \ran_{\sig}$ denotes the average over the variables $\sig$).  The 
new term in the Hamiltonian, for $eps$ sufficiently large, forces the variables $\tau$ to be near the variables $\sig$ 
and produces a \emph{quenched} disorder for the variables $\tau$.  By changing the value of $\eps$ we can to explore 
the 
phase space around a given equilibrium configuration $\sig$.  At the end we average over $\sig$ the logarithm of the 
$\sig$ dependent free energy.

\begin{figure}
 \includegraphics[width=.6\textwidth]{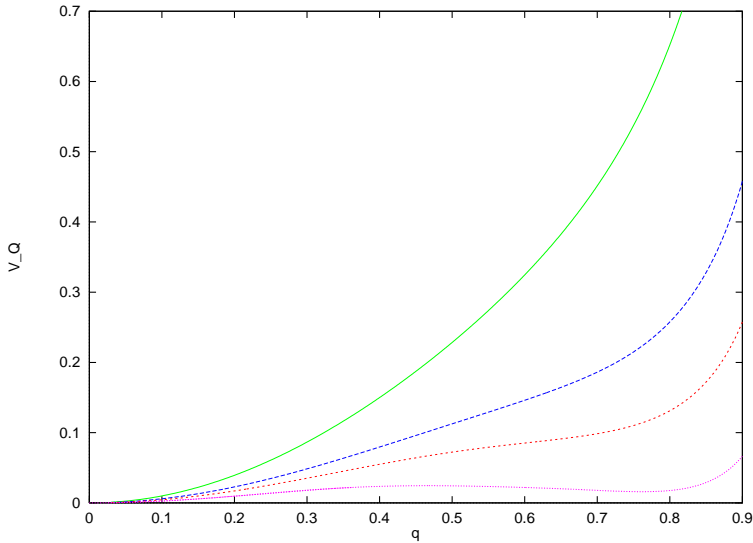}
\caption[0]{\protect\label{W}
Different shapes of the function $W$ for various temperatures: the higher curves correspond to higher temperatures.
  }
\end{figure}

The quantity $\Gamma(\eps)$ is well defined and it may be computed in also in numerical simulations.  However it is 
interesting to evaluate it in mean field models, where analytic computations are possible.  The analytic computation of 
$\Gamma(\eps)$ can be done by considering 1+$s$ replicas: the corresponding Hamiltonian is
\begin{equation}
H_{s}(\sig)\equiv H(\sig_{1})+\sum_{a=2,s} H(\sig_{a})+\eps N \sum_{a=2,1+s}q(\sig,\sig_{a})
\end{equation}
Here $\sigma_{1}$ plays the role of $\sigma$ and the  $\sigma_{a}$ (for $a=2,1+s$)  are 
$s$ replicas of the $\tau$ variables.
The quenched limit (where there is no feedback reaction of the $\tau$ variables on 
the $\sigma$ variables) is obtained in the limit $s \to 0$:
\begin{equation}
\Gamma(\eps)= \lim_{s \to  0} {\partial \over \partial s} \ln \ato 
\prod_{a=1,s}\sum_{\sig _a} \exp(- \b H_{s}(\sigma)) \cto
\end{equation}
In models where we have to perform  the average over the instances of the problems, we have to replicate $n$ times
the $s$-replicated system: we 
have to take $n \times (1+s)$ variables and to study simultaneously the limit $s\to 0$ 
and $n\to 0$. Fortunately enough, in the high temperature phase, where we are interested 
to the computation of the complexity, we do not need to break the replica symmetry (at least for the computation of the 
complexity) and we 
obtain the correct results already for $n=1$.

In the following we will study the phase diagram of the model in
the $\eps -T$ plane. Explicit computations can be found in the literature mainly for the 
p-spins spherical model, but the conclusions have a general validity \cite{FP,FRAPA}. At this end it is 
convenient to define the Legendre transform of $\Gamma(\eps)$, defined as
\bea
W(q)=\Gamma(\eps(q))-\eps(q) q \ , \\
{\partial W(q) \over \partial q} =\eps(q) \ .
\eea
The potential $W(q)$ has the meaning of the free energy with  the constraint that the 
overlap of our configuration $\tau$ with the generic configuration $\sigma$ is equal to $q$. 
As far as we are interested in studying the $q$-dependance of $W(q)$, we can set 
conventionally $W(0)=0$.

If one computes the functions $\Gamma(\eps)$ and $W(q)$ in a mean field model, one typically finds that the shape of 
the 
function $W$ is characteristic of a mean-field system undergoing a first order phase transition.  At high enough 
temperature $W$ is an increasing and convex function of $q$ with a single minimum for $q=0$.  Decreasing the 
temperature 
below a value $T_f$, where for the first time a point $q_f$ with $W''(q_f)=0$ appears, the potential looses the 
convexity property and a phase transition can be induced by a field.  A secondary minimum develops at $T_d$, the 
temperature of dynamical transition \cite{KTW}, signaling the presence of long-life metastable states.  The height of 
the secondary minimum reaches the one of the primary minimum at $T=T_s$ and thermodynamic coexistence at $\eps=0$ takes 
place.  This is the usual static transition.  In figure \ref{W} 
we show the shape of the potential in the various regions.
\begin{figure}
 \includegraphics[width=.6\textwidth]{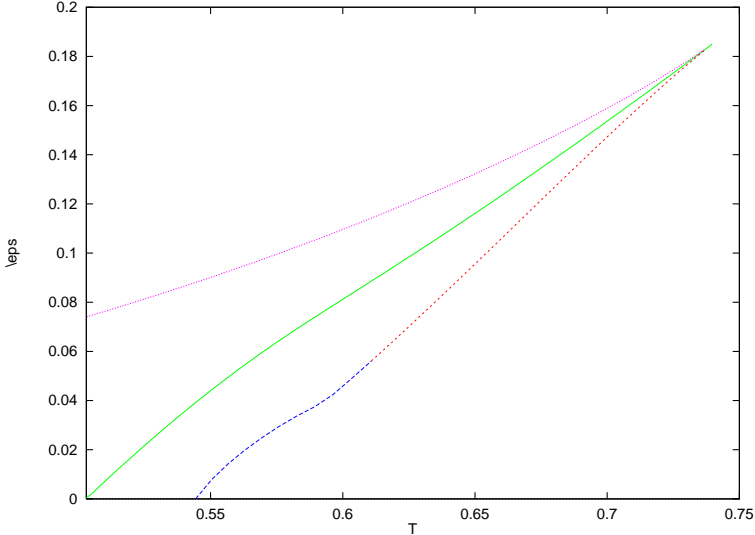}
  \caption{\protect\label{F_E1}
Phase diagram in the $T-\eps$ plane.  At the upper curve the low $q$ solution disappear, at the lower curve the high 
$q$ 
solution disappear and two locally stable solutions are present only in the region among the upper and lower curves.  
The middle curve the coexistence line where the two solutions have equal free energy.  The coexistence line touches the 
axes $\eps=0$ at $T=T_s$, while the lower curve touches it at $T=T_{D}$.  }
\end{figure}

Therefore the potential $W(q)$ has usually a minimum at $q=0$, where $W(0)=0$.  It may have a secondary minimum at 
$q=q_D$.  We have a few different situations:
\begin{itemize}
    
\item At $T>T_D$ the potential $W(q)$ has only the minimum at $q=0$.  The quantity $q_D$ cannot be defined and no 
valley 
with the equilibrium energy are present.  This is more or less the definition of the dynamical transition temperature 
$T_D$.  A more careful analysis \cite{BAFRPA} shows that for $T_D<T<T_V$ there are still valleys with energy {\sl less} 
than the equilibrium one, but these valleys cover a so small region of phase space that they are not relevant for 
equilibrium physics \footnote{In the REM limit ($p\to \infty$) the temperature $T_D$ goes to infinity.  In this limit 
the  region $T>T_D$ does not exist.  
Therefore the dynamical transition is a new feature that is not present in the REM.}.

\item Exactly at $T=T_D$ we sit at a phase transition point where some susceptibilities are divergent.  This fact 
implies (in short range models) that there is a divergent dynamical correlation length that is related to dynamical 
heterogeneities \cite{corr_length}.

\item $W(q_D)>0$.  This happens in an intermediate temperature region, above $T_c$, but below $T_D$, where we can put 
one replica $\si$ at equilibrium and have the second replica $\tau$ in a valley near it.  It happens that the internal 
energy of both the $\si$ configuration (by construction) and of the $\tau$ configuration are equal to the equilibrium 
one.  However the number of valley is exponentially large so that the free energy a single valley will be higher that 
the total free energy.  One finds in this way that $W(q_D)>0$ is given by
\begin{equation}
W(q_D)= {\ln \N _e \over N} \equiv \Sigma^{*}
\end{equation}
where $ \N _e $ is the average number of the valleys having the equilibrium energy \cite{MONA,PP} .

\item $W(q_D)=0$.  This happens in the low temperature region, below $T_c$, where we can put two replicas both at 
overlap $0$ and at overlap $q_{EA}$ without paying any prize in free energy.  In this case $q_D=q_{EA}$.
\end{itemize}

For $T<T_D$ there is a relation (\form{SV} ) among the entropy inside a valley the 
entropy of the systems and is the configurational entropy, or complexity, that is given by the value of $W$ at the 
secondary 
minimum.  This $W$ contribution vanishes at $T_c$ and becomes exactly equal to zero for $T<T_c$ \cite{KTW} .

Although the behavior of this potential function is analogous to the one found in ordinary systems undergoing a first 
order phase transition the interpretation is here radically different.  While in ordinary cases different minima 
represent qualitatively different thermodynamical states (e.g. gas and liquid), this is not the case here.  In our 
problem the local minimum appears when ergodicity is broken, and the configuration space splits into an exponentially 
large number of components.  The two minima are different manifestations of states with the same characteristics.  The 
height of the secondary minimum, relative to the one at $q=0$ measures the free-energy loss to keep the system near one 
minimum of the free energy (in configurations space).  This is just the complexity $T\Sigma$, i.e. the logarithm of the 
number of distinct valleys of the system.
\begin{figure}
    \includegraphics[width=.6\textwidth]{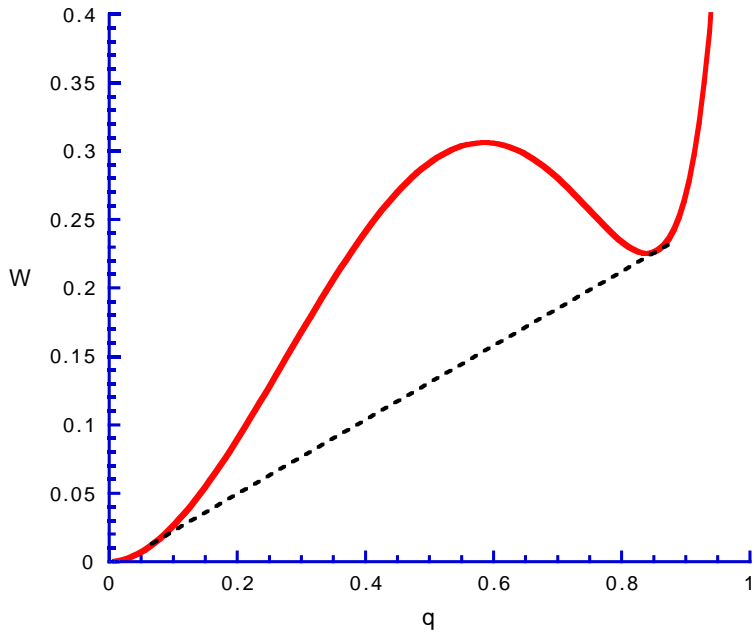}
 \caption[0]{\protect\label{MAX}
The full line is the function $W(q)$  computed in the mean field approximation. The dashed line is the correct
result (Maxwell construction).}
\end{figure}

The equation $\partial W(q)/ \partial q =\eps$ may have two stable solutions (that correspond to a local minimum of 
$W(q)-\eps q$) only in the region of the $T-\eps$ plane shown in fig. \ref{F_E1}. At the upper and low curves one of 
the 
two solutions loose its stability and it disappears: these two curves are the equivalent of the spinodal lines in 
usual first order transition. The point where the lower curve crosses the axis $\eps=0$ is the dynamical transition 
\cite{FRAPA}: 
only at lower temperatures the two systems may remain with an high value of the overlap 
without having a force that keeps them together (i.e. $\eps=0$). On the contrary the static transition is 
characterized by the fact that the coexistence line touches  the axis $\eps=0$.

\begin{figure}
     \includegraphics[width=.6\textwidth]{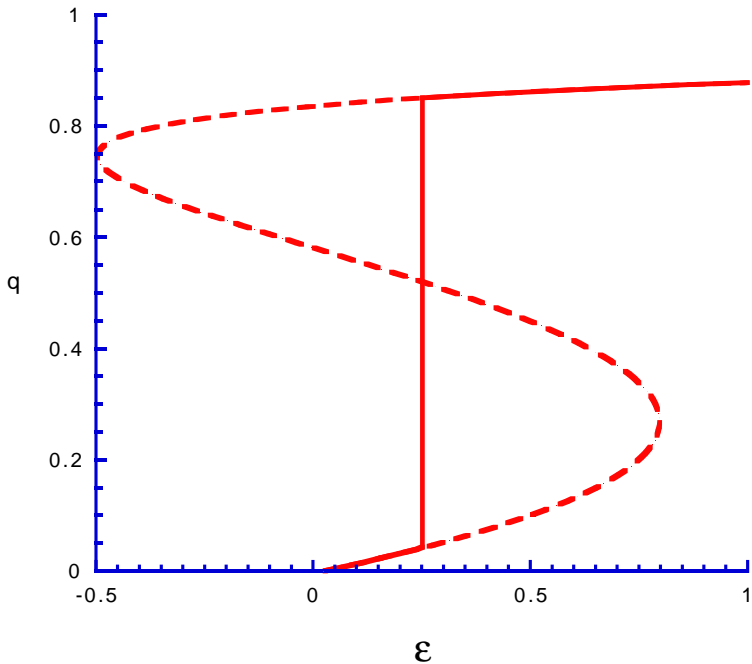}
 \caption[0]{\protect\label{qeps}
The shapes of the function $q(\eps)$ for $T>T_{c}$: the full line is the correct result and the dashed line
is the output of a mean field approximation.
  }
\end{figure}

General arguments tell us that the free energy is a convex function of the $q$, so that we the correct shape of the 
function $W$ can be obtained by the Maxwell construction (see fig.  \ref{MAX}).  In order to se the consequences of this 
fact on the definition of the complexity we can try to consider the function $q(\eps)$ for temperatures less than 
$T_{D}$ shown in fig \ref{qeps}.

As can be seen from the figures, the point where we evaluate the complexity (i.e. $\eps=0$ and high $q$) is always in 
the metastable region for $T>T_{s}$ where we equilibrium complexity is non-zero.  This causes an intrinsic ambiguity in 
the definition of complexity because the free energy in not defined with infinite precision in the metastable phase.  
However we can use the fact that the free energy is as $C^{\infty}$ function of $\eps$ near the discontinuity point to 
extrapolate the high $\eps$ free energy in the metastable region.  This ambiguity becomes smaller and smaller more we 
approach the static temperature (the amount of the extrapolation becomes smaller and smaller) and in general it is 
rather small 
unless we are very near to the dynamic phase transition.  This ambiguity is not important from practical 
purposes; however it implies that there is no sharp, infinitely precise definition of the 
equilibrium complexity.  If we forget this intrinsic ambiguity in the definition of the complexity we may arrive to 
contradictory results.

\subsection{Complexity and replicas}\label{MORE}

As we have seen we can write
\begin{equation}
Z(\beta)=
\sum_{a} \exp( -\beta N f_{a}(\beta))= 
\int d \cN(f,\beta) \exp (-\beta N f) \ ,
\end{equation}
where $f_{a}(\beta)$ is the free energy density of the valley labeled by $a$ at the temperature $\beta^{-1}$, and 
$\cN(f,\beta)$ is the number of valleys with free energy density less than $f$.

We have also seen that $\cN(f,\beta) =\exp(N \Sigma(f,\beta))$, where the configurational entropy, or complexity, 
$\Sigma(f,\beta)$ is positive in the region $f>f_{0}(\beta)$ and vanishes at $f=f_{0}(\beta)$.  The quantity $ 
f_{0}(\beta)$ is the minimum value of the free energy: $\cN(f,\beta)$ is zero for $f< f_{0}(\beta)$ \cite{FRAPA,PP,LJ}.

If the equation
\begin{equation}
{\partial \Sigma \over \partial f} =\beta
\end{equation}
has a solution at $f=f^{*}(\beta)$ (obviously this may happens only for 
$f^{*}(\beta)>f_{0}(\beta)$), we stay in the liquid (high temperature) phase.  Here the free 
density is given by
\begin{equation}
f_{eq}=f^{*}-\beta^{-1} \Sigma(f^{*},\beta)
\end{equation}
and $\Sigma(f^{*},\beta)= \Sigma^{*}(\beta)$.

Otherwise we stay in the glass (low temperature) phase and 
\begin{equation}
f_{eq}=f_{0}(\beta)\ .
\end{equation}

In order to compute the properties in the glass phase we { need} to know $\Sigma(f,\beta)$: a simple strategy to 
compute 
the complexity is the following.  We introduce the 
modified partition function
\begin{equation}
Z(\gamma;\beta)\equiv \exp ( -N \gamma G(\gamma;\beta))=\sum_{a} \exp( -\gamma N 
f_{a}(\beta)).
\end{equation}
It is evident that $Z(\beta;\beta)$ is the usual partition function and $G(\beta;\beta)$ is the usual free energy.
Using standard thermodynamical arguments it can be easily proven that in the limit $N 
\to \infty$ one has:
\be
\gamma G(\gamma;\beta)=  \gamma f - \Sigma(\beta,f),\ \ \ 
f={\partial (\gamma G(\gamma;\beta) )\over \partial \gamma}.
\ee
The complexity is obtained from $G(\gamma;\beta)$ in the same way as the 
entropy is obtained from the usual free energy \cite{MONA,LJ}:
\begin{equation}
\Sigma(\beta,f)={\partial G(\gamma;\beta) \over \partial \gamma}.
\end{equation}

A few observations are in order:
\begin{itemize}
\item In the new formalism $\gamma$, the free energy and the complexity play respectively 
the same role of $\beta$, the internal energy and the entropy in the usual formalism.  

\item In the new formalism $\beta$ only indicates the value of the temperature that is 
used to compute the free energy and $\gamma$ controls which part of the free energy 
landscape  is sampled.  

\item When $\beta \to \infty$ (at least in mean field models) we sample the energy landscape: 
\begin{equation}
Z(\gamma;\infty)=\sum_{a} \exp( -\gamma N e_{a})=\int \nu(e) de \exp (- \gamma N e)
\end{equation}
where $e_{a}$ are the minima of 
the Hamiltonian and $\nu (e)$ the density of the minima of the Hamiltonian.

\item The equilibrium complexity is given by 
$\Sigma ^{*}(\beta) = \Sigma (\beta;\beta)$. On the other hand $\Sigma (\gamma;\infty)$ 
give us information on the minima of the Hamiltonian.
\end{itemize}

In principle it is possible to get the function $\Sigma(f)$ by computing directly the number of solution of the TAP 
equations for a given value of the free energy density.  However it is simpler to obtain it by using the replica 
formalism and it is reassuring that one gets the same results with both methods \cite{MONA,FRAPA,PP,Me,CGG,CGMP}.

The computation of the modified partition function $Z(\gamma;\beta)$ can be easily done in 
the replica formalism \cite{FRAPA,PP}.  If we consider a system with $m$ replicas (with $m$ integer) and 
we constrain them to stay in the same state we find that
\begin{equation}
Z(\beta,m)=
\sum_{a} \exp( -\beta m N f_{a}(\beta))
\end{equation}
This expression coincide with $Z(\gamma;\beta)$ for $\gamma=m \beta$. Therefore there is 
a very simple way for computing $G(\gamma;\beta))$. We must consider the partition 
function of $m$ replicas that are constrained to stay in the same state, i.e. there are at 
a large value of $q$ where $q$ is chosen in a self consistent way..

Let us firstly see how this approach works in the REM. In the REM the configurations and the states practically 
coincide:  there is no $\beta$ dependence of the complexity.  The REM is defined by the property that 
$\Sigma(e)=e^{2}/2 
-\ln (2)$, so that the computation of $\Sigma(e)$ using the replicas may look pointless, however it is instructive to 
illustrate the point.

The contribution to the partition function of $m$ replicas coming from the region of phase space where all the $m$ 
configuration are identical is given by
\begin{equation}
Z(\beta,m) =2^{N}\exp(-\frac12 (\beta m)^{2})
\end{equation}
We thus get
\begin{equation}
\gamma G(\gamma;\beta) =\ln(2)-\frac12 \gamma^{2}
\end{equation}
and from the previous equation we can read back the expression for the complexity.  

In the $p$-spin model we have to find out the partition function of a systems where all the $m$ replicas are in the 
same 
block and are characterized by an high value of $q$.  Here, given the value of $m$, we have to look for a solution 
$q^{*}$ of the equation for $q$ of the replica approach:
\begin{equation}
{\partial F \over \partial q} =0 \ , \label{EqQ}
\end{equation}
with $q^{*}\ne0$, where the potential $F(q,m)$ is the replica potential introduced in the previous section.
We thus find that 
\begin{equation}
G(\beta m;\beta))=F(m,q^{*})
\end{equation}
The computation that we have done before for the statics contains the whole information needed to compute also the 
complexity and some other properties of the metastable states.  Indeed we get
\begin{equation}
\Sigma(m)={\partial F(m,q^{*})\over m} \ . \label{EqSigma}
\end{equation}
and $f(m)$ can be obtained by Legendre transform or by using the relation
\begin{equation}
f(m)={\partial m^{-1} F(m,q^{*})\over \partial m^{-1}}= F(m,q^{*}) -m \Sigma(m)\ .
\end{equation}
One finally finds the complexity as function of the free energy, by eliminating $m$.
\begin{equation}
\Sigma(f)=\Sigma(m(f))
\end{equation}

The dynamical 
temperature is is the highest temperature where eq.  \form{EqQ} has a solution at near $m=1$, while the static critical 
temperature is the first temperature where
\begin{equation}
F(q(m),m)|_{m=0}=F(q(m),m)|_{m=1}
\end{equation}
The equilibrium  transition  
temperature is given by the condition that  the equilibrium complexity satisfies the condition
\be
\Sigma^{*}\equiv{\partial F(m,q^{*})\over \partial m}|_{m=1}=0 \ .
\ee

In this way we have recovered the results of one step replica symmetry breaking (together with the mysterious condition
$\partial F(m)/ \partial m=0$) from general principles.

Although we have based our discussion on mean-field model, we expect that the qualitative features of the phase 
diagrams 
presented survive in finite dimension.  We believe that the existence of a coexistence line, terminating in a critical 
point, is a constitutive feature of systems whose physics is dominated by the existence of long lived metastable states 
like glasses.  These predictions of can be submitted to numerical test in glassy model systems as like e.g. 
Lennard-Jones or hard spheres, or polymer glasses.  For example the identification of the complexity $\Sigma$ as the 
free energy difference between the stable and the metastable phases allows an other way to measure of this quantity in 
a 
simulation.  Indeed the ending of the transition lines in a critical point implies that the metastable state can be 
reached via closed paths in phase diagram leaving always the system in (stable or metastable) equilibrium; the free 
energy difference of the two phases can be computed integrating the derivative of the free energy along such a closed 
path.

We have to study the shape of the function $F(m,q)$.  At fixed $m$ as function of $q$ it may have one of the forms 
shown 
in fig.  \ref{W}.  Below the dynamical transition near $m=1$ it has the shape of the lower curve fig.  \ref{W}.  By 
decreasing $m$ the shape of this function modifies and the secondary minimum disappears.  There is a 
temperature-dependent region of $m$ where the equation $\partial F / \partial q = 0$ has a solution at non-zero $q$: in 
this region we can compute the complexity.  It is possible that if we compute the small fluctuations in this region 
using the techniques of the next section we finds for some values of $m$ a not consistent result, i.e. e negative 
spectrum.  This phenomenon may indicate that the allowed $m$ region is smaller than that indicated by the condition of 
the existence a solution to the equation $\partial F / \partial q = 0$.  However it is also possible that mor complex 
phenomena are present, that are not fully understood at the present moment.

\subsection{A summary of the results}
I will now summarize the results. As we have seen we can distinguish a few temperature regions.
\begin{itemize}
	
\item For $T>T_f$ the only minimum of the free energy functional is given by high temperature result: we call it the 
liquid minimum (in the spins models described above it has  to zero magnetization).

\item For $T_f>T>T_D$ there is an exponentially large number of minima \cite{KPVI,BAFRPA,PP}.  For some values of the 
free energy density the complexity $\Sigma$ is different from zero, however the contribution to the free energy coming 
from these minima is higher that the one coming from the liquid solution with zero magnetization.  As discussed also in 
Cugliandolo's lectures the value $T_{D}$ coincides with the critical temperature of the mode coupling approach and in 
the glass community is called $T_{c}$.  The real critical temperature of the model, that we have called $T_{c}$ up to 
now is called $T_{K}$ for reasons that will be clear in the next section.

\item The most interesting situation happens in the region where $T_D>T>T_c$ (or 
$T_c>T>T_K$ using the glassy notation).  In this region the free energy is still given the 
high temperature solution (with zero magnetization in spin models), It is extremely surprising \cite{FP,MONA} that the 
free 
energy can be written also as the sum of the contribution of an exponentially large number of non-trivial 
minima as in eq. \ref{SUM})..

Although the free energy is analytic at $T_D$, below this temperature the the system at each given moment may stay in 
one of the exponentially large number of minima.  The time ($\tau$) to jump from one minimum to an other minimum is 
quite large and it is controlled by the height of the barriers that separate the different minima.  In the mean field 
approximation (i.e. for infinite range models) it is proportional to $\exp (A N)$ with non-zero $A$.  In short range 
models at finite dimensions we expect that the barriers are finite and $\tau\approx \tau_{0} \exp (\beta \Delta(T))$.
The quantity $\beta \Delta(T)$ is often a large number also at the dynamical temperature \cite{CGGP}  (e.g. O(10))
and the correlation time will become very large below 
$T_D$ and for this region $T_D$ is called the dynamical transition point.  The correlation time (that should be 
proportional to the viscosity) should diverge at $T_{K}$.  The precise form of the this divergence is not well 
understood.  It is natural to suppose that we should get divergence of the form $\exp(A/(T-T_{K})^{\nu})$ for an 
appropriate value of $\nu$ \cite{VF}, whose reliable analytic computation is lacking \cite{KTW,KLEIN}.  The value 
$\nu=1$ (i.e. the Vogel Fulcher law) is suggested by the experiments.

The equilibrium complexity is different from zero (and it is a number of order 1) when the 
temperature is equal to $T_D$ and it decreases when the temperature decreases and it vanishes 
linearly at $T=T_K$.  At this temperature (the so called Kauzmann temperature) the entropy of a 
single minimum becomes equal to the total entropy and the contribution of the complexity to the total 
entropy vanishes. At an intermediate temperature $T_{g}$ the correlation time becomes so large that it cannot be 
observed any more by humans.
\item
In the region where $T<T_K$ the free energy is dominated by the contribution of a few minima of the 
free energy having the lowest possible value.  Here the free energy is no more the analytic 
continuation of the free energy in the fluid phase.  A phase transition is present at $T_K$ and the 
specific heat is discontinuous here. 
\end{itemize}

\subsection{Some consideration on the free energy landscape and on the dynamics}
\begin{figure}
\includegraphics[width=.6\textwidth]{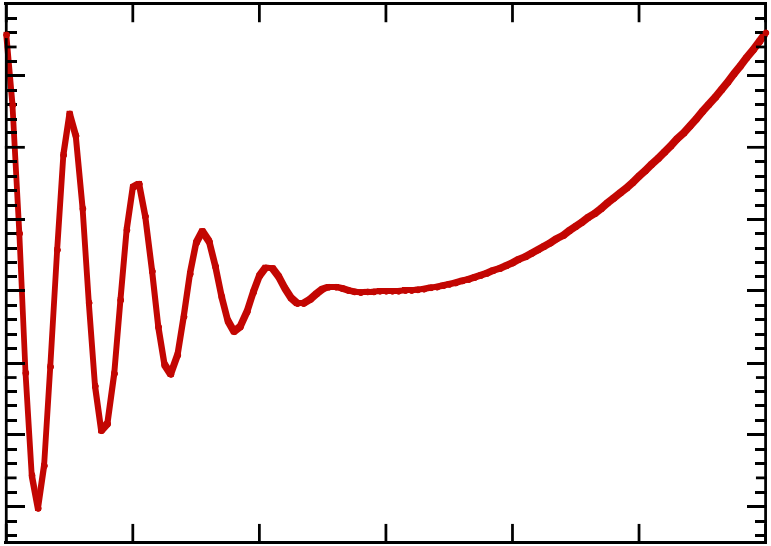}
\caption{The qualitative dependence of the free energy as function of the configuration space in 
the 
region relevant for the dynamical transition, i.e.  for $T<T_{D}$.  }\label{ART}
\end{figure}
The free energy landscape is rather  unusual; we present the following pictorial 
interpretation (fig. \ref{ART}), that is a rough simplification \cite{JL}.  At a temperature near  to 
$T_{D}$ the system stays in a region of phase space that is quite flat and correspond of 
a minimum of the total free energy.  On the contrary below $T_{D}$ the phase space is 
similar to the one shown pictorial in fig.  \ref{ART}.  The region of maxima and minima is 
separated by the region without barriers by a large nearly flat region.  The minima in the 
region at the left are still present also when $T_{f}>T>T_{D}$, but they do not correspond 
to a global minimum.

At temperatures higher than $T_D$ the system at thermal equilibrium stays in the right portion of fig.  \ref{ART}.  
When 
the temperature reaches $T_D$ the system arrives in the flat region.  The flatness of the potential causes a Van Hove 
critical slowing down that is well described by mode coupling theory \cite{MCT,BCKM} (that is exact in the mean 
field 
approximation).

In the mean field approximation the height of the barriers separating the different minima is 
infinite and the temperature $T_D$ is sharply defined as the point where the correlation time 
diverge.  The precise meaning \cite{FRAPA} of the dynamical temperature beyond mean 
field approximation has already been discussed.

\begin{figure}
\includegraphics[width=.6\textwidth]{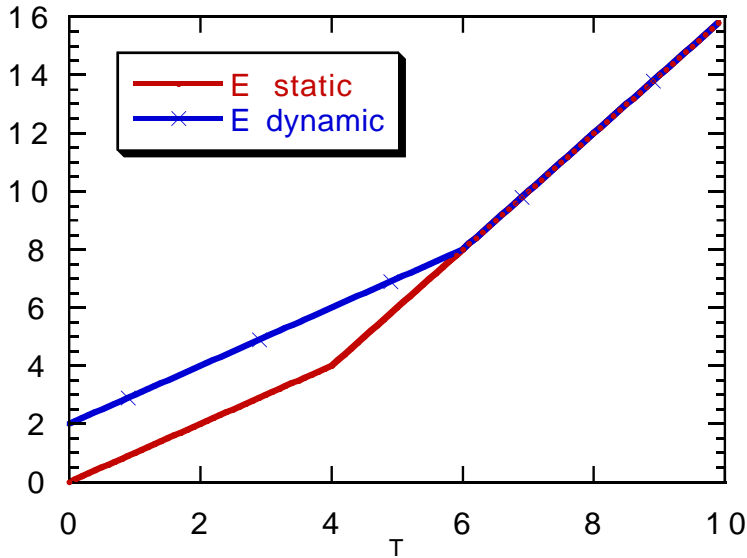} \label{EDYN}
\caption{The qualitative behaviour of the equilibrium energy and of the dynamical energy as function of 
the temperature.}
\end{figure}

Let us start from a very large system (of $N$ particles) at high 
temperature and let us gradually cool it.  It would like to go at equilibrium in the 
region with many minima.  However coming from high free energy (from the right) it cannot 
enter in the region where are many maxima; if we wait a finite amount of time 
(the time to crosses the barriers diverges as $\exp (AN)$.  the system remains confined in the flat region.  In 
this case \cite{CuKu,FM} the so called dynamical energy,
\begin{equation}
E_D=\lim_{t \to \infty} \lim_{N \to \infty} E(t,N),
\end{equation}
is higher that the equilibrium free energy.  The situation is described in fig.  \ref{EDYN}.The difference of the 
static and dynamic energy is an artifact of the mean field 
approximation if we take literarily the limit $t \to \infty$   
However it correctly describe the situation on laboratory times, where metastable states 
are observed.

In the mean field approximation  very interesting phenomena  happen below $T_D$ when 
the system is cooled from the high temperature phase due to the fact that 
the system does not really go to an equilibrium configuration but wanders in the phase space never 
reaching equilibrium.  The phenomena are the following:
\begin{itemize}
\item
The energy approaches equilibrium slowly when the system is cooled from an high energy 
configuration\cite{CuKu,BAPA}:
\begin{equation}
E(t,T)=E_{D}(T)+B(T) t^{-\lambda(T)}.\label{SLOWENE}
\end{equation}
where the exponent $\lambda(T)$ does not vanish linearly at zero temperature as 
happens for an activated process. 
\item
Aging is present, i.e.  the correlation functions and the response functions in the region of large 
time do depend on the story of the system \cite{B,POLI,FM}.
\item
In the region where aging is present the fluctuation dissipation 
theorem  is no more valid.  New generalized relations are satisfied \cite{CuKu,FM,FRARIE,MPRR,FDT,BK}, that 
replace the equilibrium fluctuation dissipation theorem.
\end{itemize}
These phenomena  will be discussed in details in Cugliandolo's lectures.

It is not clear how to compute in general the quantity $E_{D}(T)$.  There is a very simple recipe: $E_{D}$ is the 
largest energy where $\Sigma(E)\ge 0$.  According to that recipe one has to look to the smallest value of $m$ where the 
equation for $q$ has a consistent solution (as discussed in the previous section).  This last condition is equivalent 
to 
impose that the spectrum of small fluctuations of the free energy (defined in the next section) has a gap.  The value 
of 
$E_{D}$ is characterized by the fact the replica broken solution becomes unstable at $E>E_{D}$ 
(marginal stability).
This marginalistic approach correctly gives $E_{D}$ in the 
p-spin spherical model, it is not clear if it true in general \cite{BAFRPA}.

\subsection{Small fluctuations}

We have see that in the real world glassy systems have only one transition with divergent correlation time (at 
temperature $T_{K}$).  However in the idealized world of mean field theories there is a second purely dynamics 
transition $T_{D}$ at higher temperatures \cite{KTW}.  As it happens in many cases, slow relaxation is related to the 
existence of zero energy modes and this statement is true also here.  This statement can be easily verified in spin 
models where the mode coupling theory is exact and simple computations are possible.

In  spin models we  concentrate our attention on the Hessian of the free energy,  
 defined as
    \begin{equation}
    M(i,k)={\partial^{2} F[m] \over \partial m(i) \partial m(k)}\label {D2F} \ ,
    \end{equation}
where $F[m]$ is the (TAP) free energy as function of the magnetization and the 
magnetizations do satisfy the stationarity (TAP) equations:
\begin{equation}
{\partial F[m] \over  \partial m(i) }=0\label {D1F} \ .
\end{equation}
Performing the appropriate computations \cite{TAPSTA,Biroli} we finds that the spectral density of the 
Hessian  in these infinite range models has always a semicircular form:
\begin{equation}
\rho(\lambda)\propto \sqrt{(\lambda-\mu(T))(\lambda-\nu(T))}\ ,
\end{equation}
typical of random matrices \cite{Metha91}.

\begin{figure} \    \includegraphics[width=.6\textwidth]{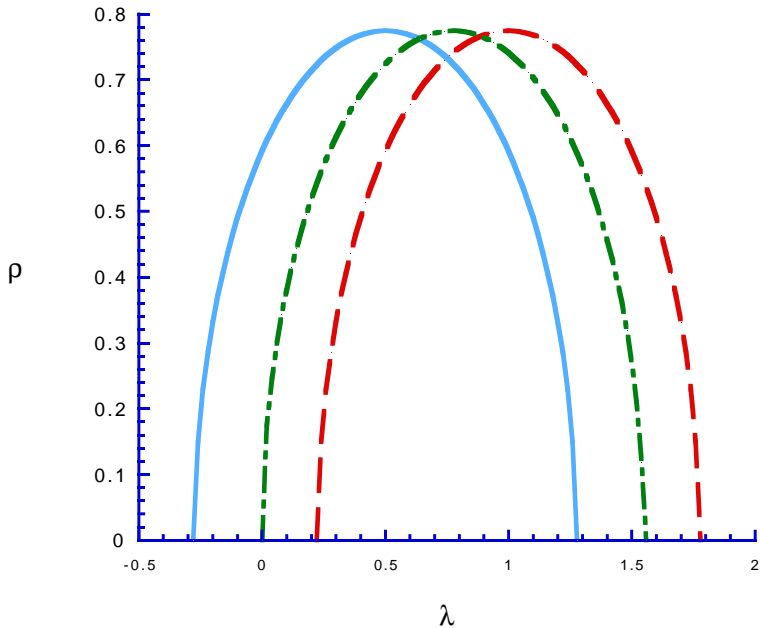}
    \caption{The qualitative behaviour of the spectrum in mean field approximation above 
    $T_{c}$, (full line), at $T_{c}$ (dot-dashed line) and below $T_{c}$ (dashed line) as 
    function of the eigenvalue $\lambda$).
    }\label{X}
    \end{figure}

\begin{itemize}
	\item At temperatures $T>T_{D}$)there are no non-trivial thermodynamically 
	relevant solutions of the equation \form{D1F} , however the dynamics is dominated by 
	quasi-solutions of the previous equations, i. e. by magnetizations such that the left 
	hand side of the previous equation is not zero, but small \cite{FV}..  
	The Hessian $M$ of the quasi-solutions has negative 
	eigenvalues and its spectrum has 
	qualitatively the shape shown in fig.  (\ref{X}).  These quasi stationary points of $F$ 
	look like saddles.
    \item
    At the dynamical transition point $T=T_{D}$ the quasi stationary points becomes real solutions of the equations 
    (\ref{D1F}).  They are essentially minima: the spectrum of the Hessian is non-negative and it arrives up to zero.  
    As it can be checked directly, the existence of these zero modes is responsible of the slowing down of the 
dynamics.  
    The different minima are connected by flat regions so that the system may travel from one minimum to an other 
    \cite{JL}.
		
    \item At low temperature the mimima become more deep, the spectrum develops a gap as shown in fig.  \ref{X} and the 
    minima are no more connected by flat regions.  In mean field models the system 
    would remains forever in one of these minima.  If the system starts from an high temperature configuration it 
cannot 
    reach these configurations. 
\end{itemize}

This picture is not so intuitive because it involves the presence of saddles with many 
directions in which the curvature is negative, and it is practically impossible to 
visualize it by making a drawing in a two or a three dimensional space.

This qualitative description can be easily verified in models where the mean field approximation is exact.  However, if 
we try to test it in finite dimensional models, we face the difficulty that the free energy functional $F[m]$ is a 
mythological object whose exact form is not known and consequently the eigenvalues of its Hessian cannot be computed.  A 
n alternative approach consists in studying the properties of the so called instantaneous normal modes (INM) 
\cite{INM1,Biroli,water} and by the saddle normal modes (SNM) \cite{C,ab,CGP,CGGP}.  For reasons of space this 
interesting point cannot be discussed any more the reader is invited to look to the original literatures.

These computations are particular relevant for glasses where the low temperature spectral density can be experimentally 
measured and one finds the famous Boson peak \footnote{The Boson peak is defined as a bump at some small but non-zero 
value of 
the eigenvalue $z$ of the spectral density $\rho(z)$ divided by the Debye density of states $z^{2}$) that has been 
observed in many material \cite{BOSE} and in numerical simulations \cite{BOSESIM}.}. In order to explain the Boson peak 
one 
has to do realistic computations, where one has to take into account the spectrum of phonons and this has been done or 
using the mode coupling formalism \cite{Gotze00}, or a microscopical approach \cite{GMPV}.

\section{Structural relations}
In this section we shall define a new equilibrium order parameter function $\tilde P(q)$, that is connected also to 
fluctuation-dissipation ratio $X(q)$ (see Cugliandolo's lectures).  This can be done by studying of the linear response 
to some special sets of perturbations of the original Hamiltonian \cite{FrMePaPe}.  This method has been recently used 
to derive 
interesting properties of the overlap distribution at equilibrium \cite{GUERRA,AI,GG,SOL}.

\subsection{Stochastic stability}

Our aim it to prove that the moments of $P(q)$  are  related to the 
response  of the system when one adds an appropriate perturbation \cite{GUERRA}.
This approach allows us to define all the relevant quantities in 
the case of single large system (in the infinite volume limit), while in the previous approach the function $P(q)$ was 
defined  as the probability distribution in an ensemble of different systems, characterized by different realizations 
of 
the disorder.  This difference is crucial if we consider the case (like glasses) where no disorder is present 
\footnote{In a glass we still have  the possibility of averaging over the total number of particles.}.

In the case of spin systems an appropriate form of the long-range perturbation
s given by:
\begin{equation}
H_p(\si)=\sum_{i_1,i_p=1,N}^{1,N}
K_{i_1,\ldots,i_p}\si_{i_1}\cdots \si_{i_p}, \label{HLR}
\end{equation}
where the couplings $K_{i_1,\ldots,i_p}$ are independent Gaussian variables with zero mean 
and variance $\ba{K_{i_1,\ldots,i_p}^2}=1/(2 N^{p-1})$.  

The Hamiltonian is
\begin{equation}
H_\epsilon=H_I+\epsilon H_p,
\end{equation}
The 
canonical average of $H_p$ 
verifies, for all values of $\epsilon$, the relation
\begin{equation}
\ba{\langle H_p\rangle} = -\beta \epsilon N \left( 1-\int
dq \; P_\epsilon(q)\, q^p \right), \label{SLR}
\end{equation}
irrespective of the specific form of $H_I$. Here the function
$P_\epsilon(q)$ is the probability  distribution of the overlap $q$ in the
presence of the perturbing term in the Hamiltonian; the average is done over the new couplings $J$ at fixed
$H_{I}$.
The derivation, involves only an
an integration by parts in a finite system. 

However the previous equation is strange. The function $P_{\eps}(q)|_{\eps=0}$ depends on the instance of the problem 
also in the infinite volume limit, while, for $\eps\ne 0$, $\langle H_p^{\rm LR} \rangle $ is a thermodynamic quantity 
shat cannot fluctuate in the infinite volume limit when we change the instance of the system.  (at least for generic 
$\eps$).  A further difficulty appears when there is an additional symmetry in absence of the random perturbation that 
is broken by the random perturbations (e.g. the spin reversal symmetry for a spin system in absence of a magnetic field 
or the translational invariance for glasses). Therefore for a large system we may expect that
\be
\tilde P(q) \ne P_{I}(q),
\ee
where $P_{I}(q)$ where is the usual overlap probability distribution computed at $\eps=0$, that may depend on the 
instance, while $|\tilde P(q)$ is the limit $\eps \to 0$ of the function $P_{\eps}(q)$ (where the limit is 
computed outside the cross-over region, i.e. $\eps>>N^{1/2}$) and it does not depend from the instance.

We need  to understand better how  the equilibrium state in the presence of the 
perturbation $\eps H_p$ is related to the equilibrium state at $\eps=0$.  Some 
complications  arise whenever the equilibrium expectation value of $H_p$ is not the same for 
all pure phases of the unperturbed systems.  Then the limit value of $H_p$ as $\epsilon\to 
0$ will be the one corresponding to the favored phases.  A simple example is the Ising 
model in the ferromagnetic phase, where one adds a 
negative magnetic field term as a perturbation.  Also in the vanishing field limit, the 
the system stays in a state with negative spontaneous 
magnetization, while the unperturbed measure corresponds to a mixture of the positive and 
negative magnetization pure states.  

In presence of many equilibrium states that enter in the Gibbs-Boltzmann measure, as it happens where replica symmetry 
is broken, the 
situation is rather complex.  Indeed, the stochastic perturbations that we have considered 
will in general reshuffle the weights of the different ergodic components in the Gibbs 
measure, or even change their nature, and this changes the $P(q)$ function to a different 
one ($\tilde P (q)$).

The principle of stochastic stability assumes that if consider an appropriate ensemble as 
for the initial random system we  have that
\be
P(q)\equiv \ba{P_{I}(q)}= \tilde P(q)
\ee
where here the bar denotes the average over the instances of the system in the appropriate ensemble.

An intuitive motivation for assuming stochastic stability is that 
perturbations we use are random  and they are not correlated with the original 
Hamiltonian.  So they should change the free energies of the various pure states of the 
original systems by random amounts.  Stochastic stability assumes in glassy systems the 
\textit{distribution} of these free energies is stable under independent random 
increments, as has been shown in mean field (in fact this property lies at the heart of 
the cavity method \cite{mpv}).  If this is the case then the two functions $\tilde P(q)$ 
and $P(q)$ will coincide.  We could also say that the equality of $\tilde P(q)$ 
and $P(q)$ indicates that the systems responds to a random perturbation  as a generic 
random system.

There are cases where stochastic stability trivially fails, i.e. when the 
original Hamiltonian has an exact symmetry, that is lifted by the perturbation.  The 
simplest case is that of a spin glass with a Hamiltonian invariant under spin inversion.  
In this case $P(q)=P(-q)$, since each pure state appears with the same weight as its 
opposite in the unperturbed Gibbs measure.  On the other hand, if we consider $H_p$ with 
odd $p$, this symmetry is lifted.  This means that in the $\eps \to 0$ limit only half of 
the states are kept.  If the reshuffling of their free energies is indeed random, then we 
shall have $\tilde P(q)=2\theta(q)P(q) \equiv \hat P(q)$.  The same type of reasoning 
applies whenever the overlap $q$ transforms according to a representation of the symmetry 
group of the unperturbed Hamiltonian $H_0$.

If this \emph{trivial} effect of exact symmetries is taken into account, for a large class of systems, the function 
$\tilde P(q)$ in the limit of small perturbations tends to the order parameter function $\hat P(q)$ of the pure system 
where the exact symmetries are lifted.  This continuity property is called \textit{stochastic stability}.  Ordinary 
systems without symmetry breaking and mean-field spin glasses are examples of stochastically stable systems.  In 
ergodic 
systems, the equality of $\tilde{P}$ and $\hat{P}$ is immediate,  both functions consist in a single delta 
function.  Thus, the problem of deriving the equality between $\tilde{P}$ and $\hat{P}$, appears only when there are 
coexisting phases are unrelated by symmetry.
Unfortunately, we are not able to characterize the class of stochastically stable systems.  In particular there is no 
rigorous proof that short-range spin glass, for which our theorem is most interesting, belong to this class.  However, 
stochastic stability has been established rigorously in mean field problems \cite{GUERRA,AI}.

If one studies more carefully the problems, one finds that stochastic stability has far reaching consequences, e.g.
\be
\ba{P_{I}(q_{1})P_{I}(q_{2})}=\frac23 P(q_{1}) P(q_{2}) +\frac13 P(q_{1}) \delta(q_{1}-q_{1})
\ee
These (and other) relations have been carefully numerically verified  in also in numerical simulations of three 
dimensional spin glasses  models \cite{MPRR} strongly suggesting the validity of stochastic stability.

\subsection{A simple consequence of stochastic stability}
Stochastic stability is a very powerful property, and it is the ingredient that allows to relate the properties of the 
low lying configurations, that dominate the Gibbs measure, to those of the configurations much higher in energy that are 
seen in the dynamics.  This is most easily explained in the usual framework of replica-symmetry breaking, considering an 
approximation with only two possible values of the overlap, $q_{0}$ among different states and $q_{1}$ among the same 
state (i.e., one-step replica-symmetry breaking).  The probability of finding a state with total free energy $F_{\al}=F$ 
is given by $\rho(\Delta F)$, where $\Delta F=F-F_0$ and $F_0$ is a reference free energy the equilibrium free energy.  
The weight of each state is given by
\begin{equation}
w_{\al}\propto \exp (-\beta F_{\al}).
\end{equation}
In one-step replica-symmetry breaking, the states that contribute to 
the Gibbs measure have nearly degenerate free energies.  We have already seen the non-extensive fluctuation of their 
free energies, corresponding to the low $\Delta F$ regime of $\rho(\Delta F)$, is given by \cite{GROMEZ,mpv_free}
\begin{equation}
\rho(\Delta F)\propto \exp(\beta m \Delta F),\label{ONESTEP}
\end{equation}
and the function $P(q)$ is given by
\begin{equation}
P(q)=m \delta(q-q_{0})+(1-m) \delta(q-q_{1})\ .
\end{equation}

Stochastic stability forces the function $\rho(\Delta F)$ to be of the form (\ref{ONESTEP}), not only when $\Delta F$ is 
finite, but also in the range where $\Delta F$ is \textit{extensive} but small (say of order $\eps N$).  Indeed it 
imposes that the form of the function $\rho(\Delta F)$ remains unchanged (apart from a possible shift in $F_{0}$) when 
one adds a small random perturbation \cite{FrMePaPe}.

Let us consider the effect of a perturbation of strength $\epsilon$ on the free energy of a state, say $\alpha$.  The 
unperturbed value of the free energy is denoted by $F_\alpha$.  The new value of the free energy $G_{\al}$ is given by 
$ 
G_{\al}=F_{\al}+\eps r_{\al} $ where $ r_{\al}$ are identically distributed uncorrelated random numbers.  Stochastic 
stability implies that the distribution $\rho(G)$ is the same as $\rho(F)$.  Expanding to second order in $\epsilon$ we 
see that this implies $d\rho/dF\propto d^2\rho/dF^2$, whose only physical solution (apart the trivial one $\rho(F)=0$, 
that corresponds to non-glassy systems) is given by eq.~(\ref{ONESTEP}) \footnote{The same conclusion could be obtained 
using the methods of reference \cite{SOL} computing the sample-to-sample fluctuations of the function $P_{J}(q)$, that 
in this case, where ultrametricity is trivially satisfied, are completely determined by the knowledge of of the 
function 
$P(q)$.}.  We see that stochastic stability fixes the form of the function $\rho$ and therefore connects in an 
inextricable way the low and the high free energy part of the function $\rho$.

This remark explain how it is possible that stochastic stability tells us something on the dynamics in the aging regime 
(as we shall see in the next section).  In the dynamical evolution from an higher temperature initial state,the 
difference between the total free energy at time $t$ and the equilibrium value will be always of order $N$, with a 
prefactor going to zero when $t$ goes to infinity.  One could argue that the dynamics probes the behavior of the 
function $\rho(\Delta F)$ at very large argument, and should not be related to the static property that depend on the 
function $\rho$ for small values of the argument.  However stochastic stability forces the function $\rho(\Delta F)$ to 
be of the form \form{ONESTEP} , also in the range where $\Delta F$ is \textit{extensive} but small, and this objection 
is no more valid.

\subsection{Fluctuation dissipation relations}
Let us now discuss the case of dynamics of a systems that at times 0 starts from a non-equilibrium configuration.  Here 
the relevant quantities are the two times correlations functions and the response functions, e.g. the correlation 
defined as
\be
C(t_{w},t)\equiv \frac{\sum_{i=1,N} \sigma_{i}(t_{w}) \sigma_{i}(t)}{N}
\ee

The finite-time response and correlation functions involved in the definition of the FDR are continuous 
functions of $\eps$ for $\eps\to 0$.  It is not evident if the limit is uniform in time, i.e. if the infinite times 
and the $\eps\to 0$ limits do commute.  The linear response regime may shrinks to zero as the 
time goes to infinity.  This possibility shows up when the perturbations favor one 
phase (as discussed in the introduction).  However  here we consider a  random perturbations and the expectations of 
$H_p$ vanishes at $\eps=0$: it is reasonable to assume that the linear response regime survives at very 
long times  \footnote{Let us stress that the existence of a linear response regime uniform in time a question 
susceptible of 
experimental investigation.}.

If  the infinite times limit and the $\eps\to 0$ limit do commute (i.e. the dynamical form of the 
stochastic stability), the dynamics is strongly constrained. For example, if we consider the 
function $X(q)$ that parametrizes the violations of the fluctuation dissipation relations in off-equilibrium 
experiments \cite{CuKu,FM}, the previous assumptions imply that:
\be
\tilde P(q)={dX \over dq} \ .
\ee. 
The proof is simple  it only involves integrations by parts. The details can be found in the original papers
\cite{FrMePaPe}.
\section{A short introduction to glasses}
Glasses \cite{glass_revue} are roughly speaking liquids that do not crystallize also at very low temperature 
(to be more precise: glasses also do not quasi-crystallize).

These liquids can avoid crystalization mainly for two reasons:
\begin{itemize}
    \item The liquid does not crystallize because it is cooled very fast: the crystallisation time 
    may become very large at low temperature (e.g. hard spheres at high pressure).  The system 
    should be cooled very fast at temperatures near the melting point; however if crystalization is 
    avoided, and the temperature is low enough, (e.g. near the glass transition) the system may be 
    cooled very slowly without producing crystalization.
    
    \item The liquid does not crystallize even at equilibrium.  An example is a binary 
    mixture of hard spheres with different radius: 50\% type A (radius $r_{A}$, 50\% type 
    B (radius $r_{A}$, where $R$ denotes $r_{B}/r_{A}$.  If $ .77 <R <.89$ (the bounds 
    may be not precise), the amorphous packing is more dense than a periodic packing, 
    distorted by defects.
   
\end{itemize}

Which of the two mechanism is present is irrelevant for understanding the liquid glass transition.

\begin{figure}
\includegraphics[width=.6\textwidth]{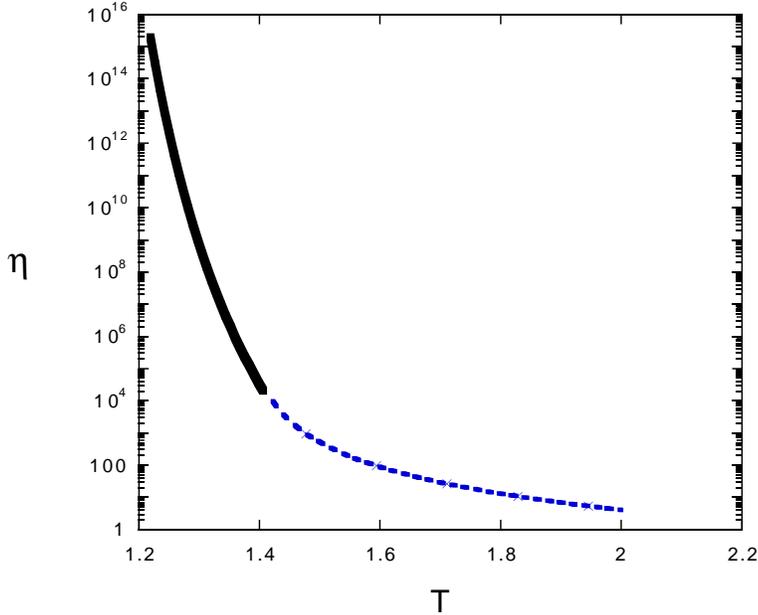}
\caption{The viscosity as function of the temperature according to the Vogel Fulcher law  
($T_{K}=1$ full line) and to the mode coupling theory  ($T_{K}=1.375$ dotted line).
 }
\label{fig_1}
\end{figure}

Other examples of glassy systems are binary mixtures: in this case we have 2 kinds of 
particles and the Hamiltonian of $N$ particles is given by
\begin{equation}
H=\sum_{a,b=1,2}\sum_{i=1,N(a)}\sum_{k=1,N(b)} V_{a,b}\op x_{a}(i)-x_{b}(k) \cp\ ,
\end{equation}
where $a=1,2$; $N(a)=N c(a)$; $\sum_{a=1,2} c(a)=1$.  In this case the values of the $N$ 
concentrations $c$ and the $3$ functions $V_{a,b}(x)$ describe the model.  Well 
studied case are:
\begin{itemize}
    
    \item A power potential (non-realistic, but simple), e.g.
 $c(1)=.5$, $c(2)=.5$, $V_{a,b}(x) =R _{a,b} x^{-12}$.

\item 
Lennard-Jones potential (more realistic), e.g. $c(1)=.8$, $c(2)=.2$, $V_{a,b}(x) =R _{a,b} x^{-12} -A_{a,b}x^{-6}$.

\end{itemize}

There are some choices of the parameters that are well studied such that the system does not crystallize. One  
has been introduced by Kob and Anderson in the L-J case \cite{KoAn}: it  corresponds to a particular choice of the 
parameters $R$ and 
$A$.

There are many other material that are glass forming, e.g.
short polymers, asymmetric molecules (e.g. OTP) \ldots

The behaviour of the viscosity in glass forming liquid is very interesting \footnote{The viscosity can be defined 
microscopically  considering a system in a box of large volume $V$: $T_{\mu,\nu}(t)$ is the total 
stress tensor at time $t$.  We define the correlation function of the stress tensor at different times:
$
\lan T_{\mu,\nu}(t) T_{\rho,\si}(0)\ran = V S_{\mu,\nu,\rho,\si} (t)
$
Neglecting indices, 
$\eta \propto \int dt S(t) \approx \tau^{-\al}$, where $\tau$ is the characteristic time of the 
system (in a first approximation we can suppose that
$\al=1$).}. There are  two regimes:
\begin{itemize}
    \item In an high temperature region the mode coupling theory \cite{MCT} is valid: it predicts 
    $\eta\propto (T-T_{d})^{-\ga}$, where $\ga$ is {\it not} an universal quantity and it 
    is O(1).
    
    \item In the low temperature region by the Vogel Fulcher law \cite{VF} is satisfied: it predicts that
    $\eta\propto \exp(A(T-T_{K})^{-1})$. Nearly tautologically fragile glasses can be defined as those glasses that 
have $T_{K}\ne 
0$; strong glasses have $T_{K} \approx 0$.

\end{itemize}

At the glass temperature ($T_{g}$) i the viscosity 
becomes so large that it cannot be any more measured. This happens after an increase 
of about 18 order of magnitude (that correspond to a microscopic time changing from 
$10^{-15}$ to $10^{4}$ seconds): the relaxation time becomes larger than the experimental 
tine.

A characteristic of glasses is the dependance of the specific heat on the cooling rate.  
There is a (slightly rounded) discontinuity in the specific heat that it is shifted at 
lower temperatures when we increase the cooling time.

\begin{figure}
    \includegraphics[width=.45\textwidth]{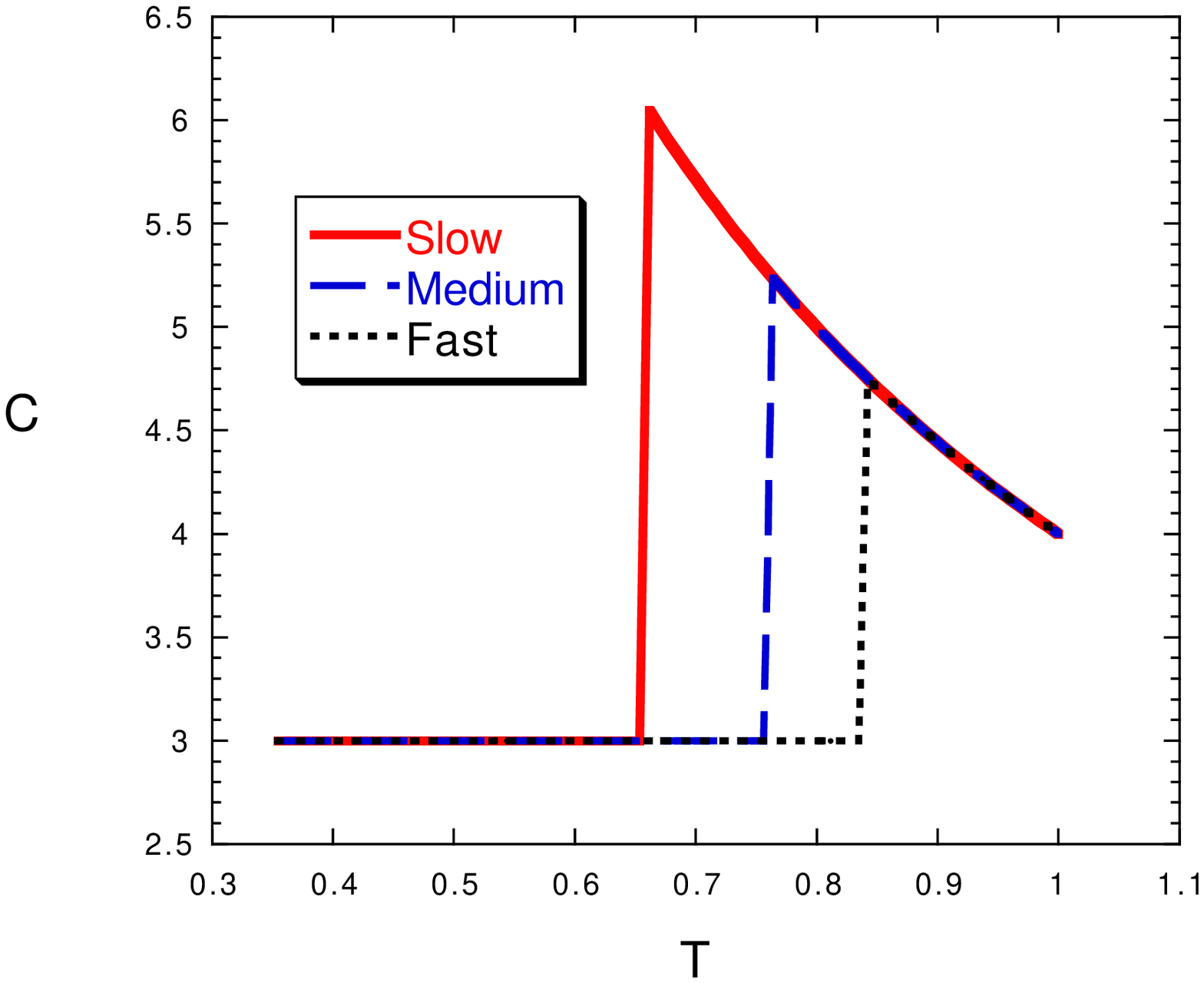}
    \includegraphics[width=.45\textwidth]{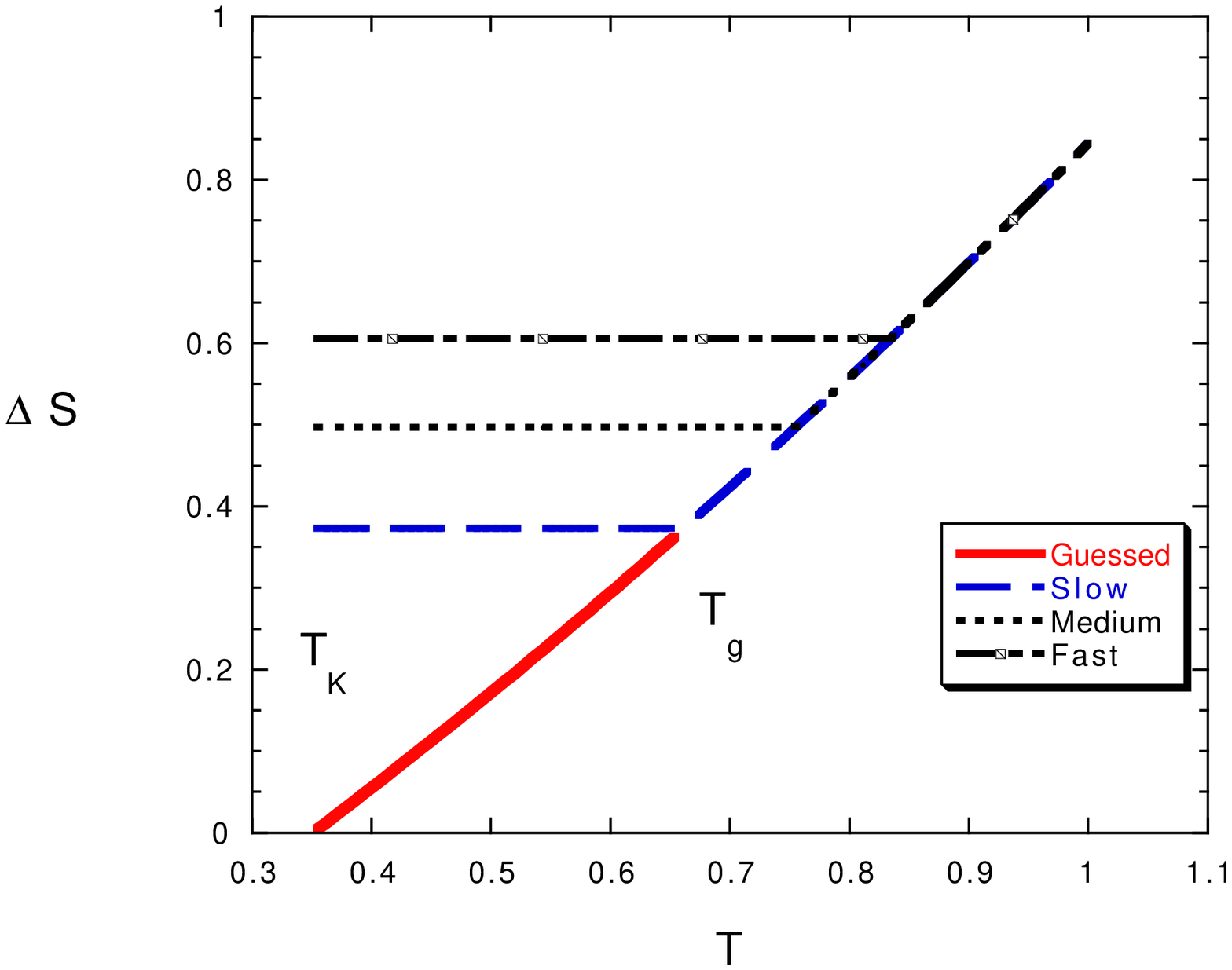}
\caption{The specific heat and the entropy excess $\Delta S$ for various cooling rates.
 }
\label{fig_2}
\end{figure}

For systems that do crystallize if cooled too slowly, one can plot $\Delta S \equiv 
S(liquid) - S(crystal)$ versus $T$ in the thermalized region.  One gets a very smooth 
curve whose extrapolation becomes negative at a finite temperature.  A negative $ \Delta S 
$ does not make sense, so there is a wide spread belief that a phase transition is present 
before (and quite likely near) the point where the entropy becomes negative.  Such a 
thermodynamic transition (suggested by Kaufmann) would be characterized by a jump of the 
specific heat.  Quite often this temperature is very similar to the temperature found by 
fitting the viscosity by the Volker Fulcher law and the two temperature are believed to 
coincide.

Usually a second order a thermodynamic phase transition induces a divergent correlation time.  However the exponential 
dependance of the correlation time on the temperature is not so common (in conventional critical slowing down we should 
have a power like behaviour); an other strange property of the glass transition is the apparent absence of an 
equilibrium correlation length or susceptibility (linear or non-linear) that diverges when we approach the transition 
point\footnote{There is a dynamical correlation length that diverge at the critical temperature that is observed 
numerically and it is predicted theoretically\cite{corr_length}.}.

\section{The replica approach to structural glasses: general formalism} 

In this section we write down the formulas corresponding to the replica approach introduced in the previous section.  
We 
keep here to the case of simple glass formers consisting of $N$ particles interacting by a pair potential $v(r)$ in a 
space of dimension $d$.

The reader may notice that in all the example that we have considered up to now a quenched disorder was present;
this feature is not present in glasses, where no random quenched variables are present in the Hamiltonian. However this 
in not a difficulty as far as there is no need of a quenched random disorder to use the replica formalism as it is 
clear 
from  the previous sections. However at the beginning it was believed that a random quenched 
disorder was  necessary in order to use the replica approach. Only much later \cite{MPR,FH,CKPR,CKMP} it was realized that the 
replica method could be used in translational invariant models where no disorder is present; some of these models 
behaves in a way very similar to real glasses: one can define both $T_{D}$ and $T_{K}$ and at low temperature the 
system 
crystallizes.

\subsection{The partition function}
The usual partition function, used e.g. in the liquid phase, is in tree dimension
\begin{equation}
Z_{1} \equiv {1 \over N!} \int \prod_{i=1}^N (d^{3} x_i) \  e^{-\beta H}
\end{equation}
We wish to study the transition to the glass phase through the onset of an
off-diagonal correlation in replica space\cite{MP,MePa1}. We use $m$ replicas and  introduce
the Hamiltonian of the replicated system:
\begin{equation}
H_m=\sum_{1 \le i<j \le N} \sum_{a=1}^m v(x^a_i-x^a_j) +
\eps \sum_{i=1 N} \sum_{a<b =m}w(x^a_i-x^b_i) 
\end{equation}
where $w$ is an attractive interaction.  The precise form of $w$ is unimportant: it should 
be a short range attraction respecting the replica permutation symmetry, and its 
strength $\eps$
that will be sent to zero in the end.  For instance one could take
\begin{equation}
w(x)=\ato  \frac{c^{2}}{x^{2+}c^{2}}\cto ^{6}
\end{equation}
with $c$ is of the order of 0.2 times the typical intermolecular distance. A positive value 
of $\eps$ forces all the particles one near to the other ones. If two systems stay 
in the same state, the expectation value of $w$ represents the self overlap and it very 
near to $1$.

The partition function of the replicated system is
\begin{equation}
Z_{m} \equiv {1 \over N!^m} \int \prod_{i=1}^N \prod_{a=1}^m (d^3 x^a_i) \  
e^{-\beta H_m}
\end{equation}

\subsection{Molecular bound states}
At low enough temperature, we expect that the particles in the different replicas may stay close to each other.  The 
role of the attractive term $w$ is to insure that all replicas fall into the same glass state,: also for small $\eps$ 
the 
particles in different replicas stay at the same place, apart from some thermal fluctuations:

Thermal fluctuations are relatively small throughout the solid phase (one can see this from the Lindeman criterion) and 
diffusion is very small, one can identify the molecules and relabel all the particles in the various replicas in such a 
way that the particle $j$ in replica $a$ always stays close to particle $j$ in replica $b$.  All the other relabelings 
are equivalent to this one, producing a global factor $N!^{m-1}$ in the partition function.

We therefore need to study a system of molecules, each of them consisting of $m$ atoms (one atom from each replica).  
It 
is natural to write the partition function in terms of the variables $r_i$ that describe the centers of masses of the 
molecules, and the relative coordinates $ u_i^a$, with $x_i^a=r_i+u_i^a$ and $\sum_a u_i^a=0$:

\bea \nn
Z_{m}&=& {1 \over N!} \int \prod_{i=1}^N \(( d^3r_i \)) 
 \prod_{i=1}^N \prod_{a=1}^m
(d^3 u_i^a) \prod_{i=1}^N \((m^3 \delta(\sum_a u_i^a)\)) \\
& & \exp\((-\beta \sum_{i<j,a} v(r_i-r_j+u^a_i-u^a_j) - \beta
\sum_i \sum_{a,b}W(u^a_-u^b_i) \))
\label{Zzu}
\eea

\subsection{ The small cage expansion}
In order to transform these ideas into a tool for doing explicit computations of the 
thermodynamic properties of a glass \cite{MePa1} we have to use an explicit method for computing the 
free energy as function of the temperature and $m$.  As is usually the case, in the liquid 
phase exact analytic computations are not possible and we have to do some approximations.  
In this section we shall use the fact that the thermal fluctuations of the particles in 
the glass are small at low enough temperature: the size of the `cage' seen by each 
particle is therefore small, allowing for a systematic expansion.  What we will be 
describing here are the thermal fluctuations around the minimum of the potential of each 
particle, in the spirit of the Einstein model for vibrations of a crystal.
 
We start from the replicated partition function $Z_{m}$ described in molecular 
coordinates in (\ref{Zzu}). Assuming that the relative coordinates $u^a_i$
are small, we can expand $w$ to leading order and write:
\bea \nn
Z_{m}(\alpha)&=& {1 \over N!} \int \prod_{i=1}^N \((d^3r_i\))  \prod_{i=1}^N 
\prod_{a=1}^m
(d^3 u_i^a) \prod_{i=1}^N \((m^3 \delta(\sum_a u_i^a)\)) \\
& & \exp\((-\beta \sum_{i<j,a} v(r_i-r_j+u^a_i-u^a_j) -
{1 \over 4 \alpha} \sum_i
\sum_{a,b} (u^a_i-u^b_i)^2 \))
\label{Zzu_exp}
\eea
In the end we are interested in the limit $\eps\equiv(1/\alpha) \to 0$.  We would like first to define the size $A$ of 
the molecular bound state, that is also a measure of the size of the cage seen by each atom in the glass, by:
\begin{equation}
{\partial \log Z_{m} \over \partial (1/\alpha)} \equiv {m (1-m) \over 2} d N A
=-{1 \over 4} \sum_i \sum_{a,b} \la (u_i^a-u_i^b)^2 \ra
\label{Adef}
\end{equation}
($d$ is the dimension that we have taken equal to 3 and $N$ is the number of particles).  We Legendre transform the 
free 
energy $\phi(m,\alpha)=-(T/m) \log Z_{m}$, introducing the thermodynamic potential per particle $\psi(m,A)$:
\begin{equation}
 \psi(m,A)= \phi(m,\alpha)+T d  { (1-m) \over 2} {A \over \alpha}
\end{equation}
What we want to see is whether there exists a minimum of $\psi$ at a finite value of $A$.

At low temperatures, this minimum should be at small $A$, and so we shall seek an expansion of $\psi$ in powers of 
$A$.  
It turns out that it can be found by an expansion of $\phi$ in powers of $\alpha$, used as an intermediate bookkeeping 
in order to generate the low temperature expansion.  

This may look confusing since we are eventually going to send 
$\alpha$ to $\infty$.  
However this method is nothing but a usual low temperature expansion in the presence of an infinitesimal breaking 
field.  
For instance if one wants to compute the low temperature expansion of the magnetization in a $d$-dimensional Ising 
model 
in an infinitesimal positive magnetic field $h$, the main point is that the magnetisation is close to one.  One can 
organise the expansion by studying first the case of a large magnetic field, performing the expansion in powers of 
$\exp(-2h)$, and in the end letting $h \to 0$.  A little thought shows that the intermediate -large $h$- expansion is 
just a bookkeeping device to keep the leading terms in the low temperature expansion.  What we do here is exactly 
similar, the role of $h$ being played by $1 / \alpha$.

\subsubsection{Zeroth order term}
We use the equivalent form:
\begin{equation}
Z_{m}(\alpha)= {1 \over N!} \int \prod_{i=1}^N \prod_{a=1}^m
(d^3 u_i^a)
\prod_i {d^3X_i \over \sqrt{2\pi \alpha \over m^2}^3}
\exp\((-\beta \sum_{i<j, a} v(x_i^a-x_j^a)-{m \over 2 \alpha} \sum_{i,a} 
(x_i^a-X_i)^2\)) \ .
\end{equation}
In the limit $\alpha \to 0$, the identity 
\begin{equation}
\exp\((- {m \over 2 \alpha} (x_i^a-X_i)^2\)) \simeq \(({2 \pi \alpha \over 
m}\))^{d/2}
\delta^3 (x_i^a-X_i)
\end{equation}
implies that:
\begin{equation}
Z_{m}^0(\alpha) = \(({2 \pi \alpha \over m}\))^{3 N (m-1)/2} 
{1 \over N!} \int \prod_i dX_i \exp\((- \beta m \sum_{i<j} v(X_i-X_j) \)) \ .
\end{equation}
In this expression we recognise the integral over the $X_i$'s as the partition function $Z_{liq}(T^*)$ of the liquid at 
the effective temperature $T^*$, defined by
\begin{equation}
 T^*\equiv T/m \ .
 \end{equation} 
Therefore the free energy, at this leading order, can be written as:
\begin{equation}
{\beta \phi^0(m,\alpha)} = {3  (1-m) \over 2 m} \log {2 \pi \alpha \over m}
-{3  \over 2 m}
\log(m)-{1 \over mN} \log Z_{liq}(T^*)
\end{equation}
The result is intuitive: in the limits where the particles of different replicas stay at the same point, the 
Hamiltonian for $m$ replicas is just the usual one, multiplied by $m$.

\subsubsection{First order term}
\label{small_cage}
In order to expand to next order of the $\alpha^{-1}$ expansion, we start from  the representation 
(\ref{Zzu_exp})
and expand the interaction term to quadratic order in the relative coordinates:
\bea \nn
Z_{m}&=& \int \prod d^3r_i d^3u_i^a \prod_i \(( m^3 \delta(\sum_a u_i^a) \))
\exp\((-\beta m \sum_{i<j}
v(r_i-r_j)\))
\\ \nn
& & \exp\((-{\beta \over 2} \sum_{i<j} \sum_{a \mu \nu} (u_i^a-u_j^a)
(u_i^a-u_j^a)
{\partial^{2} v(r_i-r_j) \over \partial r^{2}}
-{1 \over 4 \alpha} \sum_{a,b} (u_i^a-u_i^b)^2\)) \ .
\label{Zzu_quad}
\eea
(where for simplicity we have not introduced the  indices $\mu$ and $\nu$, running from $1$ to $d$, 
that denote space directions).
Notice that in order to carry this step, we need to assume that the
interaction 
potential
 $v(r)$ is smooth enough, excluding hard cores.

After some  computations one finds that 
the free energy to first order is equal to:
\bea
{\beta \phi(m,\alpha)}= \\{3  (m-1) \over 2 m} \log{1 \over \alpha} - \alpha
\beta 
 C +
{3 (1-m) \over 2 m} \log {2\pi \over m}-{3 \over 2 m } \log{m} -{1 \over mN}  
\log Z_{liq}(T^*)
\eea
where the constant $C$ is proportional to the expectation value
of  the Laplacian of the potential,
in the liquid phase at the temperature $T^*$: 
\begin{equation}
C \equiv  {1 \over 2} {1-m \over m^2} \sum_{j(\ne i)} \la \Delta v(z_i-z_j)\ra^*
\end{equation}
Differentiating the free energy  with respect to ${1 / \alpha}$ gives the equation for the size of the cage:
\begin{equation}
 {\beta }{\partial \phi \over \partial (1/\alpha)}= -{ (1-m) \over 2 m} d  
\alpha
+\alpha^2 \beta C=
 -{ (1-m) \over 2 } d  A
\end{equation}
Expanding this equation in perturbation theory in $A$ we have:
\begin{equation}
\alpha =m A -{ 2 \beta m^3 C \over 3 (m-1)} A^2
\end{equation}
The Legendre transform is then easily expanded to first order in $A$:
\bea\nn
 {\beta \psi(m,A) }= {\beta \phi}(m,\alpha) + 3 { (1-m) \over 2} {A \over 
\alpha}\\ = 
{3 (1-m) \over 2 m} \log(2 \pi A) - \beta m A C  +{3 (1-m) \over 2 m}
-{3 \over 2 m} \log m- {1 \over m N}  \log Z_{liq}(T^*)
\label{free_first_ord}
\eea

This very simple expression gives the free energy as a function of the number of replicas, $m$, and the cage size $A$.  
We need to study it at $m \le 1$, where we should maximise it with respect to $A$ and $m$.  The fact that we seek a 
maximum when $m<1$ instead of the usual procedure of minimising the free energy is a well established fact of the 
replica method, appearing as soon as the number of replicas is less than $1$ \cite{mpv}.

As a function of $A$ , the thermodynamic potential $\psi$ has a maximum
at:
\begin{equation} 
A=A_{max} \equiv
{d(1-m) \over 2 \beta m^2 } {1 \over C} = {3 \over \beta} {1 \over \int d^3r 
g^*(r) \Delta v(r)}
\label{A_first_ord}
\end{equation}
where $g^*$ is the pair correlation of the liquid at the temperature $T^*$.  A study of the potential 
$\psi(m,A_{max})$, 
that equals $\phi(m)$, as a function of $m$ then allows to find all the thermodynamic properties that we seek, using 
the 
formulas of the previous section.  This step and the results will be explained below in sect.  \ref{results}, where we 
shall  compare the results to those of other approximations.

\subsubsection{Higher orders}
The systematic expansion of the thermodynamic potential $\psi$ in powers of $A$ can be carried out easily to higher 
orders.  However the result involves some more detailed properties of the liquid at the effective temperature $T^*$.  
For instance at second order one needs to know not only the free energy and pair-correlation of the liquid at 
temperature $T^*$, but also the three points correlation.  The results for the second order, that will be discussed in 
next section, will be obtained in the framework of the hypernetted chain molecular approach that is described in 
appendix.

\subsubsection{Harmonic resummation}
One can obtain a partial resummation of the small cage expansion described above by integrating exactly over the 
relative vibration modes of the molecules.  We shall use such a procedure here, that is a kind of harmonic expansion in 
the solid phase \footnote{In some loose sense the first order in the $A$ expansion may be compared to
Einstein approximation for the specific heat and the harmonic approximation we describe he may be compared to Debye's 
jellium.}.

We work directly with $1/\alpha=0$ and start from the replicated partition function (\ref{Zzu_quad}), within the 
quadratic expansion of the interaction potential $v$ in the relative coordinates $u_i^a$.  (Clearly it is assumed that 
the $1/\alpha \to 0^+$ limit has been taken, and that its effect is to build up molecular bound states).  The exact 
integration over the Gaussian relative variables gives:

\begin{equation}
Z_m= {m^{N3/2} \sqrt{2 \pi}^{N 3 (m-1)} \over N!} \int \prod_{i=1}^N d^3r_i
\exp\((-\beta m \sum_{i<j}
v(r_i-r_j) -{m-1 \over 2} \Tr \log \((\beta M  \)) \))
\label{Z_harm_resum}
\end{equation}
where
the matrix $M$, of dimension $3N \times 3N$, is given by:
\begin{equation}
M_{(i \mu) (j \nu)}= \delta_{ij} \sum_k v_{\mu\nu}(r_i-r_k)-v_{\mu\nu}(r_i-r_j)
\end{equation}
and $v_{\mu\nu}(r) =\partial^2 v /\partial r_\mu \partial r_\nu$.
We have thus found an effective Hamiltonian for the centers of masses $r_i$ of 
the 
molecules, that basically looks like the original problem at the effective
 temperature $T^*=T/m$, complicated by the contribution of 
vibration modes that give the `Trace Log' term. 
We expect that this  should be a rather good approximation for the
glass phase. Unfortunately, even within this approximation,
it is not possible to compute the partition function exactly. The 
density of eigenstates of the matrix $M$ is a rather complicated object and we
have developed a simple approximation scheme in order to estimate it.

We thus proceed by using a \emph{quenched approximation}, i.e. neglecting the feedback of 
vibration modes onto the centers of masses.  This approximation becomes exact close to the 
Kauzman temperature where $m \to 1$.  The free energy is then:
\bea
 {\beta \phi(m,T)} =\\ -{3 \over 2 m} \log(m)- { 3 (m-1) \over 2 m } \log(2 \pi)
-{1 \over m N} \log Z(T^*) +{m-1 \over 2 m} 
\la Tr \log \((\beta M  \)) \ra^*
\eea
that involves again the free energy and correlations of the liquid at the temperature $T^*$.  Computing the spectrum of 
$M$ is an interesting problem of random matrix theory, in a subtle case where the matrix elements are correlated.  Some 
efforts have been devoted to such computations in the liquid phase where the eigenmodes are called instantaneous normal 
modes \cite{INM1}.  Here we use a simple resummation scheme that should be reasonable at high densities-low 
temperatures: it is described in the appendix \footnote{If the harmonic approximation were fully consistent, all the 
eigenvalues of $\cH$ (the so called INN, Instantaneous Normal Modes \cite{INM1}) should be positive.  This is not the 
case, however the number of negative eigenvalues becomes very small at low temperature, still in the liquid phase, 
signaling that valleys can be approximately defined in this region.}.  Using these results we can compute the 
replicated 
free energy $F_m$ only from the knowledge of the free energy and the pair correlation of the liquid at the effective 
temperature $T^*$.  The results will be discussed in section \ref{results}.

\subsubsection{Without replicas}
Up to now we have used the replica theory.  Replica theory is a very powerful tool, but it has the disadvantage that 
many of the underlying physical hypothesis cannot be seen in a clear way.  We will rederive some of the previous 
formulae without using the replica formalism \cite{MONA,LJ}.  We only suppose that at low temperatures the phase space of 
the system can be approximately divided into valleys that are separated by high barriers \cite{landscape,KST,APRV}.  In 
a first approximation each valley can be associated to one inherent structure \cite{St,HP}, i.e. one minimum of the 
potential energy and consequently there is a one to one correspondence among the valley at two different temperatures.

Let us consider a  system with $N$ particles with Hamiltonian $H(C)$, $C$ denoting the 
generic configuration of the system.  If we use the approach of the previous sections, 
the crucial step is the computation of the generalized partition function:
\begin{equation}
Z(\gamma;\beta)\equiv \exp ( -N \gamma G(\gamma;\beta))=\sum_{a} \exp( -\gamma N f_{a}(\beta)).
\end{equation}

Using the definition of the free energy in a valley
\be
\int_{C\in \alpha}  d C  \exp (-\gamma H(C))= \exp (-N\gamma 
f(\gamma,C)),
\ee
we obtain 
\bea
Z(\gamma;\beta) = 
\int  d C \exp \ato-\gamma H(C) -N \gamma  f(\beta,C) + N\gamma 
f(\gamma,C)\cto = \nonumber \\
\int  d C \exp \ato-\gamma H(C) -N \gamma  \hat{f}(\beta,C) + N \gamma 
\hat{f}(\gamma,C)\cto,
\label{MIRA}
\eea
where $\hat{f}(\beta,C)= f(\beta,C)-f(\infty,C)$ and $f(\beta,C)$ is a function that is 
constant in each valley and it is equal to the free energy density of the valley to which 
the configuration $C$ belongs. 

Before entering into the computation of $\hat{f}(\beta,C)$ it is useful to make the so called \emph{quenched 
approximation}, i.e. to make the following approximation inside the previous integral:
\begin{equation}
\exp (-A \hat{f}(\beta,C))=\exp(-A \lan \hat{f}(\beta)\ran_{\gamma}), \label{APPROX}
\end{equation}
where $\lan \hat{f}(\beta)\ran_{\gamma}$ is the expectation value of 
$\hat{f}(\beta,C)$ taken with the probability 
distribution proportional to $\exp(-\gamma H(C))$. The quenched approximation is exact if the 
temperature dependance of the energy of all 
the valleys is the same, apart from an overall shift at zero temperature. In other words we 
assume that the minima of the free energy have different values of the free energy but 
similar shapes. The quenched  approximation would be 
certainly bad if we were using the free energy ${f}(\beta,C)$ at the place of  $\hat{f}(\beta,C)$
because the zero temperature energy strongly varies when we change the minimum.

We finally find 
\begin{equation}
 G(\gamma;\beta)= F_{L}(\gamma)+ \hat{f}_{\gamma}(\beta)  
-\hat{f}_\gamma(\gamma),
\end{equation}
  $F_{L}(\gamma)$ being the free energy of the liquid ($S_{L}(\gamma)$ and $S_{\gamma}(\gamma)$ are respectively the 
entropy 
of the liquid and of a valley).  A simple algebra shows that
\begin{equation}
\Sigma(\gamma;\beta)=
S_{L}(\gamma)- S_{\gamma}(\gamma) +\hat{f}'_{\gamma}(\gamma) -
  \hat{f}'_{\gamma}(\beta), \label{FINAL}
\end{equation}
where $
\hat{f}'_{\gamma}(\beta) = {\partial \hat{f}_{\gamma}(\beta) / \partial \gamma}
$.

In the liquid phase, we find out that the  configurational entropy is given by
\begin{equation}
\Sigma(\beta)=\Sigma(\beta;\beta)=S_{L}(\beta)- S_{\beta}(\beta).
\end{equation}
The entropy of the liquid is  the entropy of the typical 
valley plus the configurational entropy.

The thermodynamic transition is characterized by the condition
\begin{equation}
\Sigma(\beta_{K})=0.
\end{equation}

In the glassy phase the free energy can be found by first computing the value of 
$\gamma(\beta)$ such that 
\begin{equation}
\Sigma(\gamma(\beta);\beta)=0\ .
\end{equation}
The quantity $\gamma(\beta)$ is the inverse of the effective temperature of the valley.  
It is easy to show  (following \cite{MONA}) that the previous formulae are completely equivalent to the 
replica approach. 

A strong simplification happens if we assume that the entropy of the valley can be evaluated in the 
harmonic approximation where we only keep the vibrational contributions.  For 
a system with $M$ degrees of freedom  the harmonic entropy of the valley near to a 
configuration $C$ is given by
\begin{equation}
S(\beta(C))=\frac{M}{2} \ln \ato {2 \pi e \over \beta}\cto -\frac12 \Tr\ato\ln (\cH(C) ) \cto,
\end{equation}
where $\cH(C)$ is an $M \times M$ Hessian matrix (e.g. if $H$ depends on the coordinates $x_{i}$ 
we have that $\cH_{i,k}=\partial^{2} H / \partial x_{i} \partial  x_{k}$).  The final 
result is just the same obtained with the replica method in the harmonic approximation if we put $m=\gamma / \beta$.

\section {The replica approach to structural glasses: some results}
\label{results}

\subsection{Three approximation schemes}

We have seen up to now three approximation schemes. 
\begin{itemize}
		\item
The small cage expansion has been carried out directly to first order in section 
\ref{small_cage}, and agree with the first order expansion within the molecular HNC 
approach.  
		\item

The second approximation scheme is the harmonic resummation method.  Again we have an 
explicit form (\ref{chain}) for the free energy per particle $\phi(m)$ only from the 
knowledge of the free energy and the pair correlation of the liquid at $T^*$.  Having this 
$m$ dependance the procedure to get the thermodynamic results is entirely the same as that 
of the first order result.
		\item

The third approximation scheme is obtained by the expansion of the molecular HNC free 
energy to second order in the cage size, as described in the appendix.  
\end{itemize}

For each of the three approximation schemes mentioned above, we need to compute the free energy and the pair correlation 
of the liquid in a temperature range close to the glass transition.  We will first consider three dimensional soft 
spheres \cite{MePa1} interacting through a potential $v(r)=1/r^{12}$.  We work for instance at unit density, since the only relevant 
parameter is the combination $\Gamma=\rho T^{-1/4}$.  In this case e have used the hypernetted chain approximation to 
get both the correlation function $g(r)$ and the free energy.  Later on we will present the results for an LJ binary 
mixture within the harmonic approximation.

\begin{figure}
\includegraphics[width=.40\textwidth,angle=-90]{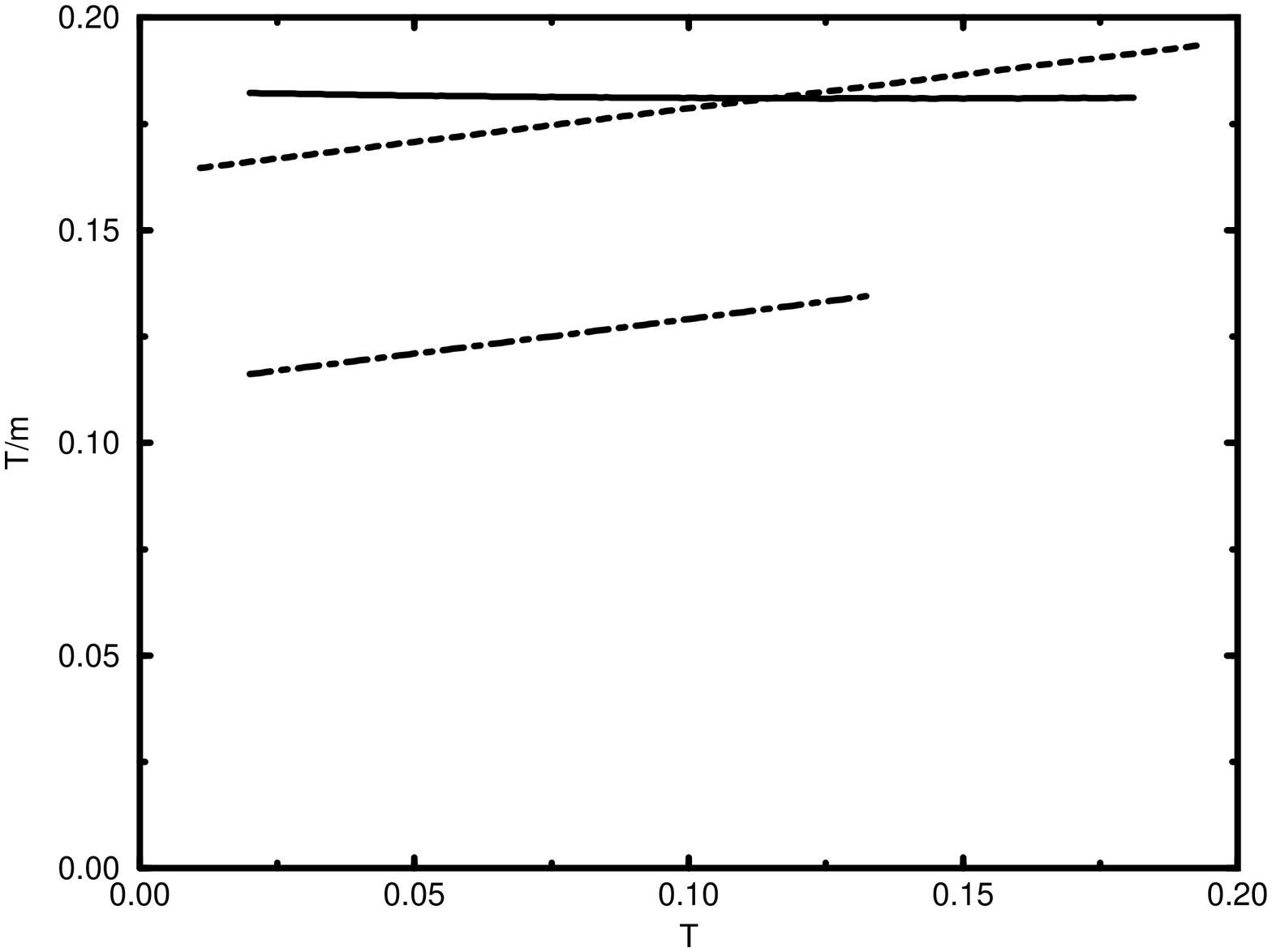}
\includegraphics[width=.40\textwidth,angle=-90]{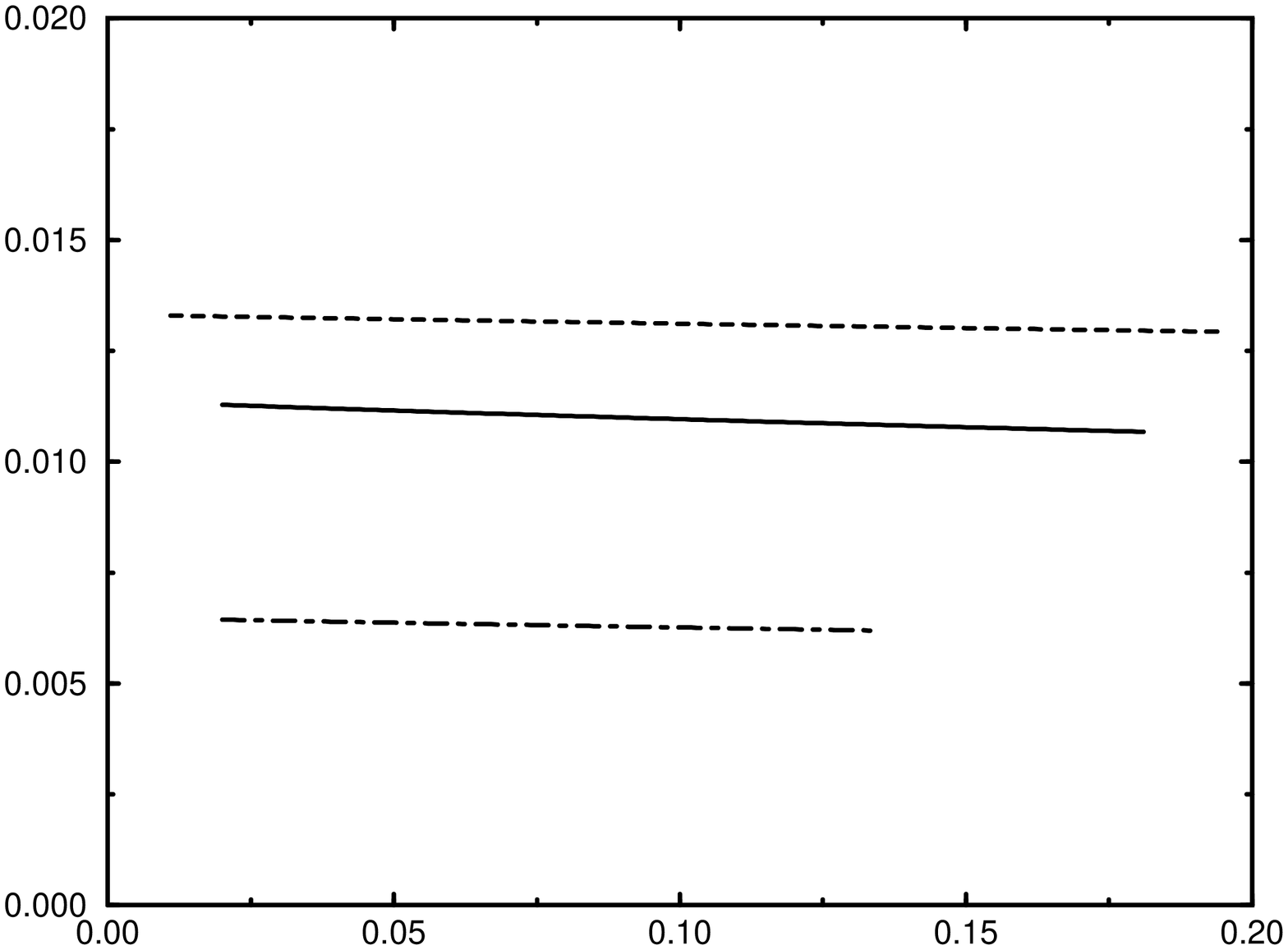}

\caption{The  effective temperature of the molecular liquid at the transition, 
$T^*= T/m^*$ (left) and the quantity $A/T$,
 versus the temperature $T$, computed in an expansion to first 
order (dashed-dotted line) and second order(full line) in the cage size $A$,
and in the harmonic resummation (dashed line). }
\label{fig_m}
\end{figure}

\subsection{Critical temperature and effective temperature}
We plot in fig.  (\ref{fig_m}) the inverse of the effective temperature $T^*$, equal to $m^*/T$, versus the temperature 
$T$ of the thermostat.  The transition temperature is given by $T^*=T$.  This gives the ideal glass transition 
temperature.  Within the first order expansion we find $T_K \simeq .14$; the harmonic resummation gives $T_K \simeq 
.19$ 
and the second order perturbation theory is $T_K \simeq .18$ We see that the two best methods, the second order and 
harmonic resummation, are in good agreement and both give a critical value of $\Gamma$ around $\Gamma \simeq 1.52$.  
This value of $\Gamma$ is in good agreement with numerical estimations of the glass transition of the soft sphere 
system, 
that range around 1.6 \cite{HANSEN}.
We also notice that the effective temperature stays relatively constant when the actual 
temperature varies.  The effective temperature  $T^*$ (that can be experimentally observed) is always near to
$T_K$, independently from the value of the temperature $T$,

\subsection{Cage size}
In replica space the cage size characterizes the size of the molecular bound state, in the approximation of quadratic 
fluctuations, as defined in (\ref{Adef}).  Its physical meaning is easily established: In the glass phase at low 
temperatures one can approximate the movement of each atom as some vibrations in a harmonic potential in the 
neighborhood of a local minimum of the energy.  The typical square size of the displacement is given by:
\begin{equation}
A= \la (r_i- \la r_i \ra)^2\ra
\end{equation}
that is the physical definition of the square size.  The cage size is plotted versus temperature in fig.  \ref{fig_m}.  
The cage size is nearly linear in temperature, as it would be in a $T$-independent quadratic confining potential.  This 
indicates that the local confining potential has little dependance on the temperature in the whole low temperature 
phase.

\begin{figure}
\includegraphics[width=.46\textwidth,angle=-90]{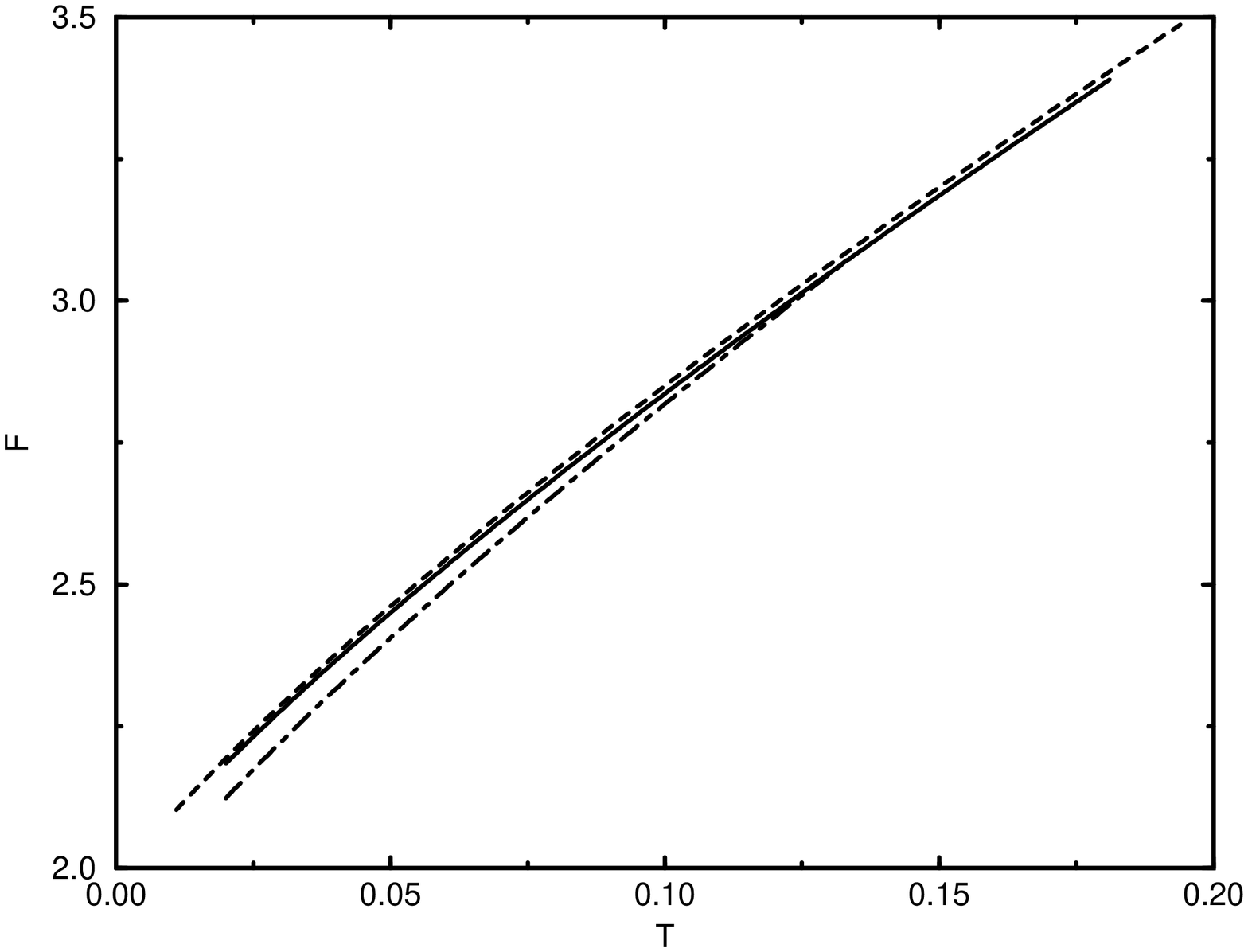}
\includegraphics[width=.46\textwidth,angle=-90]{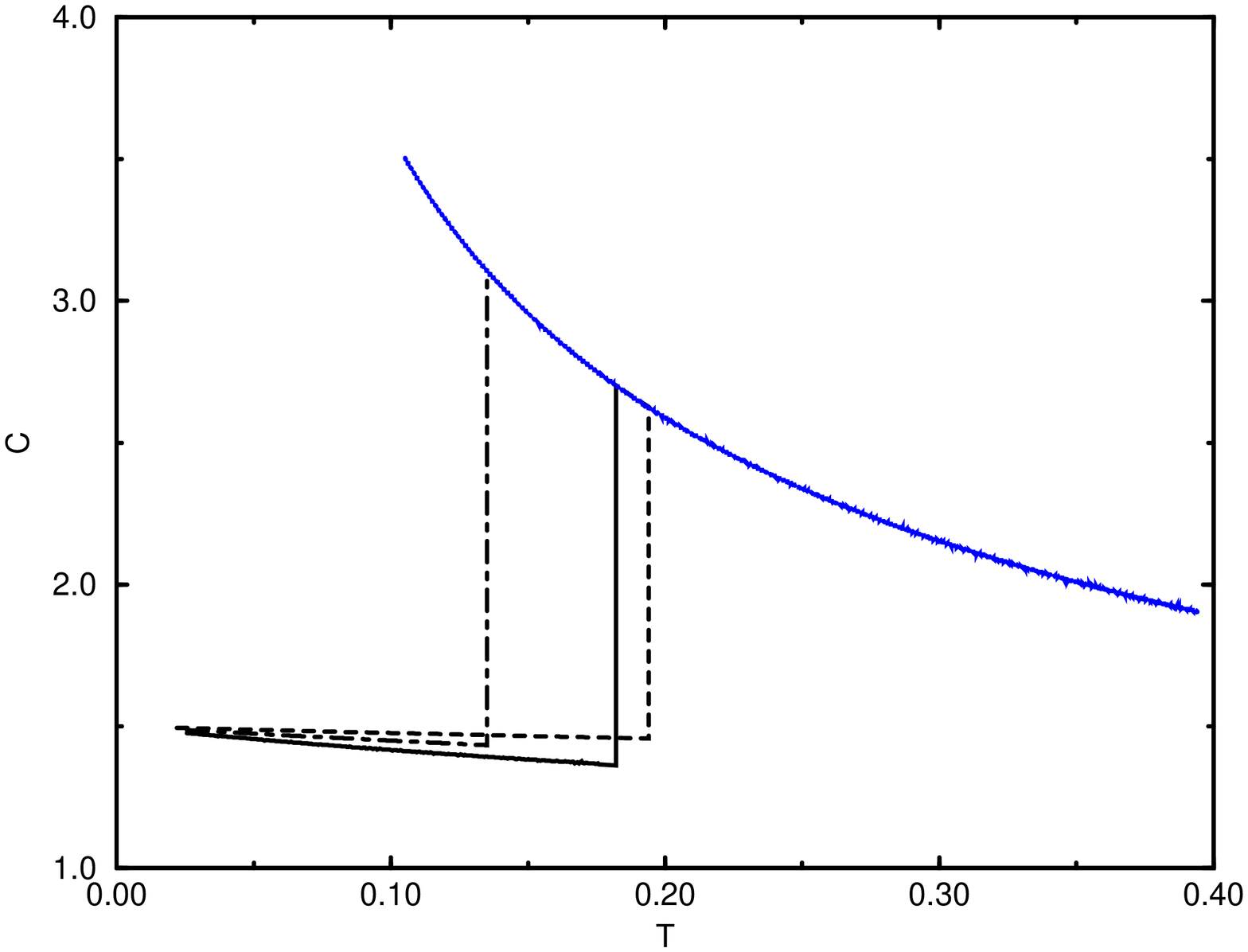}

\caption{The free energy (left) and the specific heat (right) versus the temperature,
computed in an expansion to first 
order (dashed-dotted line) and second order (full line) in the cage size $A$,
and in the harmonic resummation (dashed line). In the right panel the dotted line is the specific
heat of the liquid.}
\label{fig_C}
\end{figure}

\subsection{Free energy, specific heat and configurational entropy}
If we plot the free energy versus the temperature one would see a strong consistency between the second order term of 
the small cage expansion and the harmonic resummation .  Both data extrapolates at zero temperature to a ground state 
energy of order 1.95.  This is related to the typical energy of the amorphous packings of soft spheres.  More 
precisely, 
if we consider all the amorphous packings of soft spheres at unit density, we can count them through the zero 
temperature configurational entropy.  The lowest energy where one can find an exponentially large number of such 
packings is the ground state energy of the glassy phase that we find equal to 1.95.  However we have not taken into 
account the existence of a crystal: therefore we must first remove all crystal like configurations, i.e. configurations 
that 
correspond to a crystal with some local defects.  These configurations can be characterized by the presence of delta 
functions at the appropriate values of the momenta.  This procedure of identifying crystal like solutions has been 
explicitly done numerically in \cite{landscape}.

In fig.  \ref{fig_C} (right panel) we plot the specific heat versus temperature.  It is basically constant and equal to 3/2.  This is 
the Dulong-Petit law (we have not included the kinetic energy of the particles, that would give an extra contribution 
of 
$3/2$).  This result is very welcome: if we had treated the crystal at the same level of approximation as we considered 
here for the glass, we would have the Einstein model for which the specific heat is also given by the Dulong-Petit 
law.  
Thus the specific heat of the glass is very near to that of the crystal, has it happens experimentally.  Notice that it 
was not obvious at all a priori that we would be able to get such a result from our computations.  The fact of finding 
the Dulong-Petit law is an indication that our whole scheme of computation gives reasonable results for a solid phase.

\begin{figure}
\includegraphics[width=.6\textwidth,angle=-90]{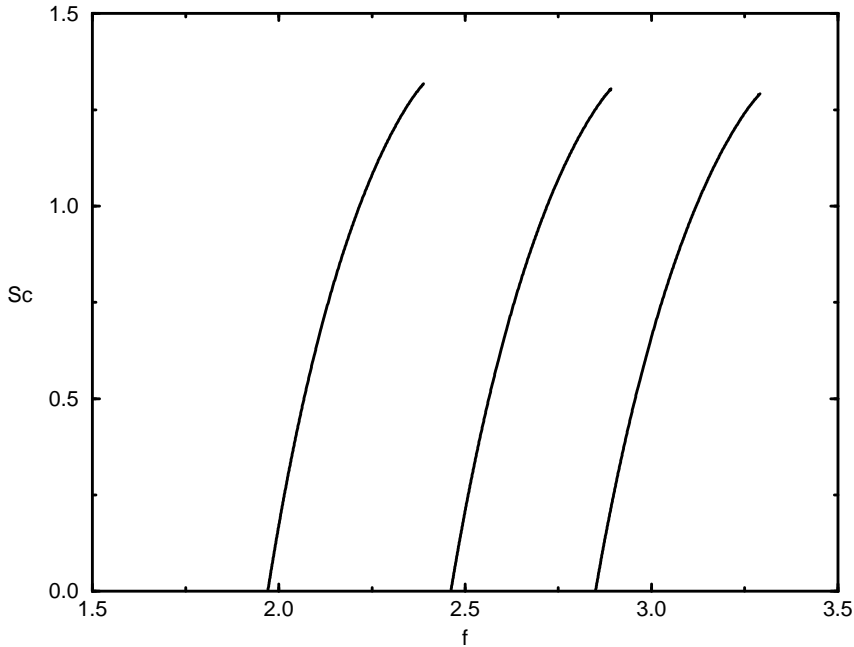}
\caption{The configurational entropy $\Sigma(f)$ versus the free energy,
computed within the harmonic resummation, 
at temperatures
$T=0.,.05,.1$ (from left to right). }
\label{fig_Sc}
\end{figure}

In fig. \ref{fig_Sc} we show the configurational entropy versus the free energy
at various temperatures, including the zero temperature case. We have included
here for simplicity only the result from the harmonic resummation procedure.

We notice that the various curves corresponding to different temperatures are not far from being just shifted one from 
another by adding a constant to the free energy.  This indicates that the main effect of temperature is to add a 
constant ($\approx 3/2 k T$) in the energies of all amorphous packings.  This correspond to the case where 
 the vibration spectrum is approximatively state independent.

\subsection{The missing dynamical transition}
We know that at the mean field level there exists a dynamical transition at a temperature $T_D$ larger than the 
thermodynamic transition temperature $T_K$.  This phase is characterized by the dynamic statement that a system will 
remain forever in the same valley, and its free energy is greater than the equilibrium one because it misses the 
contribution of the configurational entropy.  This dynamic phase is just a mean field concept, that should disappear 
when corrections, such as activated processes, due to the short range nature of the potential, are taken into account.  
However if the barriers are sufficiently high, metastable states have a very large life time and they strongly affect 
the dynamics.  

In the framework of the harmonic resummation one finds that the approximation breaks down at small but positive $\eps$ 
if the matrix of second derivatives has negative eigenvalues \footnote{ In the small cage expansion the perturbative 
method assumes that there is always a bound state: the breakdown of this assumption hardly be seen in a perturbative 
approach.} .  From this point of view the appearance of negative eigenvalues signal the dynamic transition.  
Unfortunately in the chain approximation all the eigenvalues are positive at all temperatures and no dynamic phase 
transition can be seen: the free energy is always well defined for small $\eps$.: the chain approximation may be 
reasonable at low temperature but it is certainly not good at high temperatures.  One should use a better method to 
compute the spectrum, giving reasonable results also at higher temperature.  e.g. following the approach of \cite{GMPV} 
(work in this direction is in progress).

It is clear that a study of the dynamical phase transition should be done refining the tools than the one we have 
developed here.  This is not surprising: the dynamical phase transition is present at a temperature higher than the 
static one and the approximations that we had used are especially good at low temperature.

\subsection{Lenhard-Jones binary mixtures}
There other model that is interesting to consider because they do not crystallize \cite{BIN,LJ}. Here we report the 
results fo  
is a {\sl realistic} model for glasses, is given by a binary mixture of particles (80\% 
large particles, 20 \% smaller particles) interacting via a Lennard-Jones potential, 
introduced by Kob and Andersen \cite{KoAn}.  This Hamiltonian should mimick the behaviour 
of some metallic glasses and it is one of the best studied and simplest Hamiltonian that 
do not lead to crystalization at low temperature. We first present the data at the density $\rho=1.2$.

The fact that the systems does not crystallizes implies that we can easily obtain information doing numerical 
simulations without serious difficulties \cite{LJ}.  Therefore one can use both an analytic or a numerical method to get 
information in the liquid phase.  Let us start by presenting the numerical results \cite{LJ}.  A system of $N=260$ 
particles, in a cubic box with periodic boundary conditions at density $\rho=1.2$ has been studied via Monte Carlo 
simulations.  The entropy is obtained using the formula $S(\beta)=S(0)+\int_{0}^{\beta}d\beta' (E(\beta)-E(\beta'))$.  
Given an equilibrium configuration the nearest minimum of the potential is found by steepest descendent and the 
computation of the 780 eigenvalues of $\cH(C)$ does not present any particular difficulty.

The results for the total entropy of the liquid and for the harmonic part are shown in fig.  (\ref{DUE}), left  as 
function of $T^{-.4}$ (a more detailed description of the simulations can be found in ref.  
\cite{LJ}).  The entropy of the liquid is remarkable linear when plotted versus $T^{-.4}$, as it
happens in many cases \cite{Ro}.

\begin{figure}
\includegraphics[width=.35\textwidth,angle=270]{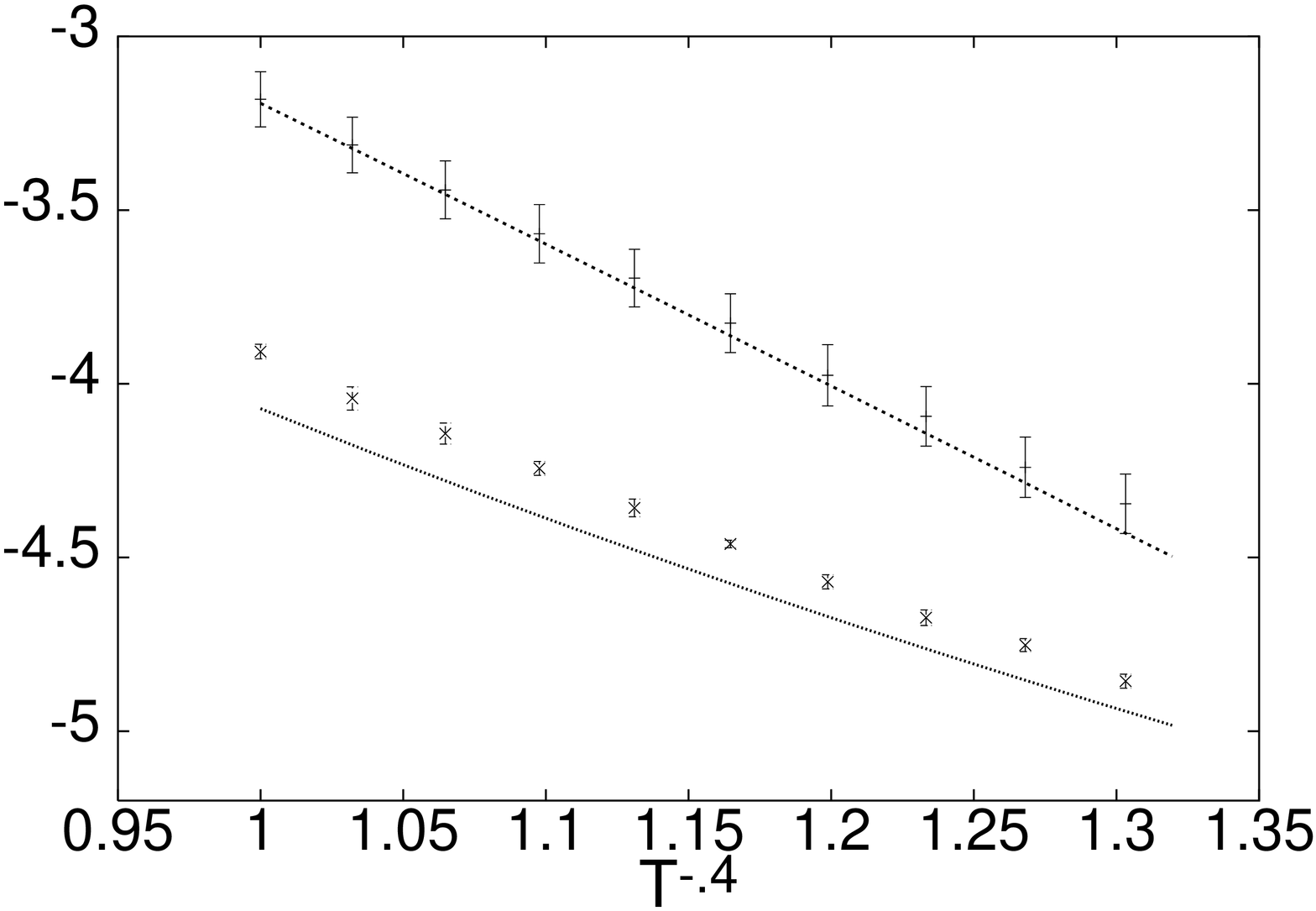} 
\includegraphics[width=.35\textwidth,angle=-90]{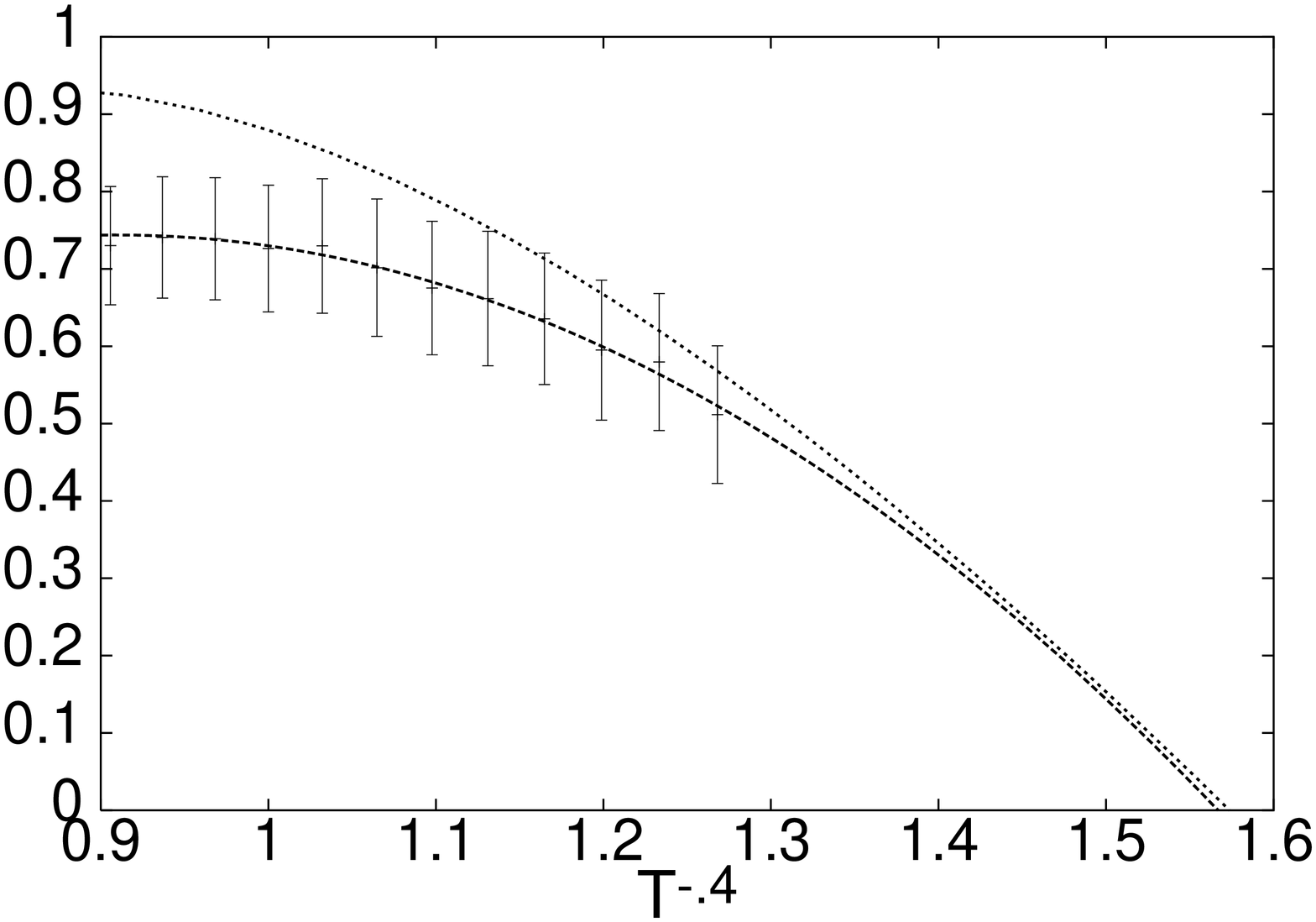}
\caption{ Left panel: analytical entropy of the liquid (upper line) compared with 
the numerical one, and analytical harmonic entropy 
(lower line) compared with the  numerical results. The horizontal axis is   
$T^{-.4}$. Right panel: analytical value of the complexity (upper line) compared with 
the numerical one ($+$ points), as functions of $\beta^{.4}$.}
\end{figure}\label{DUE}

In fig. (\ref{DUE}), right, we show the configurational entropy as function of $T^{-.4}$.  We fit it with a 
polynomial of second degree in $T^{-.4}$.  The extrapolated configurational entropy becomes zero at 
a temperature $T_{K}=.31\pm.04$, where the error contains systematic effects due to the 
extrapolation (similar conclusions have been reached in ref.  \cite{KST}).

There are many methods to compute analytically the free energy in the liquid phase that 
lead to integral equations for the correlation functions.  Unfortunately the simple HNC dos not works well. 
Here I present the results \cite{LJ} obtained by mixing the HNC  and MSA 
(mean spherical approximation) closures \cite{ZeHa}..  This technique allow us to compute 
the internal energy in the liquid with a reasonable approximation.  
The computation of the spectrum is  more involved: it can be done with the same  approach we 
used before that is described in the appendix.

We can now put everything together\cite{LJ}: the final analytic predictions for the liquid and harmonic entropy are 
shown in fig.  (\ref{DUE}), left.  The predictions for the liquid entropy turn out to be very good, while there is a minor 
discrepancy for the harmonic entropy, likely due to the rather strong approximations in the analytic evaluation of the 
spectrum.  The analytic configurational entropy is shown in fig.  (\ref{DUE}), left)  It becomes zero at $T_{K}=.32$, that 
is our analytic prediction for the thermodynamic transition.

We have in our hands all the tools to compute analytically the free energy in the low temperature case.  If one 
computes 
the specific heat,  one finds that the Dulong Petit law is extremely well satisfied in the 
low temperature region \footnote{In the harmonic approximation the Dulong Petit law would be exact if we neglect the 
$\gamma$ dependence of $S_{\gamma}(\beta)$.}. The value of $\gamma(\beta)$ weakly depends on $\beta$: its value in the 
limit $\beta \to \infty$ is only about 10\% higher that its value (i.e. $\beta_{K}$) at the transition temperature.

\begin{figure}
\includegraphics[width=.45\textwidth,angle=-90]{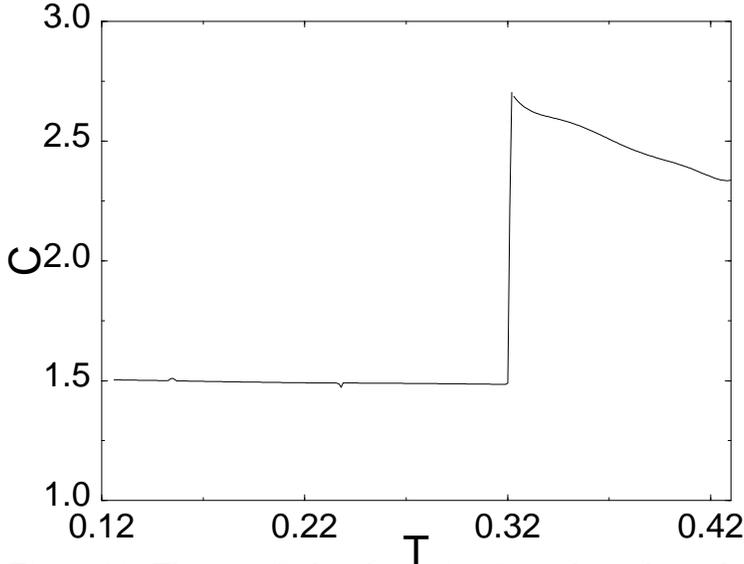}\label{TRE}
\caption{The specific heath coming from the $x$ dependent part 
of the Hamiltonian as a function of the temperature}

\end{figure}

Summarizing we have found a simple method that is able to use liquid theory method in the glasses phase 
putting in practice the old adage {\sl a glass is a frozen liquid}.  We are able to compute with a 
reasonable approximation the thermodynamics and with a little more effort we can compute the static 
and the dynamic structure functions.

\begin{figure}
\includegraphics[width=.6\textwidth,angle=-90]{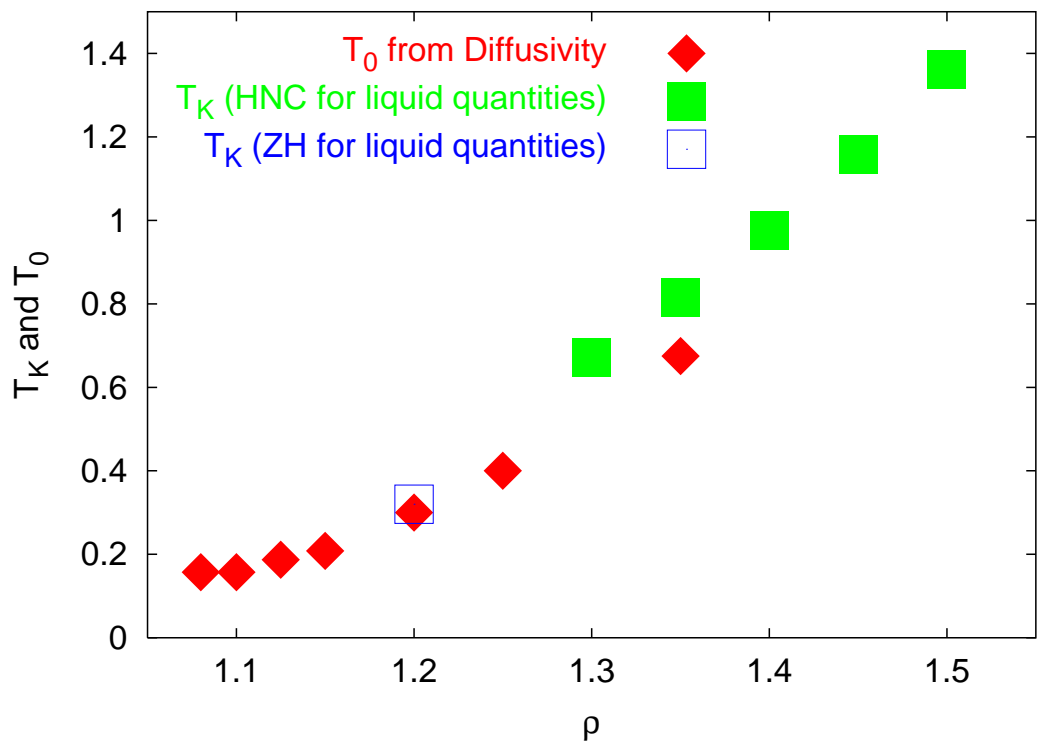}\label{COM}
\caption{The data on $T_0$, ($T_{0}$ should be near to $T_{k}$) obtained by Sastry \cite{Sastry} compared with
the analytical results \cite{BPV}.}
\end{figure}

We may wonder what happens if we change the density $\rho$. The Kauzmann  temperature should be depend on $\rho$.
A numerical estimated of the Kauzmann  temperature can be done by studying the behaviour of the diffusivity: this leads to 
a numerical of 
a temperature $T_{0}$ that would be near to $T_{k}$. The results are shown in fig. \ref{COM} together with the
analytic estimates for the same model  \cite{BPV} that are in reasonable agreement with the numerical data. 
\section{Discussion and perspectives}

Within the equilibrium framework, we have implemented so far our general strategy using rather crude methods.  These 
methods should be improved, and one should perform a more careful study of the molecular liquid. The extension to 
the case of hard spheres seem to be particularly interesting.

  There are also many other different problems that should be investigated,
\begin{itemize}
    \item A careful numerical study of the phase diagrams for coupled replicas would be welcomed in order to test the 
    correctness of the detailed theoretical predictions
    
    \item It would be very interesting to demonstrate numerically that the Kauzman transition exists beyond mean field 
    theory and that the behaviour of the system is similar to the one we expect theoretically.  Unfortunately known 
    numerical techniques are not sufficient for thermalize a glass forming liquid at low temperature in a reasonable 
    amount of CPU time: during the simulation the system is trapped in one of the exponentially large number of 
    metastable states ($O(exp(AN))$ where $A$ is a quantity of order 1),.  The introduction of an appropriate lattice 
    model, enough simple to be simulated in an very effective way, could be very useful in this respect.
    
    \item Analytic 
    and numerical method should be developed in order to compute the height of the activation barriers that should 
    dominate the dynamics at low temperature (below the mode coupling transition).  
    
    \item The analytic computation of 
    the spectrum of the instantaneous normal modes could be strongly improved.  This could br used to study the 
    properties of the Boson peak and to find analytically the value of the mode coupling and Kauzman temperature within 
    the same approach.  
    
    \item The analytic approach to the computation of the thermodynamic properties should be 
    extended to the case of hard spheres.
    
    \item 
    Least, but not the last, quantum glasses are a very wide territory that should be explore theoretically with much 
    more details.
\end{itemize}

The study of the properties at equilibrium study is to be considered as a first step before dealing the
 dynamics (equilibrium and off-equilibrium).  A very 
interesting and open problem is the computation of the time dependent correlation 
functions (and as a by-product the viscosity) in the region above $T_{K}$ and below $T_{c}$.  However a 
better understanding of activated processes in this framework is a crucial prerequisite.

It is quite possible that in the next years we shall see  progresses in some of the previously mentioned fields.

\section*{Acknowledgements}
This work has has been possible only by the help with many 
friends.  I would like to thank all of them.  A particular 
acknowledgement goes to Luca Angelani, Andrea Cavagna, Barbara Coluzzi, Leticia Cugliandolo, Silvio Franz, Irene 
Giardina, Tomas Grigera, Jorge Kurchan, Enzo Marinari, Victor Martin-Mayor, Marc M\'ezard, Federico Ricci, Felix 
Ritort, 
Juan-Jesus Ruiz-Lorenzo, Giancarlo Ruocco, Paolo Verrocchio, Miguel Virasoro and Francesco Zuliani.
\section 
*{Appendices}
\subsection*{The probability distribution of the weights}
We need to compute the probability distribution of the weights $w$ that are equal to:
\begin{equation}
w_k\equiv {\exp(- F_k) \over Z}\ ,
\end{equation}
where 
\be
Z = \sum_{k}\exp(- F_k)\ .
\ee
The probability of finding an energy $F_{k}$ in the interval $[F,F+dF]$,
is given by 
\begin{equation}
 \rho(E)=\exp (\beta  m F) \ ,
\end{equation}
and the possible values of the $F_{k}$ go from $-\infty$ to $+\infty$.

This computation  can be done with a brute force method \cite{mpv_free}. One assumes that
the index $k$ goes from 1 to $M$ and the $F$'s belong to the interval $[-\infty,F_{M}]$. One finally sends $M$ to 
infinity, keeping the quantity $M \exp(m F_{M})$ equal to a constant (the value of the constant is irrelevant).
Esplicite formulae can be written and the appropriate approximations can be done in the limit $M$ going to 
infinity using the saddle point limit. The computation is not too long, but it is not too simple.

A much more clever method can be found in \cite{RUELLETREE}. Here one tries to compute 
\be
w^{(s)}\equiv \sum _{k=1,\infty}w_k^{(s)}\ .
\ee
without introducing the cutoffs $M$ and $F_{M}$. After a few passages one finds that
\be
w^{(s)}={\Gamma(s-m)\over  \Gamma(s) \Gamma(1-m) } \ .
\ee
In this way we can go backward and we finally find that the probability of finding a $w_{k}$ in the interval
$[w,w=dw]$ is given by $\nu(w)dw$, where $\nu(w)$ is given by eq. (\ref{QUELLA}).

\subsection*{The mean field approach for a ferromagnet}
It is interesting to see the techniques of Gaussian integration and saddle point at work in a case where we already know 
the result.  The simplest case where this can be done is  the infinite range ferromagnet 
\cite{PARISISTAT}, a model that can be exactly solved..  The Hamiltonian is given by:
\be
H={J\over N} \sum_{i,k=1,N}\si(i)\si(k) - h \sum_{i=1,N}\si(i),
\ee
where the sum over  ${i,k}$ is done over all the $N(N-1)/2$ pairs of spins.  
An direct computation shows that in the limit $N\to \infty$ the mean field approximation become exact and the
magnetisation satisfies the mean field equation
\be
m=\th(\beta (J m+h)) \ . \label{MEANF}
\ee
We would like to prove this result using the saddle point method.

At this end we can rewrite the Hamiltonian as
\be
H={J\over 2N} \left( \sum_{i=1,N}\si(i) \right) ^{2} - h \sum_{I=1,N}\si(i),
\ee
where we have neglected and addictive constant equal to $-J$, that originates from the sums of the terms with $i=k$.

The argument runs as follows. In the limit $N \to \infty$, neglecting multiplicative constants that give
no contribution to the free energy density) we can write that

\bea
Z= \sum_{\{\sigma\}} \exp(-\beta H) \nonumber \\
\sum_{\{\sigma\}}\int dm \exp \left( -N \beta J m^{2}/2 +(\beta J m+ \beta h)
\sum_{i=1,N}\si(i) \right) =\\
\int dm \exp (-N f(m)) \ ,
\eea
where
\be
f(m)= \beta J m^{2}/2 -\ln\left( 2 \ch(\beta J m+ \beta h)\right) \  .
\ee

Up to now everything was exact.  The long range nature of the interaction and its homogeneity allows us to reduce the 
sum over $N$ variable to an one dimensional integral.  This integral can be estimated by various means.  In the limit 
$N\to \infty$ we can us the method of the point of maximum.  It gives
\be
Z \approx \exp(-N f(m^{*})) \ ,
\ee
where $m^{*}$ is the minimum of the function $f(m)$ and it is a solution of the equation
\be
{\partial f  \over \partial m}= 0 \ .
\ee
The previous equation coincide with the usual  mean field equation \form{MEANF} .
The correction to this result are proportional to $1/N$ (if we are not at the critical temperature) and they can be 
exactly computed by evaluating the corrections to the saddle point method.

\subsection*{The spectrum of the instantaneous normal modes}
Here I would like to explain a fast method\cite{MEPAZEE} for obtaining some analytic estimates on the spectrum of the matrix
the matrix $M$, of dimension $3N \times 3N$,  given by:
\begin{equation}
M_{(i \mu) (j \nu)}= \delta_{ij} \sum_k v_{\mu\nu}(r_i-r_k)-v_{\mu\nu}(r_i-r_j)
\end{equation}
and $v_{\mu\nu}(r) =\partial^2 v /\partial r_\mu \partial r_\nu$. The fact that the diagonal terms of $M$ do fluctuate 
complicate the analysis. This difficulty may be removed, if 
we notice that in this high density regime\footnote{Here and in what 
follows, we have not written explicitly the density.  We choose to work with density unity in order to simplify the 
formulae; however we assume that in the glassy phase the density is sufficiently high, compare to the range of the 
forces, 
that an expansion in inverse powers of the density gives the correct result. At low density te approach of 
\cite{CAGIAPA} gives the correct results.}.there are many neighbours to each point, 
and thus a good approximation is to neglect the fluctuations of these diagonal terms and substitute them by their 
average value.  We thus write:
\begin{equation}
\sum_k v_{\mu\nu}(r_i-r_k)\simeq \delta_{\mu\nu}
{1 \over 3} \int d^3r g^*(r) \Delta v(r) \equiv r_0
\label{diag_non_flu}
\end{equation}

In this approximation (\ref{diag_non_flu}) the diagonal matrix elements are all equal and can be factorized, leading to:
\begin{equation}
\la \Tr \log \(( M  \)) \ra^* =
3 N  \log (\   r_0 )+
\la Tr \log \(( \delta_{ij} \delta_{\mu\nu} - {1 \over r_0}
v_{\mu\nu}(r_i-r_k) 
\)) \ra^*
\label{tracelog}
\end{equation}
This form lends itself to a perturbative expansion in powers of
$1/r_0$ that we assume to be a \emph{small} number. The computation of the $p$-th order term in this expansion, 
\begin{equation}
{\cal T}_p \equiv {(-1)^{p-1} \over p r_0^p} \la 
\sum_{{i_1...i_p} \atop {\mu_1...\mu_p}} 
v_{\mu_1 \mu_2}(r_{i_1}-r_{i_2})...
v_{\mu_{p-1} \mu_p}(r_{i_{p-1}}-r_{i_p})
v_{\mu_p \mu_1}(r_{i_p}-r_{i_1}) \label{CATENA}
\end{equation}
still involves the $p$-th order correlation functions of the liquid at $T^*$.  We have approximated this correlation by 
introducing a simple \emph{chain} approximation involving only the pair correlation.  This chain approximation consists 
in replacing, for $p>2$, the full correlation by a product of pair correlations.  A most drastic approximation consist 
in selecting only those contributions that survive in the high density limit; systematic corrections can be computed 
following \cite{MEPAZEE,GMPV}.  Within the chain approximation, ${\cal T}_p$ is approximated by:
\bea
{\cal T}_p &=& \sum_{\mu_1...\mu_p}
\int dx_1...dx_p \  g^*(x_1,....,x_p) \  [v_{\mu_1 \mu_2}(x_1-x_2)...
v_{\mu_{p-1} \mu_p}(x_{p-1}-x_p )v_{\mu_p \mu_1}(x_p-x_1)]
\\ 
&&\simeq
\sum_{\mu_1...\mu_p}
\int dx_1...dx_p \  [g^*(x_1-x_2) v_{\mu_1 \mu_2}(x_1-x_2)]...[g^*(x_p-x_1) 
v_{\mu_p \mu_1}(x_p-x_1)] \ .
\eea
In this last form we need to compute a convolution that can be factorised
through the introduction of the Fourier transform of the pair correlation
function. We thus introduce 
the Fourier transformed functions $a$ and $b$ that are defined
from the pair correlation $g^*(r)$ by:
\begin{equation}
\int d^3 r \  g^*(r) v_{\mu\nu}(r) e^{ikr} \equiv \delta_{\mu\nu} \ a(k) +
\(( {k_\mu k_\nu \over k^2} -{1 \over 3} \delta_{\mu \nu}\)) b(k) \ .
\label{defab}
\end{equation}
In terms of these Fourier transforms, the $p$-th order term in the $1/r_0$
expansion is simply
\begin{equation}
{\cal T}_p=
\int d^3k \((a(k)+{2 \over 3} b(k)\))^p + (2) \int d^3k  \((a(k)-{1 \over 3} 
b(k)\))^p \ ,
\end{equation}
and the summation of the series over $p$ is easily done and we finally find:
\bea
\la \Tr \log \((\beta M  \))= \ra^* {d } \log (\beta r_0)
\\
\nn
+
\int d^3k \(( L_3 \(({a(k)+{2 \over 3} b(k) \over r_0}\))
+
{2} L_3 \(({a(k)-{1 \over 3} b(k) \over r_0}\)) \)) \\ 
-
\frac12 \int d^3r g(r) \sum_{\mu\nu} {v_{\mu\nu}(r)^2 \over r_0^2}
\label{chain}
\eea
where the function $L_3$ is defined as:
\begin{equation} \label{L3}
L_3(x)=\log(1-x)+x+{x^2/2} \ . 
\end{equation}

\subsection*{The hypernetted chain approximation}\label{HNC} 

We derive here the form of the HNC free energy (\ref{Ghnc}) for our molecular replicated system.  One could 
use the standard diagrammatic method \cite{HanMc}, but here we shall follow the 'cavity' like method of Percus 
\cite{percus}.  

In the phase where the replica symmetry is broken, replicas are correlated: it is convenient to 
consider 
$N$ molecules with coordinates $\x_i, i \in \{ 1,..,N\} $.  Each $\x_i$ stands for the coordinates of all atoms in 
molecule $i$: $x_i=\{ x_i^a \},a \in \{ 1,...,m\} $.  The energy of the system is given by
\begin{equation}
H=\sum_{i<j} V(\x_i,\x_j) + \sum_i U\x_i)
\end{equation}
where  $V(\x,\y)=\sum_a v(x^a-y^a)$, $v$ is the intermolecular potential and the external potential $U(\x)=\sum_a 
u(x^a)$ has been introduced for future use.

We shall need the following definitions.  The one molecule density is
\begin{equation}
\rho(\x)=\sum_i \la \prod_a \delta (x_i^a-x^a) \ra \ ,
\end{equation}
where the average $\la \cdot \ra$ is done with respect to the Boltzmann measure 
$\exp(-\beta H)$.

The two molecules correlation ($g$) is defined as:
\begin{equation}
\rho^{(2)}(\x,\y)=\sum_{i \ne j} \la \prod_a \delta (x_i^a-x^a)
 \prod_b \delta (x_i^b-x^b)\ra \equiv \rho(\x) g(\x,\y) \rho(\y)
 \end{equation}
where we have also defined the pair correlation function $g(\x,\y)$, that goes to one at large (center of mass) 
distance.  
The connected pair correlation is:
\begin{equation}
h(\x,\y) \equiv g(\x,\y)-1\ .
\end{equation}
Functional differentiation gives:
\begin{equation}
{\partial \rho(\x) \over \partial (-\beta U(\y))}=
\rho(\x) \delta(\x-\y) + \rho(\x) h(\x,\y) \rho(\y)
\end{equation}

One can also introduce the direct correlation function $c(\x,\y)$ through:
\begin{equation}
{\partial (-\beta U(\x)) \over \partial \rho(\y)} = {1 \over \rho(\x)}
\delta(\x-\y) 
-c(\x,\y) \ .
\end{equation}
The direct correlation is related to the connected pair correlation through
the Ornstein-Zernike equation $c=(1+h\rho)^{-1}h$ that reads more explicitly:
\bea
c(\x,\y)=h(\x,\y)+\int d\x_1 h(\x,\x_1)\rho(\x_1) h(\x_1,\y) \\
+\int d\x_1 d\x_2h(\x,\x_1)\rho(\x_1) h(\x_1,\x_2) \rho(\x_2) h(\x_2,y)+...
\label{ch_rel}
\eea

The idea of Percus is to compute the pair correlation by considering the one point density with a molecule fixed at one 
point.  Let us consider a problem where we have added one extra molecule, fixed at a point $\z=\{ z^1,...,z^m \} $.  
This extra molecule creates an external potential $U(\x)=V(\x,\z)$.  The one point density in the presence of this 
external potential, $\rho_U(\x)$, is related to the density $\rho(\x)$ and pair correlation $g(\x,\z)$ in the absence of 
an external potential through the conditional probability equation:
\begin{equation}
\rho_U(\x)= \rho(x) g(\x,\z) \ .
\end{equation}

The  previous equations can be used to find a perturbative expansion of the logarithm of the correlation function. 
Doing the appropriate computations one find:
\begin{equation}
\log g(\x,\z) + \beta V(\x,\z) = \int dy\  c(\x,\y) \rho(y) h(\y,\z) \ .
\label{hnc}
\end{equation}
Together with the inversion relation (\ref{ch_rel}), this defines a closed set of equations for the one and two point 
molecular densities that are the HNC closure.  In the real word the HNC equations are not exact: there are 
corrections to the r.h.s. \footnote{The first correction is proportional to $h^{5}$ and the proof of this statement can 
be done most easily using a diagrammatical approach; it is remarkable that a first order computation in $h$ may be so 
accurate.}.

In a similar way one finds that the free energy in the HNC approximation 
is a functional of the molecular density $\rho(\x)$ and the two point
correlation $g(\x,\y)$.  The 
result 
is:
\bea \nn
{\beta \psi } &= &{1 \over 2 m} \int d\x d\y \rho(\x) \rho(\y)
\[[ g(\x,\y) \log g(\x,\y)-g(\x,\y)+1+\beta v(\x,\y) g(\x,\y)\]]\\ 
&-&{1 \over 2 m} \Tr \(( \log(1+h \rho)-h \rho+{1 \over 2} h \rho h \rho \)) 
+{ 1 \over m} \int d\x \rho(\x) \log { \rho(\x) \over e}
\label{Ghnc}
\eea
In the trace term all products are
\emph{convolutions} \footnote{The trace could also be written as $\Tr (L_{3}(- h \rho)$, where $L_{3}$ is defined in eq.
(\ref{L3}).}.. 
For
instance the lowest order term in the small $\rho$ expansion of the trace is:
\begin{equation}
-{1 \over 3} \int d \x \  d\y \  d\z \ 
h(\x,\y)\rho(\y) h(\y,\z) \rho(\z) h(\z,\x) \rho(\x)
\end{equation}

We would like to optimize the thermodynamic potential $\psi$ with respect to the molecular density $\rho(\x)$ and the 
two point function $g(\x,\y)$.  We shall work at low temperatures for which $\rho$ should be nearly Gaussian.  We thus 
choose an Ansatz for $\rho$ of the type:
\bea
\rho(x)&=& \int d^3X \prod_{a=1}^m \(({\exp\((-(x^a-X)^2/(2 A)\)) \over 
(2 \pi A^{3/2}} \)) \\
&=& 
\(({ 2 \pi A \over m}\))^{3/2}
\(({ 2 \pi A }\))^{-3/2m}
\exp\((-{1 \over 4 A m} \sum_{ab} (x^a-x^b)^2\))
\label{density}
\eea
where the molecular density is parametrized by the single parameter $A$

The ideal gas contribution (last term in (\ref{Ghnc}) gives:
\begin{equation}
\int \prod_a d^3 x^a \rho(x) \log {\rho(x) \over e}= N \(( {3 \over 2} (1-m) 
\log(2 \pi A)
+{3\over 2} (1-m) -{3\over 2} \log m -1 \))
\end{equation}

The interaction term is more complicated, and it can be evaluated in the small cage regime.  At the end of the day one 
can compute the correlation function $g$ in the limit of small cage radius $A$, expanding in powers of $A$.  In this one 
we recover the zeroth and the first order of the $A$ expansion that we have obtained by a direct method.  In the same 
way after a long computation \cite{MePa1} we obtain the second order in $A$ that has been used in the main text.


\begin{thebibliography}{99}
		\begin{footnotesize}
\bi{EA}  S. F. Edwards and  P. W. Anderson {\sl Theory of Spin Glasses} J. Phys. F: Metal. Phys.
{\bf 5} 965 (1975).
		    
\bibitem{mpv} M.M\'ezard, G.Parisi and M.A.Virasoro, {\sl Spin glass theory and beyond}, World Scientific (Singapore 
1987).
\bibitem{BOOK} E. Marinari, G. Parisi and J. J. Ruiz-Lorenzo.  {\sl Numerical Simulations of
Spin Glass Systems} in {\sl Spin Glasses and Random Fields}, edited by P. Young.  Word Scientific 
(Singapore 1997).

\bi{CINQUE} E. Marinari, G. Parisi, F. Ricci-Tersenghi, J. Ruiz-Lorenzo and F. Zuliani, J. Stat.  Phys.  
{\bf 98}, 973 (2000).

\bi{AdGibbs} G. Adams and J.H. Gibbs J.Chem.Phys {\bf 43} (1965) 139; J.H. Gibbs and E.A. Di Marzio, J.Chem.Phys.  {\bf 
28} (1958) 373.  M.~Goldstein, J. Chem.  Phys.  {\bf 51}, 3728 (1969); G.~Adam and J.H.~Gibbs, J. Chem.  Phys.  {\bf 
57}, 470 (1972); F.H.~Stillinger and T.A.~Weber, Phys.  Rev.  A {\bf 28}, 2408 (1983).  F.H.~Stillinger, Science {\bf 
267}, 1935 (1995); C.A.~Angell, Nature {\bf 393}, 521 (1998).

\bi{St} F.H. Stillinger and T.A. Weber, {\em Phys.  Rev.} {\bf A 25}, 2408 (1982),F.H.~Stillinger and T.A.~Weber, Phys.  
Rev.  A {\bf 28}, 2408 (1983), F.H. Stillinger, {\em Science} {\bf 267}, 1935 (1995) C.A.~Angell, Nature {\bf 393}, 521 
(1998).

\bi{KTW} T. R. Kirkpatrick and D. Thirumalai, Phys.  Rev.  {\bf B36} (1987) 5388 ; T. R. Kirkpatrick and P. G. Wolynes, 
Phys.  Rev.  {\bf B36} (1987) 8552; R. Kirkpatrick and D. Thirumalai, Phys.  Rev.  Lett.  {\bf 58}, 2091 (1987),T. R. 
Kirkpatrick, D. Thirumalai and P.G. Wolynes, Phys.  Rev.  {\bf A40}, 1045 (1989).  A review of the results of these 
authors and further references can be found in T. R. Kirkpatrick and D. Thirumalai Transp.  Theor.  Stat.  Phys.  {\bf 
24} (1995) 927.

\bi{kauzman} A.W. Kauzman, Chem.Rev {\bf 43} (1948) 219.

\bi{CuKu} L.F. Cugliandolo and J. Kurchan, { Phys.  Rev.  Lett.} { 71}, 173 (1993); { J. 
Phys.  A: Math.  Gen.} {\ 27}, 5749 (1994).                                                                                      %
                                                                                                      %
\bi{FM} S.  Franz and M.  M\'ezard Europhys.  Lett.  {\bf 26}, 209 (1994).

\bi{pspin} A. Crisanti, H. J. Sommers, Z. Phys.  B {\bf 87} (1992) 341.  A. Crisanti, H. Horner and H. J. Sommers, Z. 
Phys.  {\bf B92} (1993) 257.

\bi{MCT} T. Geszti, J.Phys.  C {\bf 16}(1983) 5805; E.Leutheusser, Phys.Rev.  A {\bf 29} (1984) 2765; U.Bendtzelius, 
W.G\"otze and A. Sj\"olander, J. Phys.  C{\bf 17} (1984) 5915; E. Leutheusser, Phys.  Rev.{\bf A29}, 2765 (1984); T. R. 
Kirkpatrick, Phys.  Rev {\bf A31}, 939 (1985); W. Gotze and L. Sjogren, Rep.  Prog.  Phys.  {\bf 55}, 241 (1992), W. 
Gotze, {\em Liquid, freezing and the Glass transition}, Les Houches (1989), J. P. Hansen, D. Levesque, J. Zinn-Justin 
editors, North Holland. 

\bibitem{MPR}E.Marinari, G.Parisi and F.Ritort, J.Phys.A (Math.Gen.)  {\bf 27} (1994), 7615; J.Phys.A (Math.Gen.)  {\bf 
27} (1994), 7647.

\bi{METAISI}E.N.M. Cirillo and  J.L. Lebowitz, Journ. Stat. Phys. {\bf 90}, 211 (1998). and refences therin.

\bi{PARISISTAT} G.  Parisi, {\it Statistical Field Theory}, Addison Wesley (1988).

\bi{POLI} L. C. E. Struik; {\it Physical aging in amorphous polymers and other
 materials} (Elsevier, Houston 1978).

\bi{B} J.-P. Bouchaud, J. Phys. France {\bf 2} 1705, (1992). 

\bi{BCKM} J.-P. Bouchaud, L. Cugliandolo, J. Kurchan., M M\'e\-zard, Physica A {\bf 226}, 243 (1996), in \emph{Spin 
glasses and random fields}, A.P.Young editor, Worlds Scientific 1998.

\bi{NICOLA} N. Cabibbo, private comunication (1979).

\bibitem{REM} B. Derrida,  Phys. Rev. {\bf B24} (1981) 2613.

\bibitem{FPGS} J.P. Bouchaud and M. M\'ezard J. Phys.  A \textbf{30} 7997 (1997); S.Franz and G.Parisi, 
Eur. Phys. J. \textbf{B 18}, 485 (2000)

\bibitem{parisibook2} G.Parisi, {\sl Field Theory, Disorder and Simulations}, World Scientific, (Singapore 1992).

\bi{KPVI} J. Kurchan, G. Parisi and M.A. Virasoro, J. Physique {\bf 3}, 18 (1993).

\bi{CPV} M. Campellone, G, Parisi, M. A. Virasoro, (in preparation).

\bibitem{GROMEZ} D. J. Gross and M. Mezard, Nucl.  Phys.  {\bf B240} 
(1984) 431.

\bibitem{GARDNER} E. Gardner, Nucl.  Phys.  {\bf B257} (1985) 747.

\bi{GUERRANUOVO} F. Guerra \emph{Broken Replica Symmetry Bounds in the Mean Field Spin Glass Model}, cond-mat/0205123.

\bi{SK} D. Sherrington and S. Kirkpatrick, Phys.  Rev.  Lett.  {\bf 35} (1975) 1792.

\bibitem{PAR1} G. Parisi, Physica Scripta {\bf 35}, 123 (1987).

\bibitem{PAR2} G. Parisi, Phil. Mag. B  {\bf 71}, 471 (1995)

\bibitem{KASTLE} R. Haag and D. Kastler, J. Math.  Phys.  {\bf 5}, 848 (1964).

\bibitem{KASROB} D. Kastler and D. W. Roberts, Comm.  Math.  Phys.  {\bf 3}, 151 (1965).
  
\bibitem{RUELLE} D. Ruelle, {\em Statistical Mechanics} (Benjamin, Reading 1969).
  
\bi{EXP} C. Djurberg, K. Jonason and P. Nordblad, {\sl Magnetic Relaxation Phenomena 
in a CuMn Spin Glass}, cond-mat/9810314.

\bibitem{MP12} M. M\'ezard and G. Parisi: Eur.Phys.  J. B {\bf 20} (2001) 217; {\it `The cavity method at zero 
temperature'}, cond-mat/0207121 (2002).

\bi{TAP} D.~J. Thouless, P.~W. Anderson, and R.~G. Palmer, Phil.  Mag.  {\bf 35}, 593 (1977).

\bi{KLEIN} G.  Parisi, in {\sl The Oskar Klein Centenary}, ed.  by U.  Lindstr\"{o}m, World 
Scientific, (1995), Il nuovo cimento {\bf 16}, 939 (1994).

\bibitem{FP} S.Franz and G.Parisi, J. Phys.  {\bf I } (France) {\bf 5}(1995) 1401.

\bi{FRAPA}  S.Franz and G. Parisi, Phys.  Rev.  Letters {\bf 79}, 2486 (1997) and  
Physica A, {\bf 6}1,317 (1998).

\bi{MONA} E. Monasson Phys. Rev. Lett. {\bf 75} (1995) 2847. 

\bi{ACP} For a careful analysis of the free energy landscape see A.  Cavagna, I.  Giardina and
G.  Parisi, J.  Phys.  A: Math.  Gen.  1997, {\bf 30}, 7021 and references therein.

\bi{BAFRPA}A.~Barrat, R.~Burioni, and M.~M\'ezard, J. Phys.  A { 29}, L81 (1996),
A. Barrat,  S. Franz and  G. Parisi  { J. Phys. A: Math. Gen.} {\ 30}, 5593 (1997).                                      

\bi{LJ}B. Coluzzi, G. Parisi and  P. Verrocchio {\em Phys. Rev. Lett.} 
{\bf 84} 306 (2000),
{\em J. Chem. Phys.} {\bf 112} 2933 (2000); 

\bi{PP} G. Parisi and M. Potters, J. Phys. A: Math. Gen. {\bf 28} ( 1995) 5267,
 Europhys. Lett. {\bf 32} (1995) 13.

 \bi{Me} M. M\'ezard, Physica A {\bf 265} 352 (1999).
 
\bi{CGG}A. Cavagna, J. P. Garrahan and  I. Giardina J. Phys.\textbf{ A 32}, 711 (1999).

\bi{CGMP} A. Cavagna, I. Giardina, M. Mezard and G. Parisi,\emph{ On the formal equivalence of the TAP and thermodynamic 
 methods in the SK model}, cond-mat/0210665
 
\bi{CGGP} A. Cavagna, I.Giardina, T. Grigera and G. Parisi, Phys. Rev. Lett. {\bf 88}, 055502 (2002).

\bibitem{VF} H. Vogel, Phys.  Z, {\bf 22}, 645 (1921); G.S. Fulcher, J. Am.  Ceram.  Soc., {\bf 
 6}, 339 (1925).

\bi{JL} J.Kurchan and L.  Laloux, J. Phys. A {\bf 29}, 1929 (1996).

\bi{BAPA} B.  Coluzzi and G.  Parisi {\sl On the Approach to the Equilibrium and
the Equilibrium Properties of a Glass-Forming Model}, cond-mat/9712261.

\bi{FRARIE} S. Franz and H. Rieger Phys.  J. Stat. Phys.  {\bf 79} 749 (1995).

\bibitem{MPRR}E. Marinari, G. Parisi, F. Ricci-Tersenghi andJ. J. Ruiz-Lorenzo, J. Phys.  A: Math.  Gen.  {\bf 31}, 2611 
(1998).

\bi{FDT} G. Parisi, Phys. Rev. Lett. { 79}, 3660 (1997).     

\bi{BK} W. Kob and J.-L. Barrat, Phys.Rev.Lett. {\bf 79} (1997) 3660;J.-L. Barrat and W. Kob, Europhys. Lett. 46, 637 (1999).

\bi{TAPSTA}A. Cavagna, I. Giardina and G. Parisi, Phys. Rev.\textbf{ B 57}, 11251 (1998).
     
\bi{Biroli}  G. Biroli, J. Phys. Math. Gen.  {\bf 32} 8365  (1999).

\bibitem{Metha91} M.\ L.\ Metha, {\em Random matrices}, Academic Press (1991). 

\bi{FV} S. Franz and M.A. Virasoro, J. Phys.  A: Math.  and Gen.  {\bf 33}, 891 (2000).

\bi{INM1} See for instance T. Keyes, J.Phys.Chem.  A {\bf 101} (1997) 2921, and references therein.

\bibitem{water}S. Sastry, Phys.  Rev.  Lett.  {\bf 76}, 3738 (1996); F. Sciortino and P. Tartaglia, Phys.  Rev.  Lett.  
{\bf 78}, 2385 (1997); La Nave {\it et al,} Phys.  Rev.  Lett.  {\bf 84}, 4605 (2000).

\bi{C}A. Cavagna, Europhys. Lett. {\bf 53}, 490 (2001).

\bi{ab}L. Angelani, R. Di Leonardo, G. Ruocco, A. Scala, and F. Sciortino, Phys.  Rev.  Lett.  {\bf 85}, 5356 (2000) K. 
Broderix, K. K. Bhattacharya, A. Cavagna, A. Zippelius, and I. Giardina, Phys.  Rev.  Lett.  {\bf 85}, 5360 (2000),

\bi{CGP} A. Cavagna, I.Giardina and G. Parisi, J. Phys. A: Math. Gen. {\bf 34}, 5317 (2001).

\bi{BOSE} C. Masciovecchio et al., Phys.  Rev.  Lett.  {\bf 76,} 3356 (1996); F. Sette et al., Science {\bf 280,} 1550 
(1998),A. Matic et al., Europhys.  Lett.  {\bf 54,} 77 (2001), C. Masciovecchio et al., Phys.  Rev.  B {\bf 55,} 8049 
(1997); P. Benassi et al., Phys.  Rev.  Lett.  {\bf 77,} 3835 (1996), D. Fioretto et al., Phys.  Rev.  E {\bf 59,} 1470 
(1999), A. P. Sokolov et al., Phys.  Rev.  B {\bf 52,} R9815 (1995), N. J. Tao et al., Phys.  Rev.  A {\bf 44,} 6665 
(1991).

\bibitem{BOSESIM} J. Horbach et al., Eur.  Phys.  J. B {\bf 19,} 531 (2001), J. Horbach et al.  J. Phys.  Chem.  B {\bf 
103,} 4104 (1999), S. N. Taraskin and S. R. Elliot, Phys.  Rev.  B {\bf 59,} 8572 (1999), G. Ruocco et al., Phys.  Rev.  
Lett.  {\bf 84,} 5788 (2000).

\bibitem{Gotze00} W. G\"otze and M. R. Mayr, Phys.  Rev.  E {\bf 61}, 587 (2000).

\bibitem{GMPV} V. Martin-Mayor, G. Parisi and P. Verrocchio, J. Chem.  Phys.  {\bf 114,} 8068 (2001); T. S. Grigera, V. 
Martin-Mayor, G. Parisi and P. Verrocchio., J.Phys. A {\bf 14} (2002) 2167-2179, Phys.  Rev.  Lett.  {\bf 87} 
085502 (2001).

\bi{FrMePaPe} S. Franz, M. M\'ezard, G. Parisi and L. Peliti, Phys.  Rev.  Lett.  { 81} 1758 
(1998), J. Stat. Phys. {\bf 97} 459 (1999).

\bi{GUERRA} F.  Guerra, Int.  J.  Phys.  B, {\bf 10}, 1675 (1997).
 
\bi{AI} M.  Aizenman and P.  Contucci, J. Stat. Phys {bf 92}, 765 (1998).

\bibitem{GG}S. Ghirlanda and F. Guerra, J. Phys.  A: Math.  Gen.  {\bf 31} 9149 (1998).

\bi{SOL} G.  Parisi, {\sl On the probabilistic formulation of the replica approach to spin glasses},
 cond-mat/9801081.

\bi{glass_revue}  Reviews can be found in: C.A. Angell, Science, {\bf 267}, 1924 (1995) and P.De Benedetti, 
`Metastable liquids', Princeton University Press (1997).  An introduction to the theory is: J.J\"ackle, Rep.Prog.  Phys.  
{\bf 49} (1986) 171.

\bi{KoAn} W. Kob and H.C. Andersen, {\em Phys.  Rev.  Lett.} {\bf 73}, 1376 (1994).  

\bi{corr_length} G.Parisi, \emph{A divergent correlation length in off-equilibrium glasses} cond-mat/9801034, C. Donati, 
S.C. Glotzer, P.H. Poole, Phys.  Rev.  Lett.  {\bf 80}, 4915 (1998), Franz and G. Parisi, J. Phys.: Condens.  Matter 12 
(2000) 6335,C. Donati, S. Franz, G. Parisi, S.C. Glotzer, Phil.  Mag.  B {\bf 79} 1827 (1999)

\bi{FH}S. Franz, J.Hertz,  Phys. Rev. Lett. {\bf 74}, (1995) 2114.

\bibitem{CKPR} L.Cugliandolo, J.Kurchan, G.Parisi and F.Ritort,
Phys. Rev. Lett. {\bf 74} (1995) 1012.

\bibitem{CKMP} L.Cugliandolo, J.Kurchan, E.Monasson and G.Parisi, Math.  Gen.  {\bf 29} (1996) 1347.

\bi {MP}M. Mezard, G. Parisi, J. Phys. A {\bf 29}, (1996) 6515.

\bi{MePa1} M. M\'ezard and G. Parisi, {\em Phys.  Rev.  Lett.} {\bf 82} 747 (1998); M. M\'ezard and G. Parisi, {\em J. 
Chem.  Phys.} {\bf 111 No.3}, 1076 (1999).

\bi{landscape} For example see S.Sastry, P. Debenedetti, F.H.Stillinger, { Nature} { 393} 554 
(1998); F. Sciortino and P. Tartaglia, {\em Phys. Rev. Lett.}
{\bf 78}, 2385 (1997);  K. K. Bhattacharya, K. Broderix, R. Kree, A. Zippelius, cond-mat/9903120;  L. Angelani, G. 
Parisi, G. Ruocco and G. Viliani, { Phys.  Rev.  Lett.} { 81} 4648 (1998)

\bi{KST} W. Kob, F. Sciortino, P. Tartaglia, Phys. Rev. Lett. 83, 3214 (1999), W. Kob, F. Sciortino, P. Tartaglia, Europhys. Lett. 49, 590 (2000).

\bi{APRV} L. Angelani, G. Parisi, G. Ruocco and G. Viliani Phys. Rev. Lett., \textbf{81} 4648 (1998).

\bi{HP} G. Iori, E. Marinari and G. Parisi, J. Phys.  {\bf A} (Math.  Gen.)  {\bf 24} (1992) 5349, M. Fukugita, D. 
Lancaster and M. G. Mitchard, J. Phys.  Lett.  {\bf A} (Math.  Gen.)  {\bf 25} (1992) L121.

\bi{HANSEN} B.Bernu, J.-P. Hansen, Y. Hitawari and G. Pastore, Phys.  Rev.  {\bf A36} 4891 (1987), J.-L. Barrat, J-N. 
Roux and J.-P. Hansen, Chem.  Phys.  {\bf 149}, 197 (1990),J.-P. Hansen and S. Yip, Trans.  Theory and Stat.  Phys.  
{\bf 24}, 1149 (1995).
  
\bi{BIN}  B. Coluzzi, M. M\'ezard,  G. Parisi and P. Verrocchio J.Chem.Phys. {\bf 111} 9039 (1999).

\bi{Ro} Y. Rosenfeld and P. Tarazona, {\em Molecular Physics} {\bf  95}, 141 (1998).
  
\bibitem{ZeHa}G.Zerah and J.P. Hansen, {\em J. Chem. Phys.} {\bf 84} (4), 2336 (1986).

\bibitem{Sastry}S. Sastry,  Phys. Rev. Lett., \textbf{85}, 590, (2000).

\bi{BPV} B. Coluzzi, G. Parisi and P. Verrocchio, cond-mat/0007144.

\bibitem{mpv_free} M. M\'ezard, G. Parisi and M. A. Virasoro, J. Physique Lett.  {\bf 46} L21 (1985).

\bi{RUELLETREE} D. Ruelle, Commun. Math. Phys. {\bf 48}, 351 (1988).

\bi{MEPAZEE} M. M\'ezard, G. Parisi and A. Zee{\em Nuc. Phys. B} {\bf 559,} 689(1999),
V. Martin-Mayor, M. Mezard, G. Parisi, P. Verrocchio J. Chem. Phys., {\bf 114} (2001) 8068

\bi{CAGIAPA} A. Cavagna, I. Giardina and G. Parisi, {\em Phys.  Rev.  Lett.} {\bf 83}, 108 (1999).

\bi{HanMc} See for instance J.P. Hansen and I.R. Macdonald, "Theory of simple liquids", (Academic, London, 1986), or 
H.N.V. Temperley, J.S. Rowlinson and G.S. Rushbrooke, "Physics of simple liquids", NorthHolland (Amsterdam 1968).

\bi{percus} J.K. Percus, in {\it The Equilibrium Theory of Classical Fluids},
ed. H.L. Frisch and J.L. Lebowitz (New York: Benjamin; 1964).

\end{footnotesize}
\end{thebibliography}
\end{document}